\documentclass[longauth]{aa}  

\usepackage{natbib}
\bibpunct{(}{)}{;}{a}{}{,} 
\usepackage{graphicx}
\usepackage{subfigure}
\usepackage[colorlinks=True,allcolors=blue]{hyperref}
\usepackage{txfonts}
\usepackage[flushleft]{threeparttable}

\begin{document}

   \title{The ALPINE-ALMA [CII] survey: The contribution of major mergers to the galaxy mass assembly at z$\sim$5}
   \authorrunning{M. Romano et al.}
   \titlerunning{The contribution of major mergers to the galaxy mass assembly at $z\sim5$}

\author{M. Romano\thanks{E-mail: michael.romano@studenti.unipd.it}\inst{1,2}
\and
P. Cassata\inst{1,2}
\and
L. Morselli\inst{3}
\and
G. C. Jones\inst{4,5}
\and
M. Ginolfi\inst{6}
\and
A. Zanella\inst{2}
\and
M. B\'ethermin\inst{7}
\and
P. Capak\inst{8}
\and\\
A. Faisst\inst{9}
\and
O. Le Fèvre\thanks{Passed away.}\inst{7}
\and
D. Schaerer\inst{10}
\and
J. D. Silverman\inst{11,12}
\and
L. Yan\inst{13}
\and
S. Bardelli\inst{14}
\and
M. Boquien\inst{15}
\and
A. Cimatti\inst{16,17}
\and\\
M. Dessauges-Zavadsky\inst{10}
\and
A. Enia\inst{14,16}
\and
S. Fujimoto\inst{18,19}
\and
C. Gruppioni\inst{14}
\and
N. P. Hathi\inst{20}
\and
E. Ibar\inst{21}
\and
A. M. Koekemoer\inst{20}
\and
B. C. Lemaux\inst{22}
\and
G. Rodighiero\inst{1,2}
\and
D. Vergani\inst{14}
\and
G. Zamorani\inst{14}
\and
E. Zucca\inst{14}
}


\institute{Dipartimento di Fisica e Astronomia, Universit\`a di Padova, Vicolo dell'Osservatorio 3, I-35122, Padova, Italy
\and
INAF - Osservatorio Astronomico di Padova, Vicolo dell'Osservatorio 5, I-35122, Padova, Italy
\and
Associazione Big Data, via Piero Gobetti 101, 40129 Bologna, Italy
\and
Cavendish Laboratory, University of Cambridge, 19 J. J. Thomson Ave., Cambridge CB3 0HE, UK
\and
Kavli Institute for Cosmology, University of Cambridge, Madingley Road, Cambridge CB3 0HA, UK
\and
European Southern Observatory, Karl-Schwarzschild-Strasse 2, 85748, Garching, Germany
\and
Aix Marseille Univ, CNRS, CNES, LAM, Marseille, France
\and
Infrared Processing and Analysis Center, California Institute of Technology, Pasadena, CA 91125, USA
\and
IPAC, California Institute of Technology, 1200 East California Boulevard, Pasadena, CA 91125, USA
\and
Observatoire de Genève, Universit\'e de Genève, 51 Ch. des Maillettes, 1290 Versoix, Switzerland
\and
Kavli Institute for the Physics and Mathematics of the Universe, The University of Tokyo Kashiwa, Chiba 277-8583, Japan
\and
Department of Astronomy, School of Science, The University of Tokyo, 7-3-1 Hongo, Bunkyo, Tokyo 113-0033, Japan
\and
The Caltech Optical Observatories, California Institute of Technology, Pasadena, CA 91125, USA
\and
INAF - Osservatorio di Astrofisica e Scienza dello Spazio, via Gobetti 93/3, I-40129 Bologna, Italy
\and
Centro de Astronom\'ia (CITEVA), Universidad de Antofagasta, Avenida Angamos 601, Antofagasta, Chile
\and
University of Bologna, Department of Physics and Astronomy "Augusto Righi", Via Gobetti 93/2, I-40129, Bologna, Italy
\and
INAF - Osservatorio Astrofisico di Arcetri, Largo E. Fermi 5, I-50125, Firenze, Italy
\and
Cosmic Dawn Center (DAWN), Jagtvej 128, DK22000 Copenhagen N, Denmark
\and
Niels Bohr Institute, University of Copenhagen, Lyngbyvej 2, DK2100 Copenhagen \O, Denmark
\and
Space Telescope Science Institute, 3700 San Martin Drive, Baltimore, MD 21218, USA
\and
Instituto de Física y Astronomía, Universidad de Valparaíso, Avda. Gran Bretaña 1111, Valparaíso, Chile
\and
Department of Physics and Astronomy, University of California, Davis, One Shields Ave., Davis, CA 95616, USA
}

   \date{Accepted July 22, 2021}

 
  \abstract
   {Galaxy mergers are thought to be one of the main mechanisms of the mass assembly of galaxies in the Universe, but there is still little direct observational evidence of how frequent they are at $z\gtrsim4$. Recently, many works have suggested a possible increase in the fraction of major mergers in the early Universe, reviving the debate on which processes (e.g., cold accretion, star formation, mergers) most contribute to the mass build-up of galaxies through cosmic time.}
   {To estimate the importance of major mergers in this context, we make use of the new data collected by the ALMA Large Program to INvestigate [CII] at Early times (ALPINE) survey, which attempted to observe the [CII]~158~$\mu$m emission line from a sample of 75 main-sequence star-forming galaxies at $4.4<z<5.9$.}
   {We used, for the first time, the morpho-kinematic information provided by the [CII] emission, along with archival multiwavelength data to obtain the fraction of major mergers ($f_{\text{MM}}$) at $z\sim5$. By combining the results from ALPINE with those at lower redshifts from the literature, we also studied the evolution of the merger fraction through cosmic time. We then used different redshift-evolving merger timescales ($T_{\text{MM}}$) to convert this fraction into the merger rate per galaxy ($R_{\text{MM}}$) and in the volume-averaged merger rate ($\Gamma_{\text{MM}}$).}
   {We find a merger fraction of $f_{\text{MM}}\sim0.44$ (0.34) at $z\sim4.5$ (5.5) from ALPINE. By combining our results with those at lower redshifts, we computed the cosmic evolution of the merger fraction which is described by a rapid increase from the local Universe to higher redshifts, a peak at $z\sim3$, and a slow decrease toward earlier epochs. Depending on the timescale prescription used, this fraction translates into a merger rate ranging between $\sim0.1$ and $\sim4.0$~Gyr$^{-1}$ at $z\sim5$, which in turn corresponds to an average number of major mergers per galaxy between 1 and 8 in $\sim12.5$~Gyr (from $z=6$ to the local Universe). When convolved with the galaxy number density at different epochs, the merger rate density becomes approximately constant over time at $1<z<4$, including values from $10^{-4}$ to $10^{-3}$~Gyr$^{-1}$~Mpc$^{-3}$, depending on the assumed $T_{\text{MM}}$. We finally compare the specific star formation and star-formation rate density
   with the analogous quantities from major mergers, finding a good agreement at $z>4$ if we assume a merger timescale that quickly decreases with increasing redshift.}
   {Our new constraints on the merger fraction from the ALPINE survey at $z\sim5$ reveal the presence of a significant merging activity in the early Universe. Whether this population of mergers can provide a relevant contribution to the galaxy mass assembly at these redshifts and through the cosmic epochs is strongly dependent on the assumption of the merger timescale. However, our results show that an evolving $T_{\text{MM}}\propto(1+z)^{-2}$ agrees well with state-of-the-art cosmological simulations, suggesting a considerable role of mergers in the build-up of galaxies at early times.}


   \maketitle


\section{Introduction}
How galaxies grow their stellar mass through cosmic time is still one of the most puzzling questions of modern cosmology. During the last decades, two major physical mechanisms have been proposed to contribute or dominate the build-up of galaxies at different epochs of the Universe: the accretion of cold gas (both from the internal reservoir or from the outer environment of galaxies) and the merging of galaxies. 

The first process requires the supply of new cold gas to the galaxy from the filaments of the cosmic web, refreshing and/or enhancing the star formation (e.g., \citealt{Keres05,Dekel09}). Although this could be one of the main modes of galaxy assembly (e.g., \citealt{Keres09,Bouche10}), direct observations of gas flowing through the filaments are tricky while indirect evidence has increased over the last years \citep{Bouche13,Bouche16,Zabl19}. However, in situ star formation is observable through a variety of different proxies and up to the earliest epochs (e.g., \citealt{Madau14} and references therein). 
    
Galaxy mergers are also thought to play an important role in the stellar mass growth over cosmic time. They are relatively common through the history of the Universe and are observable with different methods and at different stages. The merging of galaxies is a natural prediction of the hierarchical structure formation model \citep{White91,Springel05,Klypin11} for which dark matter halos grow their mass and affect the build-up of galaxies in the Universe \citep{Kauffmann93,Khochfar05}. Not only can mergers increase the stellar mass of galaxies (up to a factor of 2 for major mergers, i.e., for galaxies of nearly equal stellar mass; \citealt{Lopez12,Oser12,Kaviraj14}), but they can also trigger starbursts and active galactic nuclei (AGN; \citealt{Silk98,Kartaltepe12,Chiaberge15}; see also \citealt{Shah20} who found that galaxy interactions do not significantly enhance AGN activity up to $z\sim3$), in some cases blowing the gas out from the galaxy which arrests, at least temporarily, the formation of new stars \citep{Silk98,Fabian06,Cattaneo09}. Therefore, observing mergers at different epochs is key to shed light on the relative contribution of the distinct processes that rule the assembly of galaxies in the Universe.

Depending on which phase the merger is passing through, there are two main ways to estimate the incidence of such events at a given epoch. On-going and post-mergers could lead to the formation of post-starburst (PSB) galaxies (at least in groups or clusters; e.g., \citealt{Wu14,Lemaux17}) and leave morphological or kinematic imprint on the merging or coalesced system such as tails of stripped material, irregular shapes, disturbed velocity fields, and/or deep absorption lines in the spectra of PSBs (e.g., \citealt{LeFevre00,Conselice03,Conselice08,Jogee09,Wild09,Casteels14}). However, observations of these features could be hampered by the spatial resolution needed to resolve the morphological details highlighting the presence of a merger (especially at high redshifts), or confused by diverse galaxy types that can mimic the kinematics or morphology of a merging system. 

On the other hand, galaxy pairs that are going to merge in a certain timescale can also be identified based on some selection criteria regarding their spatial and velocity separation (e.g., \citealt{LeFevre00,Patton2000,Lin2008,deRavel09,Lopez12,Ventou17,Mantha18,Duncan19}). Typically, close pairs must lie within a defined radius with a projected separation in the sky that could be as wide as 50~kpc and, in case of spectroscopic surveys, have a relative velocity not larger than $\Delta v=500$~km~s$^{-1}$ (i.e., $\Delta z\sim0.0017$) which excludes those pairs that are not gravitationally bound (e.g., \citealt{Patton2000}). 

Both these methods can provide an estimate of the merger fraction through cosmic time but, as they probe different phases and timescales of the merging, they can lead to discordant results if not interpreted properly. However, even by adopting the same observational method, a large scatter is present in the literature for the merger fraction and the corresponding merger rate at all redshifts. This could be attributed to several reasons: i) the different criteria used for selecting galaxy pairs; ii) the distinction between major and minor mergers for which a dual uncertainty has to be taken into account, both in the pair ratio which can be obtained from the stellar mass \citep{Lopez13,Tasca14,Man16,Ventou17,Duncan19} or the flux \citep{Bluck12,Man12,Man16,Lopez15}, and in the major (or minor) merger definition having a threshold in mass or flux ratio ranging between 2-6 \citep{Xu12,Ventou17,Ventou19}; iii) the merger timescale used to convert the pair fraction into a merger rate, which in turn depends on the method used for characterizing mergers, on their selection criteria, and on the properties of the galaxies undergoing a merger \citep{Kitzbichler08,Jiang14,Snyder17}.   

Several studies tried to constrain the fraction of major mergers and the merger rate out to $z\lesssim2$ (e.g., \citealt{deRavel09,Lopez11,Lopez13,Lotz11,Xu12}). During this time, the merger fraction is found to increase with redshift as $\propto (1+z)^m$, where $1\lesssim m \lesssim5$, depending on the luminosity, the stellar mass and the spectral type of galaxies, leading to a large scatter in the literature. Only a handful of works have extended these measurements to higher redshifts. For instance, \cite{Conselice09} used both pair counts and morphological studies to investigate the presence of mergers at $4<z<6$ and their contribution to the galaxy assembly, finding a possible peak in the merger fraction at $z\sim3$ followed by a decline which suggested a deviation from the power-law trend found at lower redshifts. However, the results by \cite{Conselice09} do not distinguish between major and minor mergers and could have been subject to large uncertainties due to photometric redshifts. A peak in the cosmic evolution of the merger fraction is also found at $1\lesssim z \lesssim 3$ by more recent observational studies (e.g., \citealt{Mantha18,Duncan19}) which, however, do not provide conclusive statements given the large above-mentioned differences on the selection and definition of the mergers (see for example the results by \citealt{Mantha18} for both stellar mass- or flux-selected pairs).

In this paper we take advantage of the recent data collected by the ALMA Large Program to INvestigate [CII] at Early times (ALPINE; \citealt{Bethermin20,Faisst20,LeFevre20}) to further constrain the importance of major mergers to the galaxy mass assembly at the end of the Reionization epoch and through cosmic time. This survey allows us to add a key additional piece of information to the wealth of multiwavelength data already available for each targeted galaxy thanks to the ALMA observations of the ionized carbon [CII] at 158 $\mu$m rest-frame and the surrounding continuum. The 3D (i.e., RA, Dec, velocity) information enclosed in each ALPINE data cube is of great importance for identifying morphological and kinematic disturbances highlighting the presence of possible merging components. In particular, \cite{LeFevre20} made a first fundamental step in the morpho-kinematic classification of the ALPINE galaxies, finding a high fraction ($\sim40\%$) of mergers (both minor and major) at $z\sim5$ among the ALPINE sample. An in-depth analysis of two merging systems found in ALPINE at $z\sim4.6$ was made later by \cite{Jones20} and \cite{Ginolfi20}, showing the strength of these observations in the identification and characterization of this kind of sources in the high-redshift Universe.

Previous works estimated the merger fraction by exploiting morphological information or direct pair counts obtained through rest-frame optical observations and photometric or spectroscopic redshifts, as derived from optical ionized gas tracers (e.g., \citealt{Conselice09,Xu12,Ventou17}). Here, for the first time, we make use of the rest-frame FIR [CII] line to compute the major merger fraction in the early Universe. As this line is less affected by dust extinction than optical tracers, morphological and kinematic analysis of this emission can reveal the presence of mergers with dusty components that are partially or completely obscured in the optical bands. Following these previous studies, we thus refine the fraction of mergers at the redshifts explored by the ALPINE survey and analyze in detail their [CII] emission combined with optical data for estimating their incidence at $z>4$ and, along with other values in the literature, through cosmic time.  

This paper is structured as follows: in Section 2 we describe the data used to estimate the merger fraction, and the previous works on the morpho-kinematic classification of the ALPINE galaxies on which we based our analysis. The criteria used for selecting mergers from the ALPINE sample, the methods employed for their characterization and the estimated parameters are stated in Section 3. The results on the merger fraction and merger rate, their evolution with the cosmic time and their contribution to the galaxy mass assembly are reported in Section 4. We discuss our results and summarize them in Section 5 and Section 6, respectively.

Throughout this work, we adopt a $\mathrm{\Lambda-CDM}$ cosmology with $H_0=70$~km~s$^{-1}$~Mpc$^{-1}$, $\mathrm{\Omega_m}=0.3$ and $\mathrm{\Omega_\Lambda}=0.7$. Stellar masses are normalized to a \cite{Chabrier03} initial mass function (IMF).

\section{Data}
The ALPINE survey was designed to detect the [CII] line at 158~$\mu$m rest-frame and the surrounding FIR continuum emission from a sample of 118 star-forming galaxies at $4.4<z<5.9$, avoiding the redshift range $4.6<z<5.1$ due to a low-transmission atmospheric window. The campaign took $\sim70$~hours of observation with ALMA in Band 7 (275-373~GHz) during cycles 5 and 6. The ALPINE targets are drawn from the Cosmic Evolution Survey (COSMOS; \citealt{Scoville17a,Scoville17b}) and Extended Chandra Deep Field South (E-CDFS; \citealt{Giavalisco04,Cardamone10}) fields and have secure spectroscopic redshifts from the VIMOS UltraDeep Survey (VUDS; \citealt{LeFevre15,Tasca17}) and DEIMOS 10K Spectroscopic Survey \citep{Hasinger18}. The rest-frame UV selection ($\mathrm{L_{UV}>0.6~L^{*}}$) ensures that these sources lie on the so-called main-sequence of star-forming galaxies (e.g., \citealt{Noeske07,Rodighiero11,Tasca15}), being thus representative of the underlying star-forming, UV-detected galaxy population at $z\sim5$ (e.g., \citealt{Speagle14}). Due to the location of the ALPINE targets in these well-studied fields, archival multiwavelength data are available from the UV to the near-IR (e.g., \citealt{Koekemoer07,Sanders07,McCracken12,Guo13,Nayyeri17}), reaching the X-ray and radio bands (e.g., \citealt{Hasinger07,Smolcic17}), allowing us to recover physical quantities such as stellar masses ($\mathrm{9\lesssim log(M_{*}/M_{\odot})\lesssim11}$) and star-formation rates (SFRs; $\mathrm{1\lesssim log(SFR/M_{\odot}yr^{-1})\lesssim3}$) through spectral energy distribution (SED)-fitting \citep{Faisst20}. 

The collected ALMA data were reduced and calibrated using the standard Common Astronomy Software Applications (CASA; \citealt{McMullin07}) pipeline and each data cube was continuum-subtracted to obtain line-only cubes with channel width of $\sim25~\mathrm{km~s^{-1}}$ and beam size $\sim1''$ (with a pixel scale of $\sim0.15''$; \citealt{Bethermin20}). A line search algorithm was then applied to each continuum-subtracted cube resulting in 75 [CII] detections (S/N$>$3.5) out of 118 ALPINE targets (of which 23 sources were also detected in continuum).

For a more in-depth description of the overall ALPINE survey, the observation and data processing, and the multiwavelength analysis see \cite{LeFevre20}, \cite{Bethermin20} and \cite{Faisst20}, respectively.

\subsection{Previous morpho-kinematic classifications}\label{sec:initial_class}
\cite{LeFevre20} performed a preliminary morphological and kinematic classification of the 75 ALPINE targets showing [CII] emission. In particular, the channel maps around the emission line, the intensity (moment-0) and velocity (moment-1) maps, the integrated [CII] spectra, the position-velocity diagrams (PVDs) along the major and minor axis of velocity maps, and the multiwavelength data at the position of the source were visually inspected by an internal team of the ALPINE collaboration that placed these galaxies in the following classes:
\begin{enumerate}
\item \textit{Rotators}. The galaxies in this class are characterized by a single source in the ancillary data and show a clear velocity gradient in the moment-1 maps which then implies a tilted (straight) PVD along the major (minor) axis. They can also exhibit a double-horned profile in the integrated spectra.
\item \textit{Mergers}. In case of interacting systems the presence of two or more components in the optical images within the ALMA field of view and/or in the moment-0 maps and PVDs is expected. A complex behavior in the channel maps and in the [CII] spectra could be visible.
\item \textit{Extended Dispersion Dominated}. These sources extend in [CII] emission over multiple ALMA beams and are typically characterized by straight PVDs along the major and minor axis and by a Gaussian line profile.
\item \textit{Compact Dispersion Dominated}. As opposed to the class 3 galaxies, these objects are unresolved in the moment-0 maps.
\item \textit{Weak}. These targets are too weak to be visually classified into one of the above classes. 
\end{enumerate}

\cite{LeFevre20} found that $\sim40\%$ of the ALPINE galaxies are undergoing a merging phase and that $\sim13\%$ of the targets are possibly forming an ordered disk at $z\sim5$. Extended and compact sources make up 20$\%$ and 11$\%$ of the sample, respectively, with the remaining galaxies (16$\%$) being too complex or weak to be classified. These results possibly suggest a large contribution of the mergers to the galaxy mass assembly at $z\sim5$. However, such a qualitative classification could be prone to different kinds of uncertainties arising from the subjectivity of the method.

On the other hand, \cite{Jones21} expanded this initial classification of the ALPINE targets by using the tilted ring fitting code $\mathrm{^{3D}Barolo}$ (3D-Based Analysis of Rotating Objects from Line Observations; \citealt{DiTeodoro15}). Their selection criteria led to a successful fit of $39\%$ of the ALPINE parent sample (i.e., 29 out of 75 sources) and to a robust classification of 14 sources. This new quantitative analysis confirmed the morpho-kinematic diversity of the ALPINE galaxies found by \cite{LeFevre20} but resulted in a somewhat different statistics with $43\%$ of rotators, $36\%$ of mergers, and $21\%$ of dispersion-dominated galaxies.

As we aim to compute the contribution of mergers to the galaxy mass assembly in the early Universe, a robust estimate of the fraction of mergers at $z\sim5$ is required. We thus perform a further and in-depth morpho-kinematic investigation of the ALPINE [CII]-detected galaxies based on the same classification criteria used in \cite{LeFevre20} and on the results by \cite{Jones21}.


\section{Methods}
To examine the merger fraction at $z\sim5$, we take advantage of the continuum and [CII] emission maps of the targets from the ALPINE data release 1 (DR1; \citealt{Bethermin20})\footnote{DR1 data are available at \url{https://cesam.lam.fr/a2c2s/data_release.php}.}. Briefly, for each source the CASA task \texttt{UVCONTSUB} was used to obtain line-only cubes. The continuum was first estimated in the uv-plane by masking the spectral channels containing line emission\footnote{For a detailed description of the procedure used to estimate the range of spectral channels containing line emission for all the [CII]-detected galaxies, see Section 6.1 in \cite{Bethermin20}.} and fitting a model to the visibilities, and it was then subtracted to the data cubes. With the CASA \texttt{IMMOMENTS} task the spectral channels that included the emission line were summed up to produce the velocity-integrated intensity (moment-0) maps, that is $M_0(x,y) = \Delta v \sum_{i}I_i(x,y)$, where $I_i$ is the intensity of the $i$-th pixel at the position $(x,y)$, $\Delta v$ is the velocity width between two consecutive channels, and the sum is over all the channels containing the spectral line \citep{Bethermin20}. By masking all the pixels below $3\sigma$ (where $\sigma$ is the rms estimated in the moment-0 map), we also compute the intensity-weighted velocity (moment-1) and velocity dispersion (moment-2) maps, defined as $M_1 = \sum_{i}I_i v_i/M_0$ and $M_2 = \sqrt{\sum_{i}I_i(v_i-M_1)^2/M_0}$, respectively. It is worth noting that in the analysis by \cite{LeFevre20} the morphology and kinematic of the [CII] line were inspected within the $2\sigma$ emission contours of the moment maps. In this work, we decide to select only the pixels above the $3\sigma$ level in order to be less affected by possible spurious emissions that could alter the effective morphology and velocity field of the targets (however, we find that this choice does not significantly change our final results).

With the CASA \texttt{IMFIT} task we fit one 2D Gaussian component to each source in the moment-0 map, retrieving best-fit parameters such as the peak intensity and integrated flux density, and morphological information like the coordinates of the [CII] emission peak, the full width at half maximum (FWHM) of the major and minor axis of the Gaussian and its position angle (PA). If the [CII] morphology shows the presence of different spatially-resolved components in the vicinity of the main ALPINE target (as in case of mergers), we fit an individual 2D Gaussian to each of them in order to retrieve the above morphological and kinematic information for all the observed sub-structures. We then make use of the morphological central position provided by the 2D Gaussian best-fit on each ALPINE target\footnote{In case of multiple spatially-resolved [CII] emissions, we use the position provided by the Gaussian fit on the global system rather than the positions obtained by fitting individual 2D Gaussians to each of the observed components.} to produce the PVDs along the major and minor axis with the CASA \texttt{IMPV} task, with an averaging width of five pixels and a pseudo-slit length of $3''$ (if not otherwise stated). Conversely to the analysis made by \cite{LeFevre20} in which the major axis was aligned with the morphological PA, we decide to produce the two PVDs with the major axis oriented along the direction of major velocity gradient in the moment-1 map, and with the minor axis perpendicular to the major one. In this way, we increase the chances of identifying possible mergers or rotating galaxies in the ALPINE sample\footnote{Two merging systems with different velocity fields or a rotating galaxy with a clear velocity gradient would be ideally represented with separated components in the PVDs and with an \lq\lq S-shape\rq\rq{} in the PVD along the major axis, respectively.}.

Lastly, we extract the spatially-integrated 1D [CII] spectrum of each target within the $3\sigma$ contours at the position of the source from the moment-0 map using the CASA task \texttt{SPECFLUX}.

\begin{figure*}[t]
\begin{center}
	\subfigure{\includegraphics[width=\columnwidth,height=15.1cm]{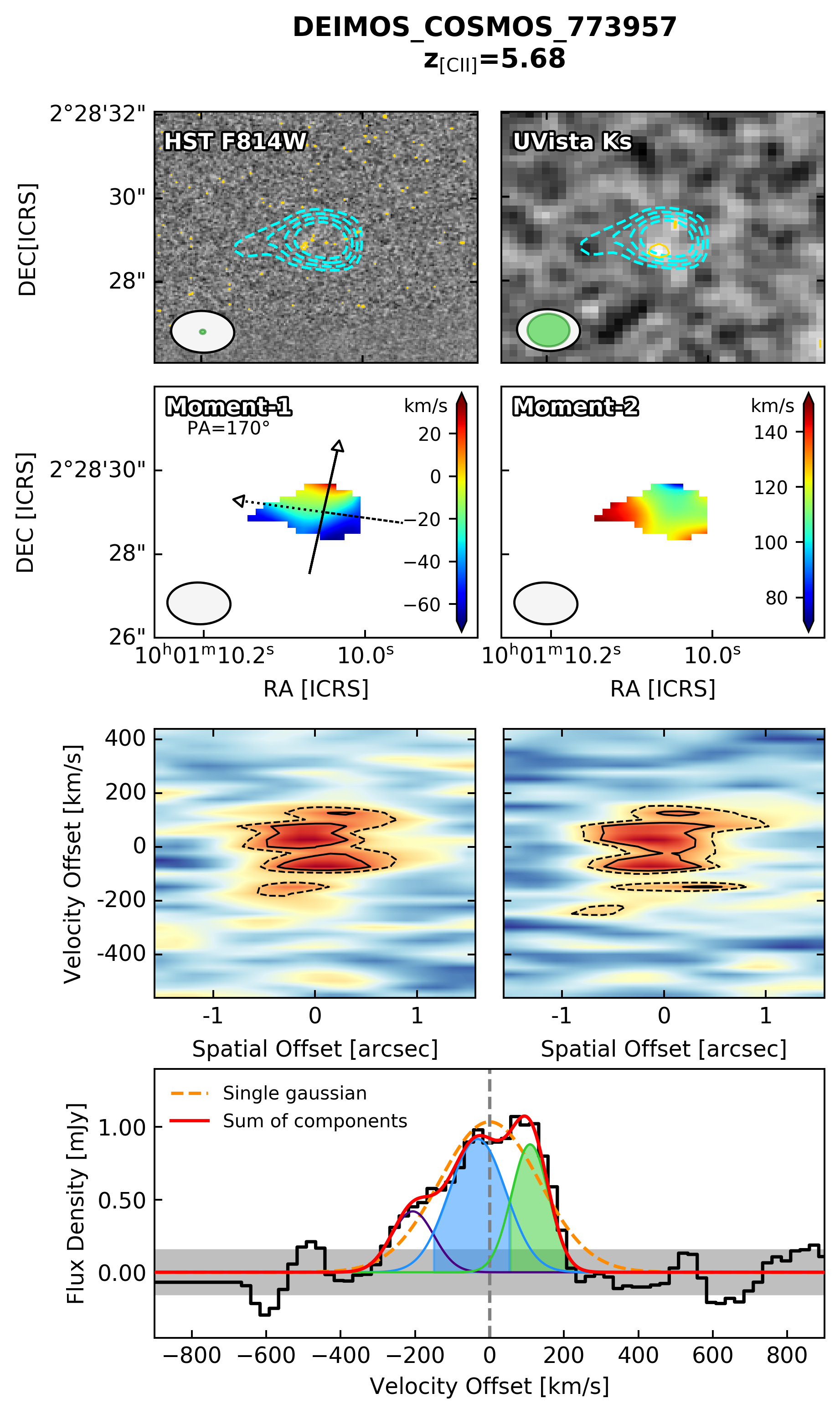}}
	\subfigure{\includegraphics[width=\columnwidth,height=15.1cm]{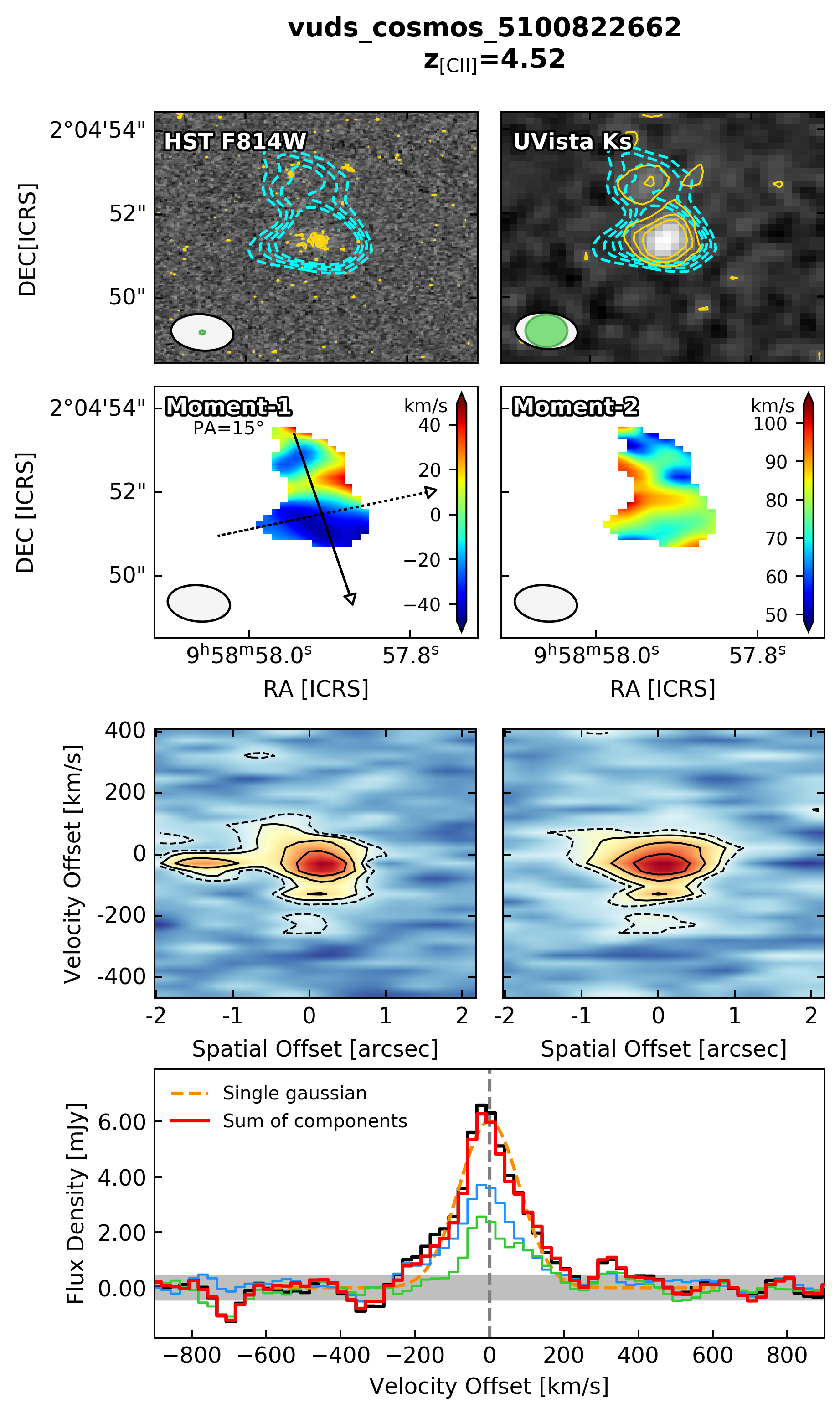}}
    \caption{\textit{Left:} Morpho-kinematic analysis of the ALPINE target $\mathrm{DC\_773957}$ at $z_{\mathrm{[CII]}}=5.68$. \textit{First row}: HST/ACS F814W (left) and UltraVISTA DR4 $K_s$-band (right) images centered on the UV rest-frame position of the target. Each cutout is $6'' \times 6''$ wide. The cyan contours show the [CII] ALMA emission starting from $3\sigma$ above the noise level. Yellow contours in the optical maps represent 3, 5 and 7$\sigma$ emission. In the lower left corner, the ALMA beam (white) and HST or UltraVISTA resolutions (green) are displayed. \textit{Second row:} moment-1 (left) and moment-2 (right) maps color-coded for the velocity and velocity dispersion in $\mathrm{km~s^{-1}}$. The velocity map reports the direction of the major (solid) and minor (dashed) axis (centered on the coordinates returned by the best-fit 2D Gaussian model on the moment-0 map) along which the PVDs are computed. \textit{Third row:} PVDs along the major (left) and minor (right) axis color-coded for the flux intensity in each pixel. Dashed contours include the $2\sigma$ emission in the maps while 3, 5 and 7$\sigma$ emission is represented by solid lines. \textit{Fourth row:} [CII] spectrum (black histogram) extracted within the 3$\sigma$ contours of the intensity map. The gray shaded band marks the 1$\sigma$ level of the spectrum while the dashed vertical line shows the zero velocity offset computed with respect to the redshift of the [CII] line. Purple, blue, and green lines are three individual possible components of the merging system, resulting in the global profile in red. The shaded areas under the curves represent the channels used to compute the [CII] intensity maps of the corresponding individual components. A single Gaussian fit is also visible with a dashed-orange line. \textit{Right:} Morpho-kinematic analysis of the ALPINE target $\mathrm{vc\_5100822662}$ at $z_{\mathrm{[CII]}}=4.52$. Same panels as in the left figure. In the bottom panel we report the [CII] spectra of the major (blue) and minor (green) merger components extracted at the positions of the two resolved emissions, as further described in the text.}
    \label{fig:merger_example}
\end{center}
\end{figure*}

\subsection{Mergers characterization}\label{sec:mergers_charact}
Figure \ref{fig:merger_example} shows two examples of the data analyzed for each [CII]-detected source. Taking as a reference the left set of panels, we display at the top the [CII] morphology superimposed on the HST/ACS F814W (left) and UltraVISTA (UVista) $K_s$ (right) cutouts centered on the UV rest-frame coordinates of the ALPINE target. We also show the $>3\sigma$ emission contours from the optical images. In the second row, the moment-1 and moment-2 maps are reported color-coded for the velocity and velocity dispersion, respectively. In the velocity field map the major and minor axis used to produce the underlying PVDs (color-coded for the flux intensity in Jy/beam) are shown. The axis are centered on the coordinates retrieved from the 2D Gaussian fit on the intensity map and are represented with the direction along which the PVDs are computed. The PA of the major axis is measured from north through east. Both the moment-1 and moment-2 maps are smoothed by using a bilinear interpolation. Finally, the lower panel reports the [CII] spectrum (in black) extracted from the continuum-subtracted data cube and modeled with a single and a multiple-component Gaussian function (in this latter case, we show also the possible single components shaping the global spectral profile).

The described panels report the morpho-kinematic analysis of one of the sources classified as merger both by \cite{LeFevre20} and by the analysis undertaken in this work\footnote{We note that \cite{Jones21} classified this target as \lq uncertain\rq{}, following their classification criteria.}, that is $\mathrm{DEIMOS\_COSMOS\_773957}$ ($\mathrm{DC\_773957}$) at $z_{\mathrm{[CII]}}=5.68$. There are many lines of evidence for considering this galaxy in a merging phase. First, the [CII] emission is elongated toward the East at $>3\sigma$, resulting in a clearly disturbed morphology. Furthermore, two peaks of emission at $3\sigma$ are resolved in the UVista $K_s$ band, which could indicate a close pair of objects or the presence of a dust screen. The velocity field is also quite disturbed, with the PVD along the major axis presenting two $3\sigma$ components (solid contours) separated in velocity by $\sim140~\mathrm{km~s^{-1}}$ and with a slight spatial offset. These two components are also visible in the [CII] spectrum with green and blue lines. A third minor component at $\sim-200~\mathrm{km~s^{-1}}$ could be instead due to a faint satellite nearby the ALPINE target and it is fairly visible in the PVDs within the $2\sigma$ contours. It is worth noting that, even though with a single Gaussian function we find an integrated flux of the line that is comparable with that obtained from the multicomponent fit, the above-mentioned three components are needed in order to reproduce the observed [CII] spectrum.

Moreover, to obtain a more robust sample of mergers we compute, for each of the analyzed merger candidate, the rms of the observed spectra and use it to compute the signal-to-noise ratio (S/N) of the possible merging components. In particular, we obtain the integrated [CII] fluxes of the two principal Gaussian components of each spectrum and divide them by the product between the rms and the corresponding FWHM. We thus consider as mergers only those systems for which the minor (in terms of the [CII] integrated flux) component has S/N$\geq3$. The S/N for the major and minor components of each system are reported in Table \ref{tab:tab_mergers}.

Another example of merging galaxies is provided on the right of Figure \ref{fig:merger_example} by the morpho-kinematic analysis of $\mathrm{vuds\_cosmos\_5100822662}$ ($\mathrm{vc\_5100822662}$) at $z_{\mathrm{[CII]}}=4.52$. This source was classified as a merger both by \cite{LeFevre20} and \cite{Jones21}. Indeed, two close and spatially resolved components are clearly visible in the moment-0 and optical maps. From the PVD along the major axis\footnote{In this case, we use a pseudo-slight of 4'' of length in order to cover all the $3\sigma$ emission from the velocity map.}, the two merging systems are at the same velocity but spatially separated by $\sim1.5''$ (i.e., $\sim10$~kpc at $z\sim4.5$). In such a case, it is not possible to compute the contribution supplied by each individual source to the global shape of the line directly from the [CII] spectrum. However, we can provide an estimate of the integrated flux from the spectrum of each single component by modeling the total [CII] emission in the moment-0 map with two 2D Gaussian functions. In this way, we can extract the spectrum of the resolved component from the continuum-subtracted data cube at the position and with the shape provided by the best-fit parameters from the model (i.e., centroid, major and minor FWHM). For this specific ALPINE target, the spectra of the two merging components are shown in the bottom row of the figure as the blue and green histograms. As can be seen, their sum is quite consistent with the shape of the original [CII] spectrum. 

A complete description of each individual merging system found from this work in the ALPINE survey is reported in Appendix \ref{app:mergers}. 

\subsection{Comparison with previous classifications}
With the methodology introduced in the previous section, we proceed for a new morpho-kinematic classification of the ALPINE targets. As the main aim of this work is to find the contribution of mergers to the galaxy mass assembly at $z\sim5$ from the ALPINE survey, we particularly focus on this class of sources, in order to obtain the final sample of mergers needed to estimate the major merger fraction in the early Universe.

As done by \cite{LeFevre20}, we visually inspect the ancillary data, the intensity maps, the velocity and velocity dispersion fields presented in Section \ref{sec:mergers_charact} to search for the presence of multiple components or disturbed morphology near the position of the targets. The channel maps, the spectra and the PVDs are checked together searching for consistent emission features. By taking into account the results of the initial qualitative classification by \cite{LeFevre20} and of the more recent quantitative analysis of a subsample of the ALPINE targets by \cite{Jones21}, we proceed with a more in-depth characterization of the [CII]-detected galaxies aimed at obtaining a robust merger fraction at $z\sim5$. Adopting the same criteria described in Section \ref{sec:initial_class} to differentiate the targets and considering the S/N of the minor merger component as described in Section \ref{sec:mergers_charact}, we find a slightly lower fraction ($\sim31\%$, 23 out of 75 [CII]-detected sources) of mergers\footnote{Some of the mergers found by \cite{LeFevre20} were also analyzed in terms of the tilted ring model fitting by \cite{Jones21} who found that the kinematic of those sources could be even compatible with rotating disks or dispersion dominated galaxies, or that the sensitivity and resolution of the data are too low for a conclusive classification (see Appendix \ref{app:mergers}).} if compared to the $40\%$ found by \cite{LeFevre20}, with 12, 20 and $7\%$ of the sample made by rotating, extended and compact dispersion dominated sources, respectively. To be more conservative in the classification of the galaxies (especially for obtaining a more robust merger statistics), we define the remaining $30\%$ of the sample as \lq uncertain\rq{}, a new category that includes the weak galaxies (as described in \citealt{LeFevre20}) and also objects that, by visual inspection, present features that are intermediate to those of various classes. This category is similar to the \lq uncertain\rq{} (UNC) class introduced in \cite{Jones21} that, based on the results of the $\mathrm{^{3D}Barolo}$ fits, contains sources they are unable to classify because of the low S/N and/or spectral resolution, or contrasting evidence in their classification criteria. Although the uncertain category is populated by a significantly larger fraction of sources with respect to the weak class ($\sim16\%$) in \cite{LeFevre20}, we recover the same qualitative morpho-kinematic distribution of the previous analysis, confirming the high fraction of rotators and mergers at these early epochs.

\begin{table*}
\begin{threeparttable}
\caption{Physical parameters of the mergers.}
\label{tab:tab_mergers}
\begin{center}
\begin{tabular}{l c c c c c c | c c c c c c}
\hline
\hline
& & & & & & & \multicolumn{3}{c}{Minor} & \multicolumn{3}{c}{Major}\\
Source & $z_1$ & $z_2$ & $\Delta v$ & $r_p$ & $\mu_{\mathrm{[CII]}}$ & $\mu_K$ & FWHM & $r_\mathrm{e}$ & S/N & FWHM & $r_\mathrm{e}$ & S/N\\
& & & [km~s$^{-1}$] & [kpc] & & & [km~s$^{-1}$] & [kpc] & & [km~s$^{-1}$] & [kpc] &\\
 (1) & (2) & (3) & (4) & (5) & (6) & (7) & (8) & (9) & (10) & (11) & (12) & (13)\\
\hline
CG$\_$38 & 5.5731 & 5.5698 & 152.3 & 2.7 & 1.8 & - & 96.9 & 1.4 & 3.8 & 165.9 & 2.0 & 3.9\\
DC$\_$308643 & 4.5238 & 4.5221 & 92.3 & 3.0 & 1.2 & - & 85.4 & 1.0 & 4.5 & 85.6 & 1.4 & 5.3\\
DC$\_$372292 & 5.1345 & 5.1374 & 144.0 & 1.8 & 2.7 & 2.9 & 82.8 & 2.2 & 4.1 & 157.9 & 3.7 & 5.7\\
DC$\_$378903 & 5.4311 & 5.4293 & 94.5 & 0.9 & 1.9 & - & 70.0 & 4.4$^{(*)}$ & 3.6 & 111.2 & 1.3 & 4.2\\
DC$\_$417567 & 5.6676 & 5.6700 & 106.7 & 5.3 & 2.1 & - & 84.8 & 1.0 & 3.1 & 163.9 & 4.0 & 3.4\\
DC$\_$422677 & 4.4361 & 4.4378 & 92.8 & 0.0 & 1.4 & - & 95.7 & 1.0 & 4.7 & 93.1 & 0.5 & 6.7\\
DC$\_$434239 & 4.4914 & 4.4876 & 206.1 & 5.5 & 6.9 & - & 103.5 & 3.9 & 4.5 & 441.8 & 3.4 & 7.2\\
DC$\_$493583 & 4.5122 & 4.5141 & 103.4 & 1.0 & 4.7 & - & 64.8 & 3.3 & 3.6 & 139.5 & 1.0 & 8.0\\
DC$\_$519281 & 5.5731 & 5.5765 & 158.0 & 0.9 & 4.5 & - & 95.7 & 0.9 & 3.1 & 163.8 & 2.3 & 8.3\\
DC$\_$536534 & 5.6834 & 5.6886 & 234.6 & 4.4 & 2.7 & - & 113.9 & 1.9 & 3.6 & 228.5 & 2.6 & 4.9\\
DC$\_$665509 & 4.5244 & 4.5261 & 95.9 & 1.0 & 2.0 & - & 70.5 & 4.5$^{(*)}$ & 3.0 & 107.0 & 2.4 & 4.0\\
DC$\_$680104 & 4.5288 & 4.5308 & 106.4 & 0.0 & 1.0 & - & 89.5 & 0.5 & 3.7 & 111.6 & 8.7$^{(*)}$ & 4.7\\
DC$\_$773957 & 5.6802 & 5.6770 & 141.2 & 3.0 & 1.6 & 5.6 & 118.0 & 1.3 & 5.9 & 179.5 & 1.7 & 6.2\\
DC$\_$814483 & 4.5823 & 4.5796 & 145.5 & 7.9 & 2.4 & - & 128.1 & 1.1 & 3.3 & 192.7 & 2.8 & 4.2\\
DC$\_$818760 & 4.5626 & 4.5609 & 92.3 & 9.9 & 1.3 & 2.6 & 127.1 & 2.4 & 10.4 & 271.0 & 2.9 & 18.3\\
DC$\_$834764 & 4.5076 & 4.5055 & 119.0 & 1.0 & 1.7 & - & 96.2 & 1.4$^{(*)}$ & 5.1 & 143.8 & 3.1 & 5.8\\
DC$\_$842313 & 4.5547 & 4.5406 & 751.5 & 11.6 & 32.9 & 1.5 & 229.9 & 4.3$^{(*)}$ & 5.1 & 1089.7 & 1.7 & 5.9\\
DC$\_$859732 & 4.5353 & 4.5315 & 205.2 & 4.9 & 3.9 & - & 55.4 & 1.7$^{(*)}$ & 3.0 & 176.4 & 3.3$^{(*)}$ & 3.7\\
DC$\_$873321 & 5.1545 & 5.1544 & 4.5 & 6.5 & 1.2 & 3.1 & 171.2 & 1.3 & 4.9 & 219.3 & 2.8 & 7.6\\
vc$\_$5100541407 & 4.5628 & 4.5628 & 1.9 & 13.8 & 1.6 & 1.4 & 203.8 & 4.4$^{(*)}$ & 3.5 & 166.5 & 1.2 & 6.9\\
vc$\_$5100822662 & 4.5210 & 4.5205 & 22.3 & 10.9 & 1.6 & 1.7 & 189.6 & 4.0 & 5.1 & 192.7 & 2.0 & 8.0\\
vc$\_$5101209780 & 4.5724 & 4.5684 & 217.3 & 10.8 & 4.1 & 2.5 & 126.1 & 2.3 & 4.4 & 303.9 & 6.6 & 7.4\\
vc$\_$5180966608 & 4.5294 & 4.5293 & 8.9 & 7.2 & 3.0 & 3.7 & 243.6 & 1.8 & 3.4 & 231.6 & 4.3 & 10.9\\
\hline
\end{tabular}
\begin{tablenotes}
\small
\item \textbf{Notes.} The right side of the table shows the measurements of FWHM (in km~s$^{-1}$), [CII] emission size (in kpc), and S/N for both the minor and major components, intended as such in terms of their integrated [CII] fluxes. The asterisk near the size of some of the sources indicates a bad $r_e$ estimate from the best-fit parameters of the 2D Gaussian fit on the moment-0 maps of the individual components.
\end{tablenotes}
\end{center}
\end{threeparttable}
\end{table*}

\begin{figure*}
    \begin{center}
	\includegraphics[width=1.7\columnwidth]{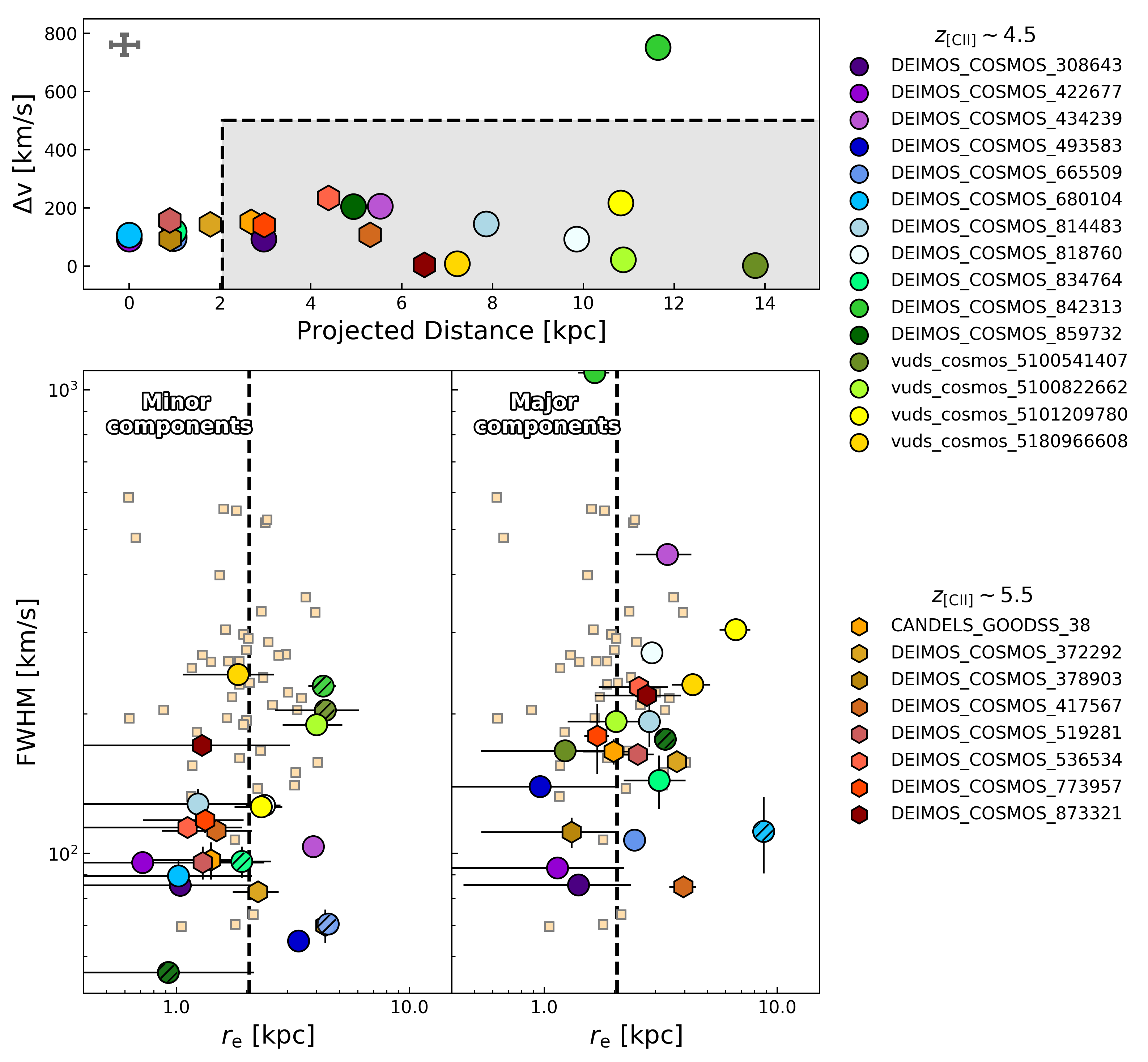}
	\end{center}
    \caption{\textit{Top panel:} Velocity separation between the components of each ALPINE merger as a function of their projected distance. Each target is represented by its own color while the symbols are different for $z\sim4.5$ (circles) and $z\sim5.5$ (hexagons) galaxies. The gray shaded region shows the area of the plot populated by galaxies which are more likely to be real mergers, with a projected distance larger than $>2.1$~kpc and $\mathrm{\Delta v<500~km~s^{-1}}$. The error bar on the top-left corner represents the typical error on each value. \textit{Bottom panel:} FWHM as a function of $r_{\mathrm{e}}$ for the minor (left) and major (right) components of each merger. The colors and symbols are the same as in the top panel. Hatched markers identify the sources for which the best-fit size estimate is not reliable, after a visual inspection of the residuals. The dashed black lines mark the average [CII] size of the ALPINE targets (as estimated in this work), represented with small light squares in the background.}
    \label{fig:phase_space}
\end{figure*}

\subsection{Physical parameters estimate}\label{sec:physical_params}
For each source classified as merger we measure several physical quantities that help us to estimate the merger fraction at $z\sim5$. From the spectra, we compute the [CII] flux ratio between the major and minor components of the merger\footnote{In the case the merging system is composed by more than two sources, we only consider the two major components in terms of their [CII] fluxes.}, that is $\mu_\mathrm{[CII]} = F_\mathrm{1,[CII]}/F_\mathrm{2,[CII]}$, finding a good agreement with the corresponding ratio computed from the PVDs. To obtain the integrated fluxes of the individual galaxies, we use a Gaussian decomposition of the global [CII] spectrum, as described in Section \ref{sec:mergers_charact}. When allowed by the resolution, we also take the UVista $K_s$-band flux ratio of the two merging sources, namely $\mu_{K}\equiv F_{1,K}/F_{2,K}$, where $F_{1,K}$ and $F_{2,K}$ are the integrated fluxes obtained by fitting a 2D Gaussian function to the major and minor components of the merger in the considered band, respectively. Then, from the spectral features, we also obtain the separation in velocity ($\Delta v$) between the intensity peaks of the merger components, and the FWHM both for the individual objects and single (overall spectrum) source. 

We measure the projected distance ($r_p$) between the centers of the merger components as $r_p = \theta \times d_A(z_m)$, where $\theta$ is the angular separation in the sky (in arcsec) between the two galaxies and $d_A(z_m)$ is the angular diameter distance (in kpc~arcsec$^{-1}$) computed at the mean redshift $z_m$ of the two sources. In case the two components are not spatially resolved in [CII], we consider the distance between the two best-fit centroids returned by the 2D Gaussian fit on their moment-0 maps. Similarly to what has been done in \cite{Ginolfi20}, the latter maps are obtained by collapsing the channels including the emission of the individual components. Such a case is shown, for instance, in Figure \ref{fig:merger_example} for the target $\mathrm{DC\_773957}$. The shaded blue and green areas under the curves in the spectrum mark the collapsed channels used for building the [CII] intensity maps of the two components. As it can be seen, the velocity channels assigned to each component are not overlapping with each other. If instead the two objects are spatially resolved but at the same velocity (as for $\mathrm{vc\_5100822662}$), we are not able to create the [CII] moment-0 maps of the individual components (i.e., the two sources emit at the same frequency and we cannot disentangle their contribution in [CII]) and we measure the distance directly from the [CII] intensity map or from the PVDs.

Finally, we compute the size of the single components by fitting a simple 2D Gaussian model to the [CII] intensity maps obtained as described above. Following \cite{Fujimoto20}, we refer to the merging component size as the circularized effective radius $r_{\mathrm{e}}$ defined as $r_{\mathrm{e}} \equiv \sqrt{a~b}$, where $a$ and $b$ are the best-fit semi-major and semi-minor axis, respectively. When the source is resolved, we use the best-fit beam-deconvolved $a$ and $b$ parameters, otherwise we take the beam-convolved sizes provided by the fit and correct them for the ALMA beam (the average beam minor axis of ALPINE is $\sim0.8''$).  It is worth noting that \cite{Fujimoto20} measured the [CII] effective radii by fitting the line visibilities with the CASA task \texttt{UVMULTIFIT} and assuming an exponential-disk profile. However, by applying our more simplistic size measurements to the ALPINE galaxies analyzed by \cite{Fujimoto20} and considering the mergers as single components, we found a good agreement with their results. A larger scatter is found only for some of the objects classified as mergers whose size measurements are flagged as unreliable by \cite{Fujimoto20}, likely because of the complicated [CII] morphology.   

From now on, we refer to $r_{\mathrm{e}}$ when talking about the size of a source. All the information presented in this section is reported in Table \ref{tab:tab_mergers}.


\subsection{Description of the merger sample}\label{sec:size_mergers}
We show in Figure \ref{fig:phase_space} (upper panel) the difference in velocity between the merger components as a function of their projected distance $r_p$ for all the possible merging systems in ALPINE. In the upper-left corner, we report the uncertainties assumed for these data. The error on $\Delta v$ is obtained by the sum in quadrature of the average uncertainty on each spectral element (i.e., 25~km~s$^{-1}$). To estimate the typical error on the projected distance, we consider two random centroids corresponding to the initial positions of the minor and major components of the merger. Then, we obtain the distribution of the errors on the centroids as provided by the 2D Gaussian best-fits on the intensity maps of the merging components, and extract from that $\mathrm{N}=1000$ values used to perturb the initial positions. At each iteration, we re-compute the projected distance between the perturbed positions of the two components. We thus estimate the average uncertainty on $r_p$ as the standard deviation of the distribution of the N projected separations (corresponding to $\sim0.3~$kpc).

To derive a robust major merger fraction ($f_\mathrm{MM}$), we decide to consider as mergers only those systems in which the components are separated by no more than $\mathrm{\Delta v\leq500~km~s^{-1}}$, assuming that this is the limit for a system to be gravitationally bound (e.g., \citealt{Patton2000,Lin2008,Ventou17}). As shown, all the mergers in our sample satisfy this constraint except one source. As analyzed in detail in Appendix \ref{app:mergers}, this system is composed by the ALPINE target $\mathrm{DC\_842313}$ and a peculiar neighbor source with a large FWHM and [CII] flux, and could even be a merger of three galaxies with a $\Delta v$ between the target and the further third component smaller than 500~km~s$^{-1}$. We thus account for this object in the computation of the merger fraction. 

Also, we consider as reliable mergers only those systems with a projected distance larger than the typical [CII] sizes of individual galaxies at these redshifts. Indeed, closer components could just represent clumps of star formation within the same galaxy, faking the presence of a merger by affecting the morphology and kinematics of the [CII] line. From the size measurements of the ALPINE galaxies we find an average [CII] size of $\sim2.1$~kpc, that is in good agreement with the median value found by \cite{Fujimoto20} and with the typical distance between galaxies and satellites or mergers (>2 kpc) at $z>4$ (e.g., \citealt{Carniani18,Whitney19,Zanella21}). Therefore, we use this value as a threshold to classify the secure merging systems. 

The two selection criteria on the distance and velocity separations define the gray region in Figure \ref{fig:phase_space} which thus includes the sources that are classified as robust mergers. However, we cannot exclude a priori the systems with $r_p<2.1$~kpc as they could be structures in an advanced stage of merging that we are not able to resolve with ALMA because of their close proximity and observational limitations (i.e., large synthesized beam, limited sensitivity). It is worth noting that, in some cases, spatially close clumps are observed with HST within the typical ALPINE beam size. However, the lack of redshift information for these sources prevent us from considering them as associate galaxies experiencing a merging. This could have an impact on our results, possibly reducing the estimated major merger fraction (and, consequently, the merger rate) at $z\sim5$. For these reasons, we also show in Figure \ref{fig:phase_space} (bottom panels) the individual sizes of the minor and major components of each of the possible mergers as a function of their FWHM. The colors and marker symbols are the same as in the top panel, while the hatched markers highlight those objects with a bad $r_\mathrm{e}$ estimate from the fit, after a visual inspection of the residuals. We also show for comparison the data of the individual ALPINE galaxies (excluding the mergers) as light squares. As it can be seen, the sizes of the individual components lie, within the uncertainties, among those of the ALPINE targets. This suggests that all the individual components we are analyzing could be merging galaxies and that, with the current data, we cannot exclude them from the computation of the merger fraction at $z>4$.

Finally, we provide an estimate of the stellar mass range covered by our merger components. We only have information on the stellar mass of the merging system as a whole through SED-fitting measurements \citep{Faisst20} because the majority of the merger candidates are not resolved in most of the optical and near-IR bands. We make the assumption that $M_{*,1}+M_{*,2}=M_{*}$ and define the mass ratio as $\mu=M_{*,1}/M_{*,2}$, where $M_{*,1}$ and $M_{*,2}$ are the stellar masses of the primary and secondary galaxy (i.e., $M_{*,1}>M_{*,2}$) and $M_{*}$ is the total mass of the system from the SED-fitting. Then, we set $\mu=4$ as a threshold to separate major ($1\leq \mu \leq4$) from minor ($\mu>4$) mergers (e.g., \citealt{Lotz11,Mantha18,Duncan19}) and use this value to solve the above equations for $M_{*,1}$ and $M_{*,2}$. With the assumptions above, we obtain a stellar mass distribution of the major components ranging between $\mathrm{log}(M_{*,1}/M_{\odot})\sim9.0$ and 10.7, with a mean value $\mathrm{log}(M_{*,1}/M_{\odot})\sim10$ (reducing by $\sim0.2$~dex if assuming $\mu=1$). We want to stress that this is just an exercise that is intended to provide an estimate of the stellar mass range of the merger components we are analyzing and that helps us in the comparison of our estimates with those of previous works. Indeed, we are not considering that some of the merger components are resolved in the $K_s$ band and have their own mass ratio (see Section \ref{sec:frac_derivation}). Therefore, following the above considerations, we can conclude that the close pairs of our sample are characterized by $r_p<20$~kpc, $\Delta v<500$~km~s$^{-1}$, and average stellar masses $\mathrm{log}(M_{*}/M_{\odot})\sim10$.


\subsection{Accounting for completeness}\label{sec:completeness}
When deriving the merger fraction, we have to consider the possibility of missing a certain number of mergers because of the faintness of the minor component. Indeed, mergers with a principal component near the threshold of the observable [CII] flux may imply the presence of a secondary component that could be lost by instrumental limitations. To account for this, we must correct for the incompleteness of our sample.

Previous works typically assumed corrections based on the limiting flux or stellar mass of the corresponding survey (e.g., \citealt{Patton2000,Lopez11,Mundy17,Duncan19}). ALPINE is not a flux-limited survey as each individual ALMA pointing reaches different depths (e.g., \citealt{Bethermin20}). For this reason, we adopt the classical completeness corrections found in the literature but considering for each object its limiting flux. In particular, we apply a weight to each minor component of our sample defined as
\begin{equation}
    w_\mathrm{comp}(L,z) = \frac{\int_{L_2}^{L_1} \Phi(L,z) \,dL}{\int_{L_\mathrm{lim}}^{L_1} \Phi(L,z) \,dL},
    \label{eq:weight}
\end{equation}
where $\Phi(L, z)$ is the [CII] luminosity function derived from the UV-selected central ALPINE targets \citep{Yan20}, $L_\mathrm{lim}$ is the luminosity corresponding to the limiting flux of each ALPINE pointing, $L_1$ is associated to the flux of the primary component $F_\mathrm{1,[CII]}$, and $L_2$ corresponds to $F_\mathrm{2,[CII]} = F_\mathrm{1,[CII]}/4$. We compute the [CII] luminosities by following \cite{Solomon92} with the relation
\begin{equation}
L_\mathrm{[CII]} = 1.04 \times 10^{-3}~F_\mathrm{[CII]}~D^2_\mathrm{L}(z)~\nu_{obs}(z)~[L_{\odot}],
\label{eq:L_CII}
\end{equation}
where $F_\mathrm{[CII]}$ is the integrated flux (in Jy~km~s$^{-1}$), $D_\mathrm{L}(z)$ is the luminosity distance (in Mpc) at the redshift of the galaxy, and $\nu_\mathrm{obs}(z)$ is the observed frequency of the [CII] line (in GHz). We also assume that $w_\mathrm{comp}=1$ when $L_2 > L_\mathrm{lim}$. Such a correction corresponds to the ratio between the number density of galaxies with [CII] flux higher than $F_\mathrm{2,[CII]}$, and the number density of galaxies with a flux higher than the flux limit of each ALMA pointing.


\section{Results}\label{sec:results}
\subsection{The fraction of major mergers in ALPINE}\label{sec:frac_derivation}
\begin{figure}
    \begin{center}
	\includegraphics[width=\columnwidth]{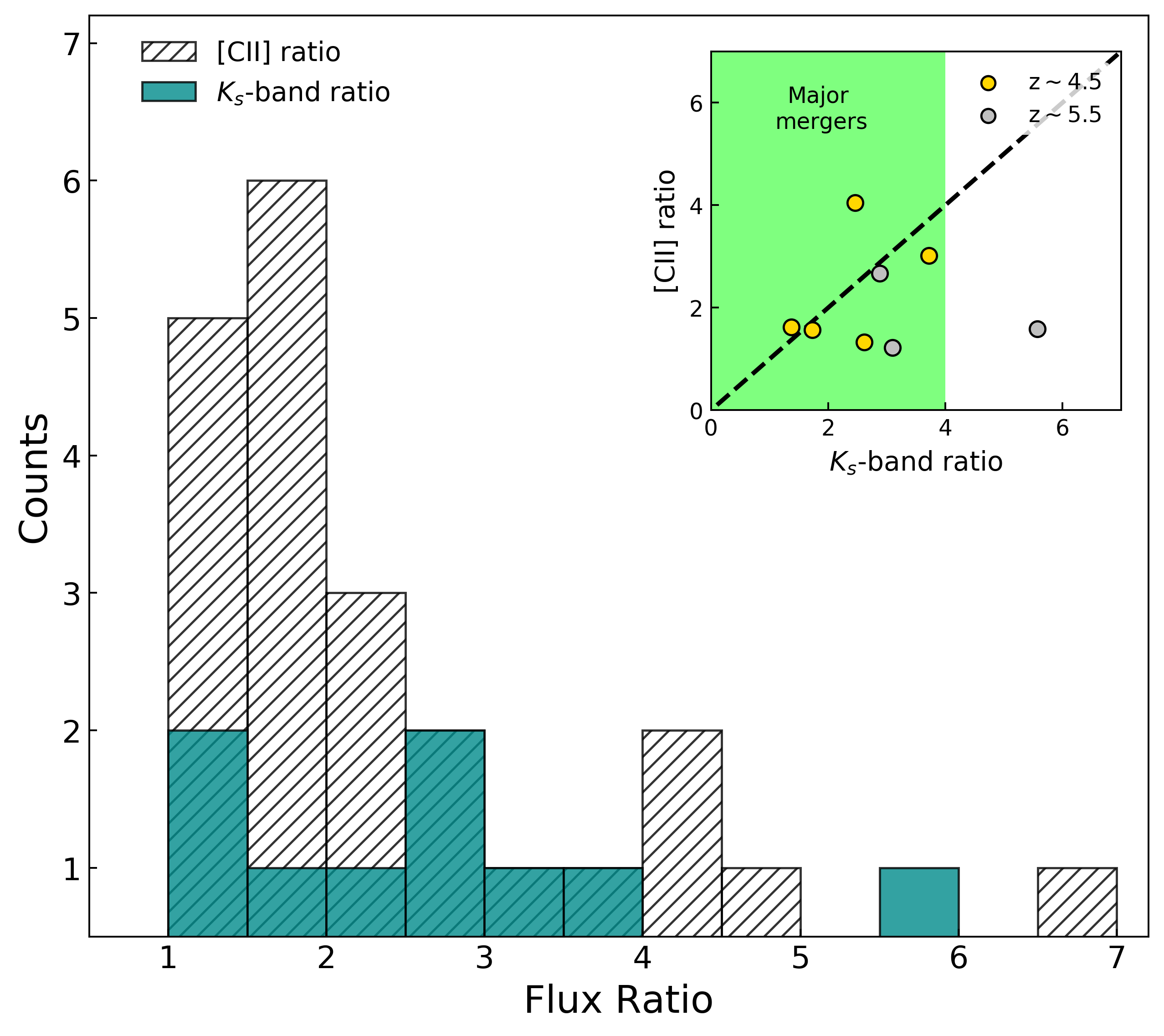}
	\end{center}
    \caption{Distributions of the $K_s$-band (turquoise histogram) and [CII] (hatched histogram) flux ratios between the components of each merging system. \textit{Inset:} comparison between the [CII] and $K_s$-band ratios for the sources having this information in common, both at $z\sim4.5$ and $z\sim5.5$ (represented as yellow and gray circles, respectively). The green area marks the region of the figure with $\mu_K<4$. The dashed black line reports the 1:1 relation between the two quantities.}
    \label{fig:CII_K_ratio}
\end{figure}

The aim of this work is to derive the fraction of galaxies undergoing a major merger at $z\sim5$ and to estimate their contribution to the galaxy mass assembly through cosmic time. Although the cosmic evolution of minor mergers is still poorly constrained at these redshifts, several studies suggest that they are more frequent in the nearby Universe showing a decreasing fraction for $z\gtrsim3$ \citep{Lopez11,Lotz11,Ventou19}. We thus have to take into account a possible contamination from them in our sample.   

First, as we do not have the stellar mass information for the majority of our merger components, we define as major mergers those candidates for which the $K_s$-band flux ratio is smaller than 4, that is $1<\mu_{K}<4$. The $K_s$ band is indeed a good tracer of the stellar mass of galaxies up to $z\sim4$ (e.g., \citealt{Laigle16}). At higher redshifts, the rest-frame $K_s$ band samples the emission below the Balmer break ($K_s$-band corresponds to rest-frame $\sim3600~\AA$ at $z\sim5$, taking as a reference the mean wavelength of the filter) which is no more directly linked to the stellar mass of the galaxy. In these cases, mid-IR bands could be better tracers of the galaxy stellar masses than the $K_s$ filter, but their worst resolution do not allow us to spatially resolve the close galaxies for which we have the $K_s$ fluxes. For comparison, \cite{Lopez13} used the $i$-band ratios to identify major mergers. At the redshifts explored in their work ($0.9<z<1.8$), the observed $i$ band corresponds to $\sim3000-3500~\AA$ rest-frame (which is quite similar to the wavelength range covered by the $K_s$ band at the highest redshift of our sample). They compared these ratios with those obtained from the $K_s$ band (sampling the emission between 8000 and 10000~$\AA$ rest-frame, thus tracing better the stellar mass content of the galaxies), finding no significant changes in their major merger classification, and then supporting the $i$-band results. Therefore, following these arguments, we consider $\mu_{K}$ reasonably comparable to the stellar mass ratio of our sources, treating cautiously these estimates for those galaxies at the highest redshifts of our sample.

The $K_s$-band ratio is available for 9 out of 23 mergers, while for the remaining sources we only have the [CII] flux ratio, from which we cannot draw conclusions about the nature of the merger. These two ratios are shown in Figure \ref{fig:CII_K_ratio} where it is evident that the majority of the sources have both $\mu_{K}$ and $\mu_{\mathrm{[CII]}}$ smaller than 4. Furthermore, in the inset we compare the [CII] and $K_s$-band flux ratios for the objects having these two measures in common\footnote{Both in the main figure and in the inset we do not report the measure of the [CII] flux ratio for the merger $\mathrm{DC\_842313}$ (i.e., $\mu_{\mathrm{[CII]}}\sim33$). We also do not consider this source in the statistics described in the text because, as elaborated in Appendix \ref{app:mergers}, this is a peculiar source.}. There is a good agreement between the two ratios for some of the objects in the figure, but others show a large scatter from the 1:1 relation, suggesting that complex physical processes could take place in these systems. However, we find that 7 out of 8 sources (i.e., $\sim88\%$ of this subsample) have $\mu_K < 4$ while only one system shows a larger $K_s$-band flux ratio. Although a larger statistical sample would be needed, we can interpret this as an indication of the minor merger contamination of our sample. Indeed, if we consider that only the sources outside the green region contribute to the minor merger contamination (thus lying at $\mu_{K}>4$), we can assume that the 12\% of all the objects having only the $\mu_{\mathrm{[CII]}}$ information is affected by minor mergers (or, equivalently, that only the 88\% of that sample contains major mergers). We note that, when excluding from this statistics the sources at $z\sim5.5$ (whose $K_s$-band ratio is less reliable as a tracer of their stellar mass), all the objects in the inset of Figure \ref{fig:CII_K_ratio} have $\mu_K < 4$, raising the fraction of major mergers in this subsample to 100\%. This is taken into account later, when estimating the uncertainty on the merger fraction. 

We define the major merger fractions in the two ALPINE redshift bins (at mean redshift $z\sim4.5$ and 5.5, respectively) as
\begin{equation}
    f_{\text{MM}} = 0.88~\frac{\sum_{j=1}^{\mathrm{N_p}} w_\mathrm{comp}^{j}}{\mathrm{N_g}},
    \label{eq:fmm}
\end{equation}
where $\mathrm{N_p}$ represents the number of all the mergers in our sample, $\mathrm{N_g}$ is the number of ALPINE galaxies in the considered redshift bin, and $w_\mathrm{comp}^{j}$ is the weight associated to each merger to correct for incompleteness (see Section \ref{sec:completeness}). The factor 0.88 accounts for the statistical uncertainty on the real merger nature (the fact that the 12\% of the $\mu_{\mathrm{[CII]}}$-only subsample could be affected by the presence of minor mergers). Note also that, as detailed in Section \ref{sec:size_mergers}, we are not excluding any source with $r_p<2$~kpc because the individual sizes of these merger components are comparable with those of the ALPINE galaxies.

\begin{table}[t]
\caption{Number of sources at $z\sim4.5$ and $z\sim5.5$ from the ALPINE survey used for the computation of the major merger fraction.}
\label{tab:fmm_info}
\begin{center}
\begin{tabular}{c c c}
\hline
\hline
 & $z\sim4.5$ & $z\sim5.5$\\
\hline
$\mathrm{N_g}$ & 46 & 29\\
$\mathrm{N_p}$ & 15 & 8\\
$\mathrm{N_{p,corr}}$ & 23 & 11\\
\hline
$f_{\text{MM}}$ & $0.44^{+0.11}_{-0.16}$ & $0.34^{+0.10}_{-0.13}$\\
\hline
\end{tabular}
\end{center}
\end{table}

To attribute an uncertainty to $f_{\text{MM}}$ we consider two extreme cases. At first, as an upper limit to the major merger fraction, we count all the mergers in each redshift bin and divide them by $\mathrm{N_g}$, independently of their components projected distance and flux ratio. This uncertainty includes also the case in which our sample is not contaminated by minor mergers (i.e., all the analyzed mergers are major). As a lower limit instead, we only take the mergers whose components have $r_p>2$~kpc, assuming that all the individual components with a smaller projected separation are not major mergers (see discussion above). Then, given the small statistics on both the total number of ALPINE galaxies and the subsample of mergers, we compute the corresponding Poissonian errors and add them in quadrature to the above-mentioned uncertainties.

\begin{figure}
    \begin{center}
	\includegraphics[width=\columnwidth]{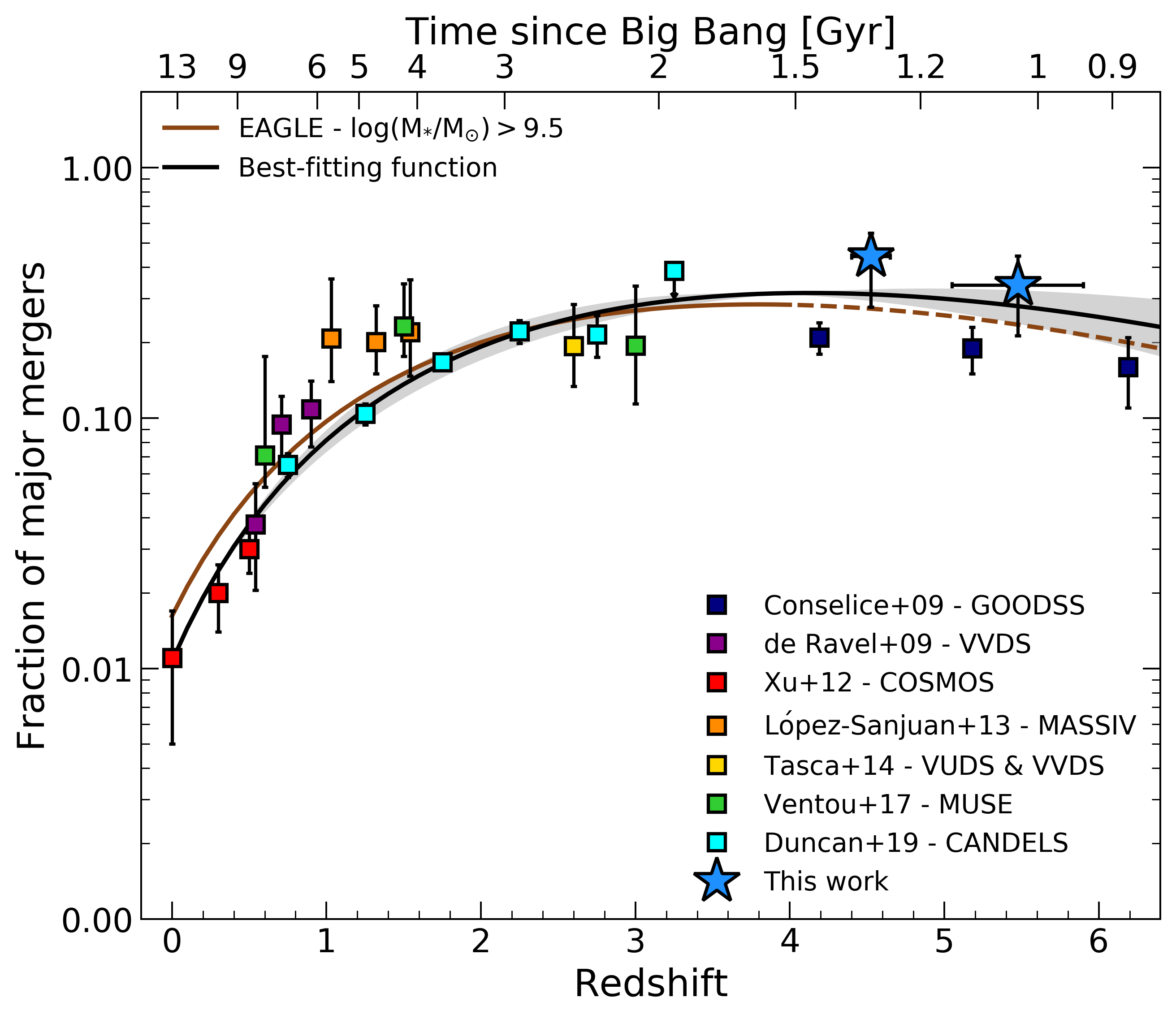}
	\end{center}
    \caption{Cosmic evolution of the major merger fraction $f_{\text{MM}}$ from the local to the early Universe. Colored squares show the data collected from the literature at different cosmic times through pair counts, both with spectroscopic \citep{deRavel09,Lopez13,Tasca14,Ventou17} and photometric \citep{Xu12,Duncan19} redshifts, and/or morphological studies \citep{Conselice09}. Blue stars are the $f_{\text{MM}}$ estimates found in this work. The solid black line and the shaded region are the best-fit to the data with a combined power-law and exponential function and the associated 1$\sigma$ error, respectively. Finally, the solid brown line illustrates the parameterized redshift evolution (up to $z=4$) of the major merger fraction from the EAGLE simulation for galaxies with log$(M_{*}/M_{\odot})>9.5$. The dashed line is just an extension of that curve to higher redshifts.} 
    \label{fig:merger_frac}
\end{figure}

With these assumptions and from Equation \ref{eq:fmm} we obtain $f_{\text{MM}}^{z_4} = 0.44^{+0.11}_{-0.16}$ and $f_{\text{MM}}^{z_5} = 0.34^{+0.10}_{-0.13}$, at $z\sim4.5$ and $z\sim5.5$, respectively. These results are shown in Figure \ref{fig:merger_frac} and reported in Table \ref{tab:fmm_info} along with the numbers used for their computation at both redshifts. We note that, by using Equation \ref{eq:fmm}, we recover a merger fraction $\sim30\%$ higher than obtained if not accounting for completeness corrections.
It is also interesting that, assuming that 12\% of the mergers are minor would imply a minor merger fraction of at least $\sim5\%$, in agreement with the estimates by \cite{Ventou19} at $z>3$ (i.e., $0.08^{+0.07}_{-0.05}$). It should be noted that this fraction could be even higher. Indeed, the relatively low spatial resolution of the ALPINE survey does not allow us to constrain the minor merger fraction at $z\sim5$ \citep{LeFevre20}, as we are not able to detect faint satellites that are instead expected to be in the neighborhood of $z>4$ galaxies by simulations (e.g., \citealt{Pallottini17,Kohandel19}).

Finally, it is worth mentioning that $\sim10\%$ of the ALPINE targets are part of a massive proto-cluster of galaxies (PCI J1001+0220) at $z\sim 4.57$ located in the COSMOS field \citep{Lemaux18}. In principle, high-density environments like those in high-$z$ proto-clusters, or groups and low-mass clusters at lower redshift, could enhance the merging activity, resulting in a merger rate that could be several times larger than what expected in lower-density regimes (e.g., \citealt{Lin10,Lotz13,Tomczak17,Pelliccia19}). To account for this possible caveat, we compare the ALPINE members of the proto-cluster with the mergers in Table \ref{tab:tab_mergers}, finding that only two of them are part of the observed over-density. However, by removing these two sources from our analysis, we do not find any significant change in the merger fraction at $z\sim4.5$ and, consequently, in the estimate of the merger rate.

\begin{table*}[t]
\begin{threeparttable}
\caption{Description of the selection criteria and sample properties in previous pair counts studies and in this work.}
\label{tab:literature}
\begin{center}
\begin{tabular}{c c c c c c}
\hline
\hline
Literature & Redshift range & Major/Minor separation & $\Delta v$ & $r_p$ & Primary Galaxy Mass\\
& & & [km~s$^{-1}$] & [kpc] &\\
(1) & (2) & (3) & (4) & (5) & (6)\\
\hline
\cite{deRavel09} & $0.5<z<0.9$ & $F_1/F_2\leq4.0$ & $\leq500$ & $<20$ & $\mathrm{log}(M_{*}/M_{\odot})\geq9.5$\\
\cite{Xu12} & $0\leq z \leq0.6$ & $M_1/M_2\leq2.5$ & $\leq500^{(*)}$ & $5-20$ & $9.8<\mathrm{log}(M_{*}/M_{\odot})\leq10.2$\\
\cite{Lopez13} & $0.9<z<1.8$ & $F_1/F_2\leq4.0$ & $\leq500$ & $<20$ & $9<\mathrm{log}(M_{*}/M_{\odot})<11$\\
\cite{Tasca14} & $1.8\leq z \leq4.0$ & $M_1/M_2\leq4.0$ & $\leq500$ & $<25$ & $9<\mathrm{log}(M_{*}/M_{\odot})<11$\\
\cite{Ventou17} & $0.2\leq z \leq4.0$ & $M_1/M_2\leq6.0$ & $\leq500$ & $<25$ & $\mathrm{log}(M_{*}/M_{\odot})\geq9.5$\\
\cite{Duncan19} & $0.5\leq z <3.5$ & $M_1/M_2\leq4.0$ & $\leq500^{(*)}$ & $5-30$ & $9.7<\mathrm{log}(M_{*}/M_{\odot})<10.3$\\
This work & $4.4\leq z <5.9$ & $M_1/M_2\leq4.0$ & $\leq500$ & $<20$ & $9\lesssim \mathrm{log}(M_{*}/M_{\odot})\lesssim11$\\
\hline
\end{tabular}
\begin{tablenotes}
\small
\item \textbf{Notes.} The asterisks in column 4 indicate the works in the literature that made use of photometric redshifts to define the selection criterion on the velocity separation between pairs.
\end{tablenotes}
\end{center}
\end{threeparttable}
\end{table*}

\subsection{Cosmic evolution of the major merger fraction}\label{sec:merger_frac}
During the last years, a large number of studies focused on the estimate of the merger fraction at different redshifts and on its evolution through cosmic time (e.g., \citealt{LeFevre00,Conselice03,Conselice09,Lopez13,Mundy17,Ventou17,Mantha18,Duncan19}). However, the choice of various selection criteria on the spatial and velocity separation for the pair counts, the different mass ranges probed by each author and the diverse threshold in $\mu$ to distinguish between major and minor mergers make a direct comparison between the many results found in the literature quite problematic. Therefore, we now compare the major merger fraction found from this work at $4<z<6$ with those computed from samples of galaxies with similar stellar masses and merger selection criteria.  

At $z<1$, \cite{Xu12} evaluated the major merger fraction from a sample of close pairs drawn from the COSMOS survey \citep{Ilbert09}. From that work we select only the results in the stellar mass bin $9.8<\mathrm{log}(M_{*}/M_{\odot})\leq10.2$, to be consistent with the average stellar mass of our sample. At this epoch we also account for the merger fraction computed by \cite{deRavel09} who analyzed a sample of spectroscopically confirmed galaxy pairs from the VIMOS VLT Deep Survey (VVDS; \citealt{LeFevre05}) with $\mathrm{log}(M_{*}/M_{\odot})>9.5$. At $1<z<4$, \cite{Lopez13} and \cite{Tasca14} exploited the VUDS, VVDS and the Mass Assembly Survey with SINFONI in VVDS (MASSIVE; \citealt{Contini12}) surveys to estimate $f_{\text{MM}}$ from merging systems with spectroscopic measurements and average stellar masses $\mathrm{log}(M_{*}/M_{\odot})\sim10$. \cite{Ventou17} observed the \textit{Hubble} Ultra Deep Field \citep{Beckwith06} and the \textit{Hubble} Deep Field South \citep{Williams00} with the Multi Unit Spectroscopic Explorer (MUSE) to constrain the galaxy major merger fraction up to $z\sim6$. However, we take only their data points with $\mathrm{log}(M_{*}/M_{\odot})\geq9.5$ which extend up to $z\sim3$. Finally, we get the pair fraction computed by \cite{Duncan19} in a mass-selected ($9.7<\mathrm{log}(M_{*}/M_{\odot})<10.3$) sample of galaxies drawn from the CANDELS survey \citep{Grogin11,Koekemoer11} using a probabilistic approach based on their photometric redshifts. All these works assume a projected separation $r_p<30$~kpc between the merger components and a difference in velocity $\mathrm{\Delta v\leq500~km~s^{-1}}$. Most of them adopt $\mu<4$ to identify the major merger population, with $\mu$ computed either as the ratio between the stellar mass of the primary and secondary galaxy in the system or as the difference in magnitude between the two components. The only exceptions are from \cite{Xu12} and \cite{Ventou17} that make use of a mass ratio of 2.5 and 6, respectively (however, as stated by \citealt{Ventou17}, adopting a mass ratio limit of 4 implies the loss of only a few sources in their sample, not significantly altering the final statistics). Table \ref{tab:literature} summarizes the main features of the samples from which the observational data at $z<4$ introduced above are taken. 

As shown in Figure \ref{fig:merger_frac}, we combine these data with our measurements at $z\sim5$ to provide the cosmic evolution of the major merger fraction. Traditionally, this is parameterized with a power-law fitting formula of the form
\begin{equation}
    f_{\text{MM}}(z) = f_0~(1+z)^m,
    \label{eq:fmm_powerlaw}
\end{equation}
where $f_0$ is the merger fraction at $z=0$ and $m$ is the index that rules the redshift evolution. Our data suggest a slight decline of the merger fraction at $z>4$, with a possible peak at lower redshift, as also previously found by other studies (e.g., \citealt{Conselice09,Ventou17,Mantha18,Duncan19}). Therefore, we make use of a combined power-law and exponential function to fit our observational data, such as
\begin{equation}
    f_{\text{MM}}(z) = \alpha~(1+z)^m~\mathrm{exp}(\beta(1+z)),
    \label{eq:fmm_evol}
\end{equation}
where $\beta$ controls the exponential side of the curve and $f_0=\alpha~\mathrm{exp}(\beta)$. By using a non linear least square algorithm, we fit our data both with Equation \ref{eq:fmm_powerlaw} and Equation \ref{eq:fmm_evol}. Taking advantage of the Bayesian information criterion (BIC), we find a significant statistical evidence ($\Delta \mathrm{BIC}>10$) in favor of the combined power-law and exponential form for describing the evolution of the merger fraction through cosmic time. These results are in contrast with those by \cite{Duncan19} who, also relying on the BIC, found that there was no strong statistical evidence in favor of one of the two above-described parameterizations. This difference could be attributed to the fact that, for stellar masses $\mathrm{log}(M_{*}/M_{\odot})\sim10$, they measured the major merger fraction up to $z\sim3$, while in this work our new ALPINE data allow us to constrain this quantity up to earlier epochs (i.e., $z\sim5$). 

\begin{table*}[t!]
\caption{Best-fit parameters and 1$\sigma$ uncertainties obtained by fitting the merger fractions and merger rates with a power-law plus exponential functional form of the type $\alpha(1+z)^{m}~\mathrm{exp}(\beta(1+z))$.}
\label{tab:fit_params}
\begin{center}
\begin{tabular}{c c c c}
\hline
\hline
$T_{\text{MM}}$ & $\alpha$ & $m$ & $\beta$\\
\hline
\multicolumn{4}{c}{Merger fraction $f_{\text{MM}}$}\\
- & $0.024\pm0.003$ & $4.083\pm0.501$ & $-0.797\pm0.189$\\\\
\multicolumn{4}{c}{Merger rate $R_{\text{MM}}$}\\
\cite{Kitzbichler08} & $0.020\pm0.003$ & $4.282\pm0.488$ & $-1.036\pm0.182$\\
\cite{Jiang14} & $0.015\pm0.002$ & $4.350\pm0.492$ & $-0.740\pm0.184$\\
\cite{Snyder17} & $0.010\pm0.001$ & $6.083\pm0.501$ & $-0.797\pm0.189$\\

\hline
\end{tabular}
\end{center}
\end{table*}

For these reasons, we show in Figure \ref{fig:merger_frac} the best-fitting function to our data obtained through Equation \ref{eq:fmm_evol}, and use the results from this parameterization for the rest of the work. In particular, the parameters of this fit are $\alpha = 0.024\pm0.003$, $m=4.083\pm0.501$ and $\beta = -0.797\pm0.189$ which are in good agreement with those found by \cite{Duncan19} when fitting the CANDELS data points for galaxies with stellar mass $9.7<\mathrm{log}(M_{*}/M_{\odot})<10.3$. For comparison, we also show the results obtained by \cite{Conselice09} exploiting morphological analysis and pair counts for a sample of Lyman-break drop-out galaxies at $4<z<6$ with masses $\mathrm{log}(M_{*}/M_{\odot})>9-10$. Their data points are lower than our results but still comparable to them within the uncertainties. Anyway, as they do not differentiate between major and minor mergers, we do not include them in the computation of the cosmic evolution of the merger fraction. 

We show in Figure \ref{fig:merger_frac} the best-fit parameterization of the major merger fraction found by \cite{Qu17} for galaxies with $\mathrm{log}(M_{*}/M_{\odot})\geq9.5$ from the Evolution and Assembly of Galaxies and their Environments (EAGLE) hydrodynamical simulation \citep{Schaye15}. They adopted a stellar mass ratio $\lesssim4$ to identify major mergers and the same parameterization as in Equation \ref{eq:fmm_evol}, obtaining a cosmic evolution of $f_{\text{MM}}$ that is similar to ours up to $z\sim4$ (the higher redshift reached by the simulation). Extending their prediction to earlier epochs, a slightly milder redshift evolution with respect to our results is found. However, the two trends are still in good agreement with each other.

Finally, it is worth noting that we are recovering the cosmic evolution of the merger fraction by comparing galaxies with a constant stellar mass range at all redshifts. We are aware that, in this way, we are not necessarily tracing the same galaxy population through cosmic time, as instead achieved with constant cumulative comoving number density selections \citep{Papovich11,Conselice13,Ownsworth14,Mundy15,Torrey15}\footnote{Although this is no more assured in case of major mergers or strong changes in star formation (e.g., \citealt{Leja13,Ownsworth14}).}. Nevertheless, as the majority of the observational and theoretical works in the literature make use of the constant stellar mass selection to estimate the fraction and rate of major mergers through cosmic time, such a choice allows us to make a fair comparison to these works and provides a quite easily quantifiable observational benchmark for future studies.


\begin{figure*}[t]
    \begin{center}
	\subfigure{\includegraphics[width=\columnwidth]{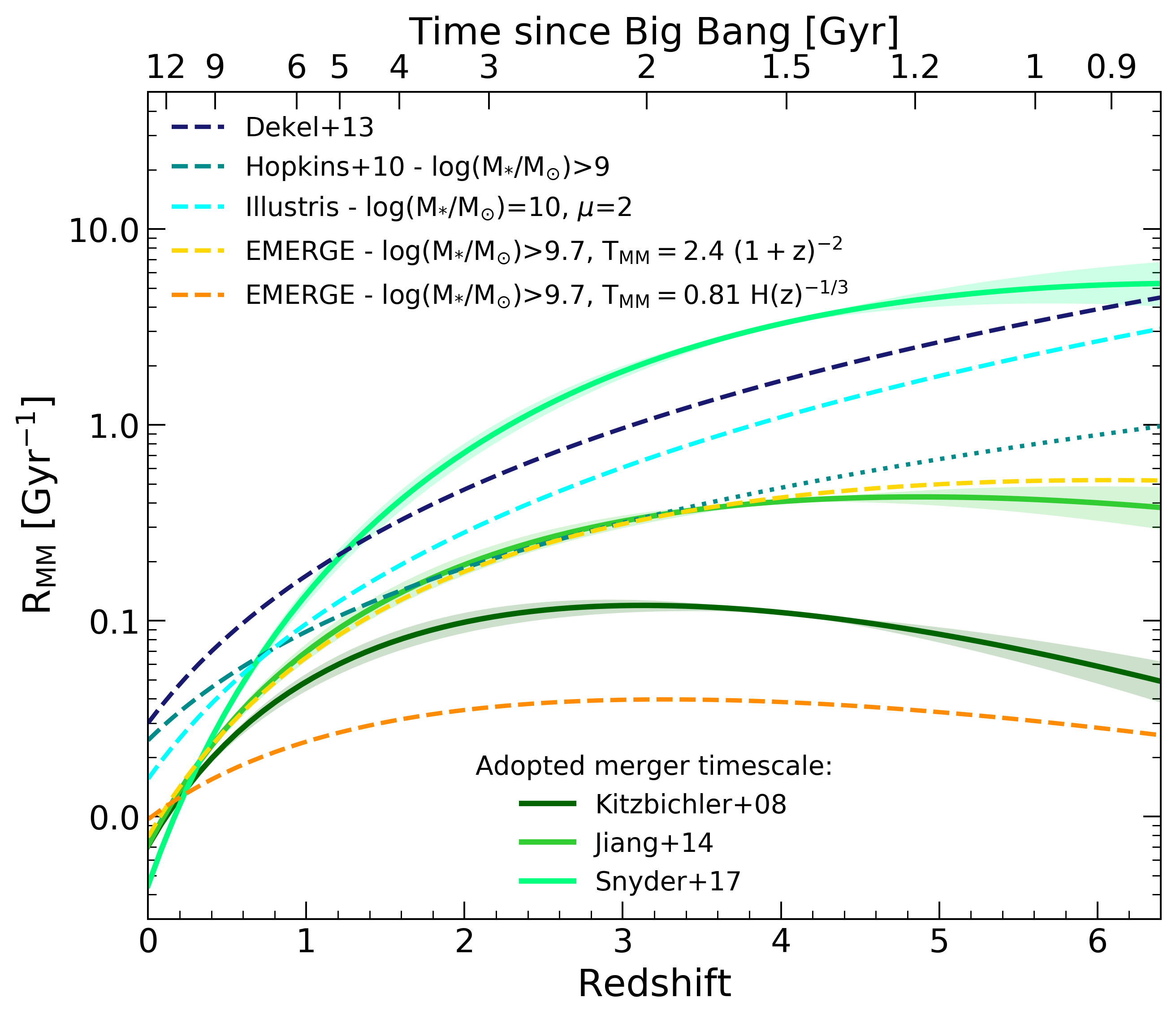}}
	\subfigure{\includegraphics[width=\columnwidth]{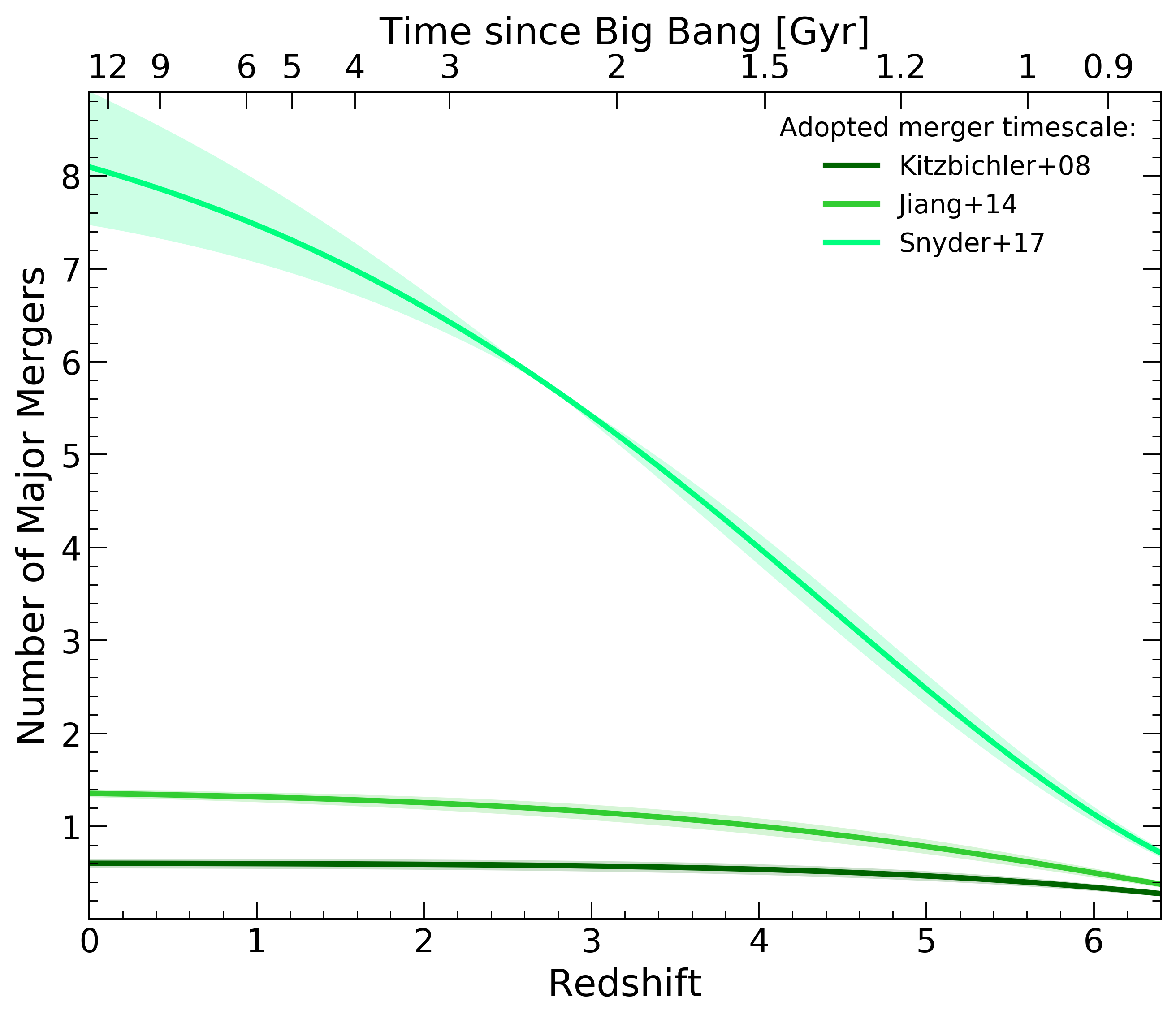}}
    \caption{\textit{Left panel:} Redshift evolution of the major merger rate. The solid lines with the shaded regions represent the best-fitting functions to the data assuming three different merger timescales and their associated 1$\sigma$ uncertainties, respectively. The cosmic evolution of $R_{\text{MM}}$ computed from the halo-halo merger rate by \cite{Dekel13}, from the empirical model by \cite{Hopkins10}, and from the Illustris \citep{Rodriguez15} and EMERGE \citep{Oleary21} simulations are also shown with dashed dark blue, turquoise, cyan, yellow and orange lines. The dotted turquoise line shows the extrapolation of the \cite{Hopkins10} merger rate to higher redshifts. \textit{Right panel:} Cumulative number of major mergers per galaxy over cosmic time. Solid lines represent the cumulative distributions obtained by integrating Equation \ref{eq:Nmm} adopting the merger timescale prescriptions of \cite{Kitzbichler08}, \cite{Jiang14} and \cite{Snyder17}. The shaded regions are the associated uncertainties computed from the best-fit errors on the merger rate cosmic evolution.}
    \label{fig:merger_rate}
\end{center}
\end{figure*}



\subsection{Galaxy major merger rate}\label{sec:merger_rate}
The major merger fraction can be translated into the merger rate $R_{\text{MM}}$ (i.e., the number of mergers per galaxy and Gyr) as
\begin{equation}
    R_{\text{MM}}(z) = \frac{C_{\text{merg}}~f_{\text{MM}}(z)}{T_{\text{MM}}(z)},
    \label{eq:Rmm_evol}
\end{equation}
where $f_{\text{MM}}(z)$ is the pair fraction at redshift $z$ estimated in Section \ref{sec:merger_frac}, and $C_{\text{merg}}$ is the fraction of close pairs that will eventually merge into a single system in a typical timescale $T_{\text{MM}}(z)$. This last term represents the major source of uncertainty in the merger rate computation and is usually obtained through simulations based on the dynamical friction timescales affecting the merging components (e.g., \citealt{Kitzbichler08,Lotz10,Jiang14,Snyder17}). 

\cite{Kitzbichler08} applied a semi-analytic model to the Millennium simulation \citep{Springel05} outputs finding an average merging time that depends linearly on the redshift and projected distance of the pair and weakly on the stellar mass of the main galaxy, namely $T_{\text{MM}}\propto r_p~M_{*}^{-0.3}~(1+z/8)$. This relation is valid at $z\leq1$, while at higher redshift it becomes
\begin{equation}
    T_{\text{MM}}^{-1/2} = T_0^{-1/2}+f_1z+f_2(\mathrm{log}M_{*}-10),
    \label{eq:Tmm_kitz}
\end{equation}
where $T_0$ is the merging time at $z=0$, $M_{*}$ is the stellar mass of the primary galaxy, while $f_1$ and $f_2$ are two coefficients of the parameterization. A mass dependence similar to the $z\leq1$ \cite{Kitzbichler08} timescale ($T_{\text{MM}}\sim M_{*}^{-0.2}$) was also found by \cite{Lotz10} for mergers in the stellar mass range $9.7<\mathrm{log}(M_{*}/M_{\odot})<10.7$ but in a smaller sample. More recent works explored a different redshift evolution of the merger timescale. For instance, \cite{Jiang14} took into account the mass loss due to dynamical friction of the virial masses of galaxies experiencing a merger and, by using a high-resolution $N-$body cosmological simulation, found $T_{\text{MM}}\propto H(z)^{-1/3}$, where $H(z)$ is the Hubble parameter at redshift $z$. An even stronger redshift evolution was found by \cite{Snyder17} by comparing mass-selected close pairs with the intrinsic galaxy merger rate from the Illustris simulations \citep{Genel14,Vogelsberger14}. They found $T_{\text{MM}}=2.4~(1+z)^{-2}$ but for primary galaxies with stellar mass range $10.5<\mathrm{log}(M_{*}/M_{\odot})\leq11$. It is better to specify that, in the case of \cite{Snyder17}, $T_{\text{MM}}$ does not correspond to the merger timescale of Equation \ref{eq:Rmm_evol}, rather it is the merger observability timescale which already includes the effects of the $C_{\text{merg}}$ term. Nevertheless, we refer to that as $T_{\text{MM}}$ for simplicity.

For the rest of this work, we compare the results obtained with the merger timescales provided by \cite{Kitzbichler08}, \cite{Jiang14} and \cite{Snyder17}, in order to highlight the large differences in the final outcome due to the choice of this parameter. About the probability for a pair to merge over a certain timescale (i.e., $C_{\text{merg}}$), it is usually fixed to 0.6 (e.g., \citealt{Lotz11,Mantha18}). However, as in most cases this quantity is already included in the merger timescale prescriptions provided in different works (e.g., \citealt{Duncan19}), we set $C_{\text{merg}}=1$ throughout.

With these assumptions, we present in Figure \ref{fig:merger_rate} (left panel) the cosmic evolution of the merger rate, as obtained by combining our new $z>4$ ALPINE data with those at lower redshifts from the literature. For each data point we compute $R_{\text{MM}}$ from Equation \ref{eq:Rmm_evol}, adopting the three different merger timescales defined above\footnote{When adopting the merger timescale by \cite{Jiang14}, we just consider the $T_{\text{MM}}$ redshift evolution found in their work (i.e., $T_{\text{MM}}\sim H(z)^{-1/3}$) and normalize it to $T_{\text{MM}}(z=0)$ from \cite{Kitzbichler08}. This is also similar to setting a constant merger timescale of $\sim1$~Gyr.} and fit the resulting data with the same functional form used for the merger fraction. We show the best-fitting functions to these data (and their $1\sigma$ uncertainties) in Figure \ref{fig:merger_rate} and report the corresponding parameters in Table \ref{tab:fit_params}.

As evident, all the three functions show a steep increase of the merger rate from the local Universe to $z\sim1$. However, at higher redshifts they lead to large differences about the incidence of these sources at early epochs. When considering the \cite{Kitzbichler08} timescale, we find a redshift evolution which is similar to what obtained in Section \ref{sec:merger_frac} for the merger fraction, with a peak at $z\sim2.5$ and a quite fast decrease at higher redshifts. Both the merger rate evolutions obtained with $T_{\text{MM}}(z)$ by \cite{Jiang14} and \cite{Snyder17} show instead a milder redshift attenuation at $z\gtrsim4$. In particular, the increasingly smaller merger timescale found by \cite{Snyder17} involves a very large merger rate in the early Universe compared to the other functions, possibly implying a significant role of major mergers in the galaxy mass assembly at $z\sim5$.

We then compare our results with the cosmic evolution of several expected merger rates from simulations. At $z<3$, \cite{Hopkins10} found that $R_{\text{MM}}\propto (1+z)^{1.8}$ for galaxies with $\mathrm{log}(M_{*}/M_{\odot})>9$ and $\mu<3$. This redshift evolution is comparable to our best-fitting function with $T_{\text{MM}}(z)\propto H(z)^{-1/3}$ at $1\lesssim z \lesssim 3$. However, if we extrapolate their results up to $z\sim6$, we find that their predicted major merger rates could be larger (by $\sim0.5$~dex) with respect to our observations. A stronger and increasing redshift dependence was found by \cite{Rodriguez15} for the merger rate from the Illustris simulations. By exploiting their parametric fitting formula with a stellar mass\footnote{It is worth noting that, as also stated by \cite{Snyder17} and \cite{Oleary21}, the mass ratio defined by \cite{Rodriguez15} is evaluated at the time when the secondary galaxy has achieved its maximum stellar mass. This prevent us to make a direct comparison among the results of this and the other simulations.} of $\mathrm{log}(M_{*}/M_{\odot})=10$ and an average mass ratio $\mu=2$ over cosmic time (see Section \ref{sec:specific_mar}), the simulation tends to over-predict our observed merger rates at $z\sim0$, while it underestimates them at intermediate redshifts. We also show the merger rate predictions found by \cite{Oleary21} using the results from the Empirical ModEl for the foRmation of GalaxiEs (EMERGE; \citealt{Moster18}). In particular, they modeled the merger fraction evolution as in Equation \ref{eq:fmm_evol} and then convolved it with a typical merger timescale to obtain $R_{\text{MM}}$ as a function of redshift. We consider their best-fit parameters describing $f_{\text{MM}}$ for galaxies with $\mathrm{log}(M_{*}/M_{\odot})\geq9.7$ and projected distance $r_p<30$~kpc. When convolving this $f_{\text{MM}}(z)$ with a merger timescale $T_{\text{MM}}\propto H(z)^{-1/3}$ and with a power-law scaling as $T_{\text{MM}}\propto (1+z)^{-2}$ we obtain a lower and milder $R_{\text{MM}}$ cosmic evolution with respect to our major merger rate results obtained with a \cite{Kitzbichler08} timescale, and a comparable trend with the \cite{Jiang14}-based evolution, respectively. Lastly, we show the predicted halo-halo merger rate by \cite{Dekel13} which over-predicts the observed merger rates at low redshifts but is in good agreement with the \cite{Snyder17}-based $R_{\text{MM}}$ at $z\sim6$. 

We compute the average number of mergers a galaxy undergoes between $0<z<6$ by integrating the merger rate over time as
\begin{equation}
    N_{\text{MM}} = \int_{z_1}^{z_2}{\frac{R_{\text{MM}}(z)}{(1+z)H_0E(z)}dz},
    \label{eq:Nmm}
\end{equation}
where $E(z)=\sqrt{\mathrm{\Omega_m}(1+z)^3+\mathrm{\Omega_\Lambda}}$, assuming a flat Universe (i.e., $\mathrm{\Omega_k}=0$). We show in Figure \ref{fig:merger_rate} (right panel) the cumulative number of major mergers as a function of redshift adopting three different timescales, with the 1$\sigma$ uncertainties derived from the errors on the major merger rate cosmic evolution. The average number of mergers computed with the \cite{Kitzbichler08} and \cite{Jiang14} prescriptions is low, ranging between 0.5 and 1.4 down to $z=0$. This is consistent, for instance, with the results by \cite{Mundy17} who found that galaxies with $\mathrm{log}(M_{*}/M_{\odot})>9.5$ at $z\sim3.5$ (selected at constant number densities, thus representing the progenitors of $\mathrm{log}(M_{*}/M_{\odot})>11$ galaxies at $z\sim0$) undergo $\sim0.5$ major mergers between $0<z<3.5$. However, if we assume an evolving merger timescale as $2.4~(1+z)^{-2}$ \citep{Snyder17}, the average number of mergers since $z=6$ increases significantly, reaching $N_{\text{MM}}\sim8$ at the present day. Interestingly, this is similar to the result obtained by \cite{Conselice09} that found $N_{\text{MM}}\sim7$ by integrating Equation \ref{eq:Nmm} from $z=6$ to 0, and assuming a constant merger timescale $T_{\text{MM}}=0.35$ (which is equal to the average timescale by \citealt{Snyder17} at $z<6$; see their Figure 15). When examining our results at higher redshifts, we find that the number of mergers diminishes quickly. In particular, at $z>4$, $N_{\text{MM}}\lesssim1$ if we consider the \cite{Kitzbichler08} and \cite{Jiang14} merger timescale evolution, suggesting that, in these scenarios, not every galaxy will undergo a merger during this epoch.


\subsection{Merger rate density}
\begin{figure}
    \begin{center}
	\includegraphics[width=\columnwidth]{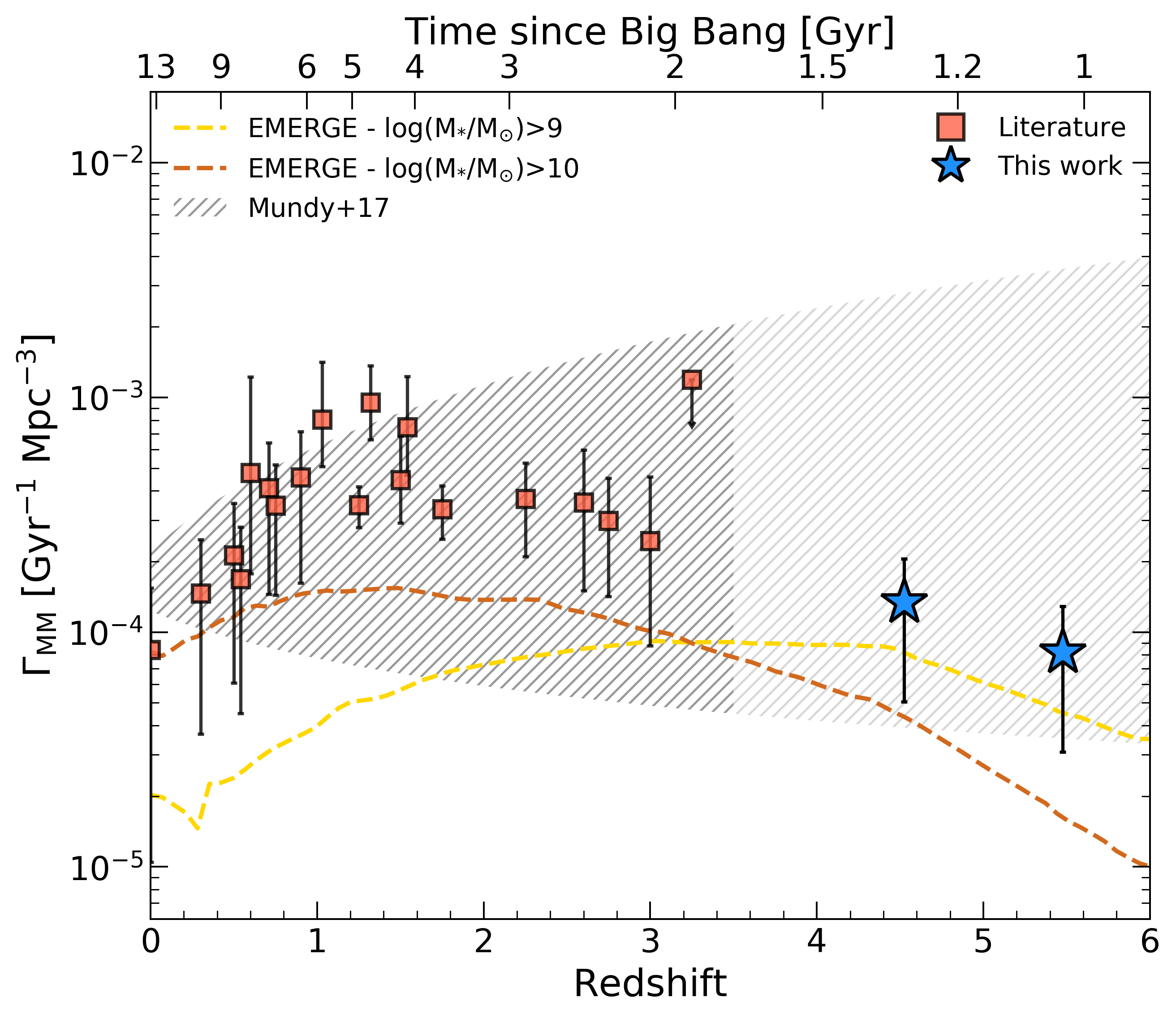}
	\end{center}
    \caption{Merger rate density as a function of redshift. The squares represent the volume-averaged merger rates computed from the literature data at $z<4$ adopting the \cite{Kitzbichler08} prescription for the merger timescale. The blue stars are the ALPINE data at $z\sim5$ from this work. The error bars are obtained by combining in quadrature the uncertainties on the merger fractions and on the number densities. The darker hatched area reports the results including errors by \cite{Mundy17} at $z<3.5$. The lighter area only shows their extrapolation to higher redshifts. The dashed yellow and brown lines display the evolution of $\Gamma_{\text{MM}}$ found with the EMERGE simulation \citep{Oleary21} for galaxies with $\mathrm{log}(M_{*}/M_{\odot})>9$ and $\mathrm{log}(M_{*}/M_{\odot})>10$, respectively.} 
    \label{fig:merger_rate_dens}
\end{figure}

As discussed in Section \ref{sec:merger_rate}, the merger rate traces the number of merger events per galaxy at a given mass and redshift. A more informative quantity is the volume-averaged merger rate which is defined as
\begin{equation}
    \Gamma_{\text{MM}}(z) = \frac{f_{\text{MM}}(z)~n(z)}{T_{\text{MM}}(z)},
    \label{eq:Gamma}
\end{equation}
where $f_{\text{MM}}$ and $T_{\text{MM}}$ are the previously defined merger fraction and merger timescale, and $n(z)$ is the number density of galaxies in Mpc$^{-3}$. The latter is computed by integrating the galaxy stellar mass function (GSMF) in a certain redshift bin and between a minimum and maximum stellar mass limit in the range $M_{*}^\text{min}<M_{*}<M_{*}^\text{max}$. In particular, we exploit the GSMFs best-fit parameters from \cite{Mortlock15}, \cite{Santini12} and \cite{Davidzon17} at $z<3$, $3<z<4$ and $z>4$, respectively, after converting into a \cite{Chabrier03} IMF. However, it is worth noting that, as stated in \cite{Mundy17}, making use of different GSMF parameterizations does not significantly change the final results. We thus integrate these functions from $\mathrm{log}(M_{*}/M_{\odot})=9$ to $\mathrm{log}(M_{*}/M_{\odot})=11$, in order to include the full possible range of galaxy masses for the considered data at different redshifts. We estimate the errors on $n(z)$ by perturbing the corresponding GSMF with the associated uncertainties on its best-fit parameters at each redshift. We repeat this procedure 10$^{4}$ times, recomputing the integrated number densities at each step. From the perturbed distributions we then obtain the 16th and 84th percentiles as the 1$\sigma$ errors on $n(z)$. 

Figure \ref{fig:merger_rate_dens} shows the merger rate density computed with Equation \ref{eq:Gamma} for each of the data introduced in Section \ref{sec:merger_frac} and for the two $z>4$ ALPINE bins. For the sake of simplicity, in this case we report only the data points obtained by adopting the \cite{Kitzbichler08} prescription for the merger timescale. When using the $T_{\text{MM}}(z)$ redshift evolution from \cite{Jiang14} and \cite{Snyder17}, we obtain similar trends but shifted to higher merger rate densities. The error bar on each point is computed by propagating the merger fraction and number density uncertainties on Equation \ref{eq:Gamma}. At $1\lesssim z \lesssim 5$, \cite{Duncan19} found that the volume-averaged merger rate is quite constant. On the other hand, from our data we find a slight decrease both at $z<1$ and at $z>4$. At these early epochs, this difference could be caused by the poor constraints on the GSMFs adopted by \cite{Duncan19} which result in large uncertainties on their data, making it impossible to draw robust conclusions at $z\sim5$. Nevertheless, our derived merger rate densities are in agreement, within the uncertainties, with other results derived in the literature. For example, \cite{Mundy17} studied the evolution of the merger rate density up to $z\sim3.5$ for a large sample of $\mathrm{log}(M_{*}/M_{\odot})>10$ galaxies. They found that, for close pairs with projected separations $5<r_p<20~$kpc, the $\Gamma_{\text{MM}}(z)$ evolution is better described by a power-law of the form $\Gamma_0(1+z)^{\gamma}$, with $\Gamma_0 = 1.64^{+0.58}_{-0.41} \times 10^{-4}$ and $\gamma = {0.48}^{+1.00}_{-1.15}$. We report their results, along with the uncertainties, in Figure \ref{fig:merger_rate_dens} and extrapolate them to the redshifts explored by ALPINE. As evident, our data points are comparable with the results by \cite{Mundy17}. If we also fit our merger rate densities (derived assuming the \cite{Kitzbichler08} merger timescale) with a power-law function we obtain $\Gamma_0 = (2.40\pm0.63) \times 10^{-4}$ and $\gamma = -0.16\pm0.23$, that are consistent with the above outcomes. The power-law fit on the data computed with the timescales by \cite{Jiang14} is in agreement with the \cite{Mundy17} findings, as well. If we instead consider the results obtained with the \cite{Snyder17} timescale, we find a rapid increase of the merger rate density with redshift, which departs from the upper envelope of \cite{Mundy17} already at $z\gtrsim1$.

Finally, we also show the results from the EMERGE simulation \citep{Oleary21} for galaxies in two different stellar mass bins. The two curves in the figure represent the intrinsic merger rate from the simulation for major mergers (i.e., $\mu<4$) selected basing on their main progenitor mass, thus describing a population of galaxies that will undergo a merger after a certain time. Galaxies with $\mathrm{log}(M_{*}/M_{\odot})>9$ are characterized by an increase of the merger rate density up to $z\sim1$, an almost constant $\Gamma_{\text{MM}}$ until $z\sim4$ and a slow decrease to the highest redshifts. At larger stellar masses, a slighter increase of $\Gamma_{\text{MM}}$ is present at low redshifts, while a steeper decrease in the early Universe is found. Both these trends are comparable to the one obtained in this work after exploiting data at different redshifts and including the new estimates at $z>4$ from the ALPINE survey. The fact that they are systematically smaller than our results could be ascribed to several factors, such as the lower merger fraction evolution found in their simulation or the measure of the intrinsic merger rate which is computed when the merging process is already happened (thus possibly enhancing the merger timescales with respect to our assumptions).


\subsection{The major merger specific mass accretion rate}\label{sec:specific_mar}
\begin{figure*}
    \begin{center}
    \subfigure{\includegraphics[width=\columnwidth]{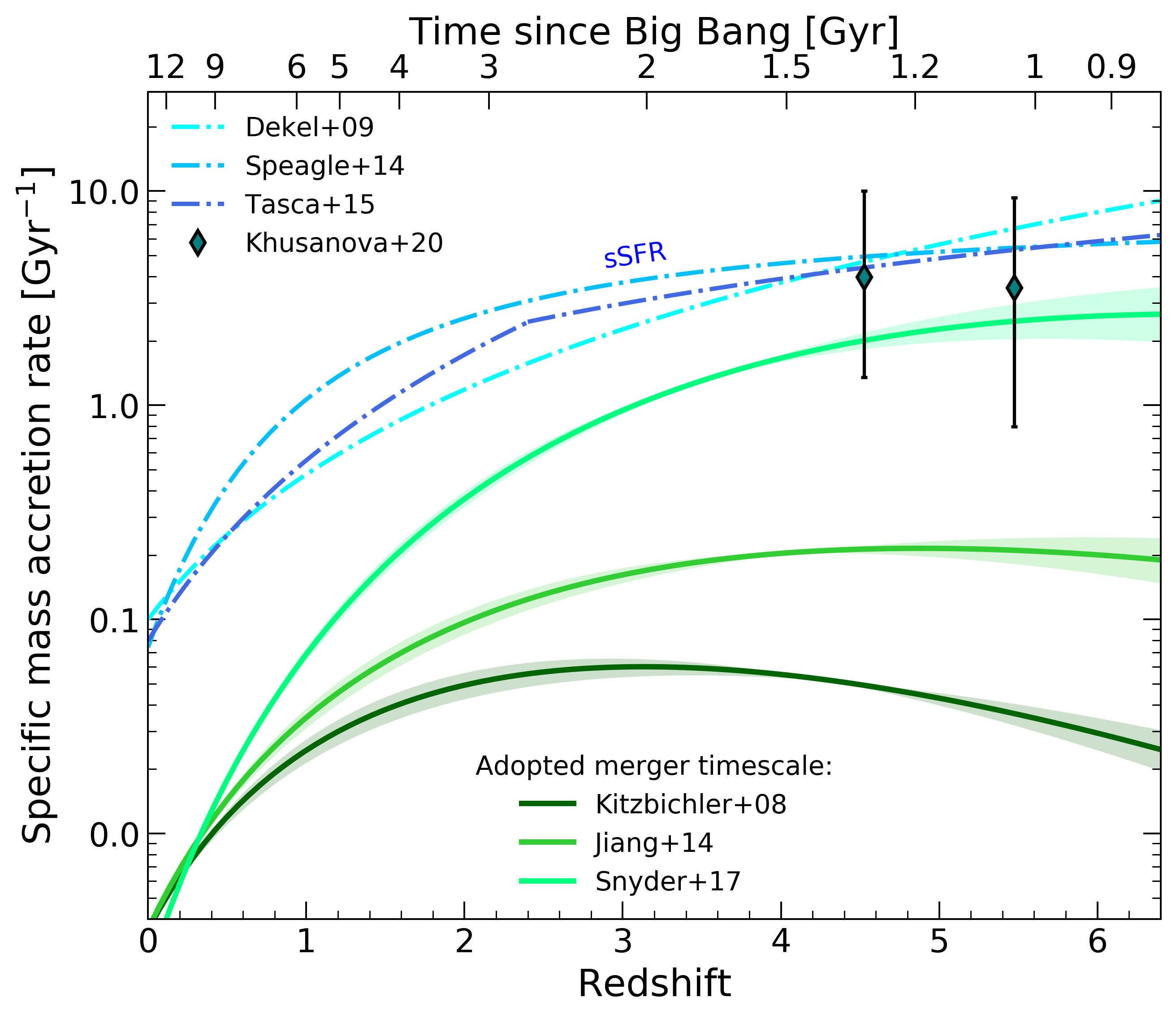}}
	\subfigure{\includegraphics[width=\columnwidth]{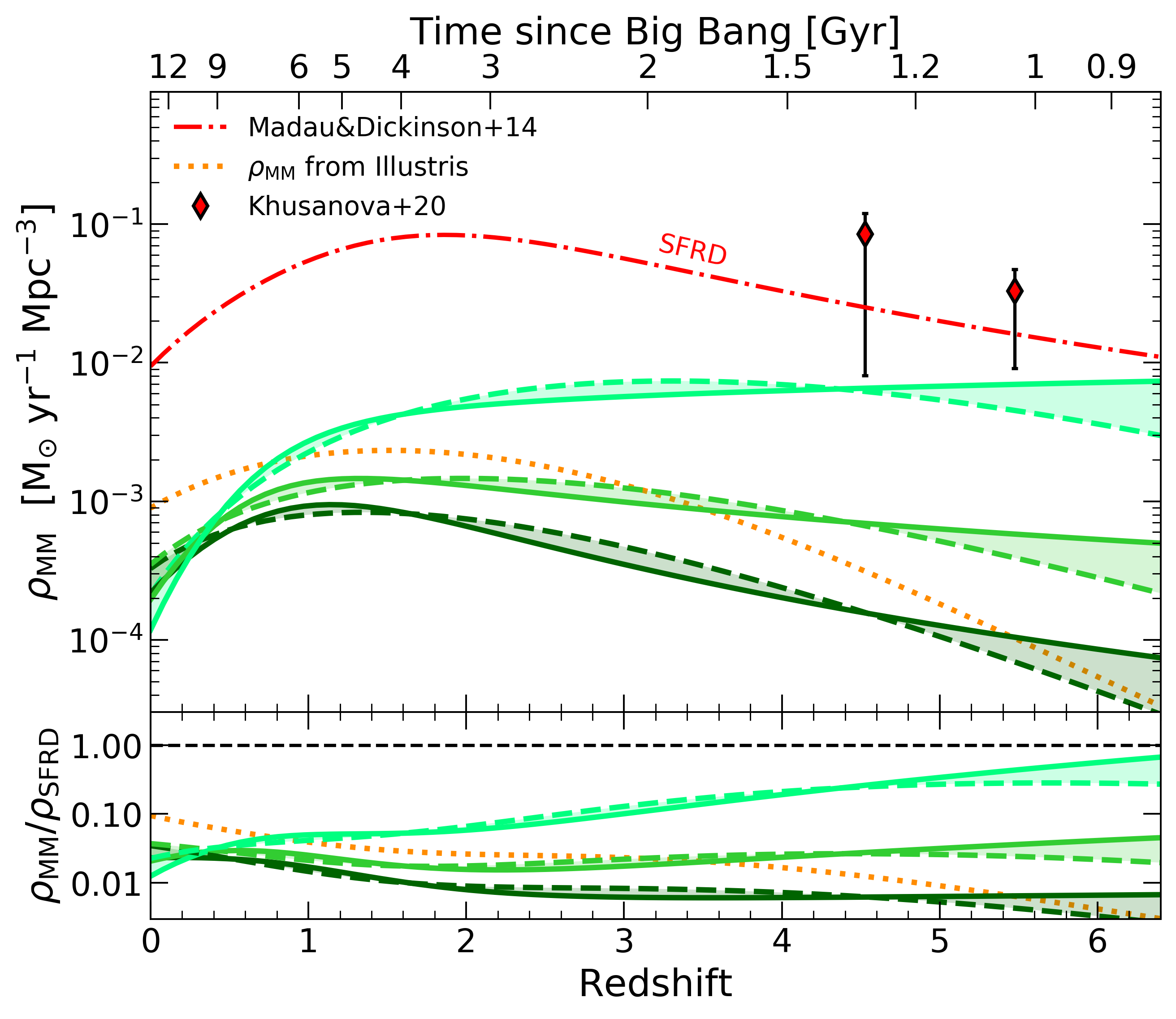}}
	\end{center}
    \caption{\textit{Left:} Comparison between the redshift evolution of the specific mass accretion rate (as derived from $R_{\text{MM}}$) and the specific star formation rate. The solid lines with the shaded regions are the best-fitting functions (and corresponding uncertainties) to the same $R_{\text{MM}}$ data of the left panel of Figure \ref{fig:merger_rate} at different merger timescales divided by the average mass ratio $\overline{\mu}$ through cosmic time. The dot-dashed lines are several specific star formation rates from the literature \citep{Dekel09,Speagle14,Tasca15}. The sSFR by \cite{Dekel09} is normalized to sSFR$(z=0)=0.1$, as in \cite{Tasca15}. The dark cyan diamonds are the values by \cite{Khusanova21} from the ALPINE survey. \textit{Right:} Stellar mass accretion rate density ($\rho_{\text{MM}}$) as a function of redshift (top panel). Solid and dashed lines represent the best-fits to the data assuming double power-law, and combined power-law and exponential functions, respectively. The shaded regions highlight the uncertainties resulting from the choice of the fitting form. The colors correspond to alternative merger timescales and are the same as in the left panel. The dotted curve reports the mass accretion rate density, $\rho_{\text{ill}}$, obtained from the Illustris simulation, as described in the text. The dot-dashed red line shows the SFRD ($\rho_\mathrm{SFRD}$) by \cite{Madau14}. The red diamonds are the total SFRD values obtained from the ALPINE survey \citep{Khusanova21}. The bottom panel displays the ratio between $\rho_{\text{MM}}$ and $\rho_\mathrm{SFRD}$ as a function of cosmic time. The dashed horizontal line marks a ratio equal to 1.} 
    \label{fig:merger_ssfr}
\end{figure*}

By taking advantage of the merger rates and merger fraction deduced in the previous sections, we now provide an estimate of the average stellar mass gained by a galaxy during a merger event over cosmic time. To this aim, we compute the major merger specific mass accretion rate as $R_{\text{MM}}~\overline{\mu}^{-1}$, where $\overline{\mu}$ is the average mass ratio. We find that, for each data sample used to compute the merger rate at different redshifts, $\overline{\mu}^{-1}\sim0.5$, including the average mass ratio from our sample of major mergers for which we know $\mu_K$. Therefore, we just take the best-fitting functions to the cosmic evolution of the merger rate and multiply them by a factor $\sim0.5$ to obtain the specific mass accretion rate in Gyr$^{-1}$. Our results are reported in Figure \ref{fig:merger_ssfr} along with the specific star formation rate (sSFR) evolution obtained by different authors \citep{Dekel09,Speagle14,Tasca15,Khusanova21}. As shown, star formation seems to be the dominant mode of mass growth at all epochs. However, if we assume a merger timescale $T_{\text{MM}}\propto (1+z)^{-2}$ \citep{Snyder17}, the specific mass accretion rate reaches the sSFR at $z>3$, being comparable to it within the uncertainties. Therefore, we cannot exclude that major mergers may have significantly contributed to the galaxy mass assembly in the early Universe. In particular, if the galaxy growth is dominated by cold gas accretion, the sSFR should evolve with redshift as $(1+z)^{2.25}$ (e.g., \citealt{Dekel09}). Some authors find instead a flattening of the trend or even a possible decrease with redshift as from the ALPINE data \citep{Tasca15,Faisst16,Khusanova21}, leaving space to other mechanisms that could regulate the assembly of galaxies at high redshifts, such as major mergers (e.g., \citealt{Tasca15,Faisst16}).


\subsection{The major merger mass accretion rate density}
Similarly to what done in Section \ref{sec:specific_mar} for the specific mass accretion rate, we exploit here the major merger rate density to obtain an estimate of the mass accreted through major mergers per unit time and volume for galaxies in a given stellar mass range, $\rho_{\text{MM}}(z)$. This quantity can be compared to the other mechanisms of mass accretion, such as the mass gained through the process of star formation, and is then of fundamental importance to understand the role of mergers in the Universe. 

Following \cite{Duncan19}, we assume that the increase in stellar mass for each merger and at each redshift is given by $\overline{M}_{*}~\overline{\mu}^{-1}$, where $\overline{\mu}$ is the average mass ratio defined in Section \ref{sec:specific_mar} and $\overline{M}_{*}$ is the average stellar mass calculated through the GSMF. The latter is computed as
\begin{equation}
    \overline{M}_{*}(z) = \frac{\int_{M_{*}^\text{min}}^{M_{*}^\text{max}}{\Phi(z,M_{*})~M_{*}~dM_{*}}}{\int_{M_{*}^\text{min}}^{M_{*}^\text{max}}{\Phi(z,M_{*})~dM_{*}}},
    \label{eq:average_M}
\end{equation}
where $\Phi(z,M_{*})$ represents the shape of the GSMF in a certain redshift bin. We estimate the uncertainties on $\overline{M}_{*}(z)$, as previously done for the number density, by recomputing it N times after perturbing the corresponding GSMF with its associated errors. At each redshift, we obtain a typical uncertainty of $\sim0.5~$dex. In this way, the stellar mass accretion rate density can be estimated as
\begin{equation}
    \rho_{\text{MM}}(z) = \overline{\mu}^{-1}~\overline{M}_{*}~\Gamma_{\text{MM}}(z).
    \label{eq:rho_mm}
\end{equation}
We report the cosmic evolution of $\rho_{\text{MM}}$ in Figure \ref{fig:merger_ssfr} (right panel, on top). The dashed lines show the best-fitting curves to the data assuming a combined power-law and exponential function, as done for the merger fraction and merger rate, for different merger timescales. As shown, at $z>1$ the curves start to decrease quite rapidly to lower mass accretion rate densities.

Following \cite{Mundy17}, we then provide the major merger accretion rate density computed with the results from the Illustris simulation, $\rho_{\text{ill}}(z)$. We convolve the specific mass accretion rate by \cite{Rodriguez16} with the number density of galaxies by \cite{Torrey15} as
\begin{equation}
    \rho_{\text{ill}}(z) = \int_{M_{*}^\text{min}}^{M_{*}^\text{max}}\int_{\mu_\text{min}}^{\mu_\text{max}}
    \label{eq:rho_ill}{\Phi_{\text{ill}}(z,M_{*})~{\dot{m}}_\text{accr}(z,M_{*},\mu)~d\mu~dM_{*}},
\end{equation}
where $\Phi_{\text{ill}}$ is the shape of the GSMF at redshift $z$ and stellar mass $M_{*}$, and $\dot{m}_\text{accr}$ represents the average amount of mass accreted by a single galaxy when the merger occurs. As done for $\rho_{\text{MM}}$, we integrate the GSMF in the range $9<\mathrm{log}(M_{*}/M_{\odot})<11$ and the mass accretion rate for mass ratios between $1<\mu<4$, accounting for major mergers. The result of the integration is reported in Figure \ref{fig:merger_ssfr} as a function of redshift. As can be seen, the computed $\rho_{\text{ill}}$ is comparable with our findings, lying between the best-fit curves obtained by adopting the \cite{Kitzbichler08} and \cite{Snyder17} prescriptions for the merger timescale, respectively. Moreover, the shape of the curve is quite similar to those deduced by our analysis, with a fast decrease at high redshifts.

To compare the relative contribution of mergers and star formation to the mass assembly of galaxies through the cosmic time, we also show the star-formation rate density (SFRD) from \cite{Madau14}, after converting for a Chabrier IMF\footnote{To convert SFRs from \cite{Salpeter55} to Chabrier IMF we multiply by 0.63 (e.g., \citealt{Madau14}).}. Moreover, we find that our data are also well fit by a double power-law of the form assumed by \cite{Madau14} for the SFRD, although we find no significant differences between the two functional forms when comparing them with a BIC statistics. The results of these fits are reported in the figure (solid lines), showing a flatter trend to the early Universe with respect to the combined power-law and exponential curves. In the bottom region of the figure we show the ratio between the major merger and star formation contributions to the stellar mass accretion in galaxies at different epochs. Looking at both curves, if we assume the merger timescales by \cite{Kitzbichler08} and \cite{Jiang14}, major mergers seem to contribute less than $10\%$ to the galaxy mass assembly compared to the star formation mechanism at all redshifts. For comparison we also report, in the top panel, the total (UV+IR) SFRD obtained with the ALPINE data \citep{Khusanova21}. These measurements indicate a possible $z>4$ evolution of the SFRD that is shallower than previously thought, further decreasing the importance of major mergers to the mass assembly of galaxies at these epochs. On the other hand, adopting the \cite{Snyder17} timescale prescription, a larger contribution of mergers in the early Universe is in place which becomes comparable to that provided by the star formation at $z\gtrsim6$.


\section{Discussion}
\subsection{The importance of the merger timescale}
As evidenced by the results presented in Section \ref{sec:results}, one of the major sources of uncertainty in the investigation of the contribution of major mergers to the mass assembly of galaxies through cosmic time is the typical time needed for a pair of objects to coalesce with each other. Indeed, this parameter is critical for converting the pair fraction into a merger rate, affecting all the derived quantities as well. Because of this, it could represent the main reason of disagreement between models and observations noticed in many works. 

Previous studies often assumed $T_{\text{MM}}$ as a constant over time. However, the emerging picture of a possibly decreasing (as found in this work) or nearly flat merger fraction through high redshifts \citep{Conselice09,Ventou17,Mantha18}, compared to the increase of the merger rate found in most simulations over the same epoch (e.g., \citealt{Hopkins10,Rodriguez15}), suggests that a redshift-dependent merger timescale is more suitable to reconcile models and observations. Here we show the results derived from three different merger timescales \citep{Kitzbichler08,Jiang14,Snyder17}, as introduced in Section \ref{sec:merger_rate}, and compare them with state-of-the-art simulations. \cite{Kitzbichler08} found a merger timescale which slightly depends on the primary galaxy mass and that increases with redshift, leading to lower values of the merger rate when moving to early epochs. On the other hand, \cite{Jiang14} and \cite{Snyder17} found a slow and fast decrease of $T_{\text{MM}}$ over time, respectively, implying a larger contribution from mergers to the galaxy mass assembly at $z>4$. 

Looking at Figure \ref{fig:merger_rate} (left panel), the merger rates obtained from the \cite{Kitzbichler08} timescale prescription are not easily reproducible by models. In fact, although we find values of $R_{\text{MM}}$ comparable to those of the EMERGE simulation \citep{Oleary21} at $z\gtrsim5$, the latter are computed with a \cite{Jiang14}-like evolving timescale which can provide $T_{\text{MM}}$ values up to 5 times lower than the \cite{Kitzbichler08} ones at these epochs.
A good agreement with the $ T_{\text{MM}}\propto H(z)^{-1/3}$ merger rates is found by convolving the EMERGE pair fraction evolution with the decreasing \cite{Snyder17} timescale (again using smaller merging times in the simulation with respect to the observations) and with the outputs of the \cite{Hopkins10} simulations, at least at $z<3$ (indeed, they predicted larger merger rates with respect to our findings at earlier epochs). However, in this case, we note that the merger rates from \cite{Hopkins10} are properly computed up to $z\sim3$ and that the high-redshift extrapolation could be different from the real outputs of their simulation. Moreover, the observed shift between our findings and the EMERGE results due to the different choice of the merger timescale could be ascribed to the merger fraction evolution which, from their simulation, is smaller than ours. Finally, the fast-decreasing merger timescale by \cite{Snyder17} lead to merger rates which are more than an order of magnitude larger than those from the \cite{Kitzbichler08} and \cite{Jiang14} prescriptions. Adopting this formalism, we find that our merger rates are larger than those predicted by the Illustris simulations \citep{Rodriguez15} (with an average mass ratio $\sim2$) and comparable with the merger rates of $M_\text{halo}\sim10^{12}~M_{\odot}$ by \cite{Neistein08,Dekel13} at $z\sim6$.

\cite{Duncan19} found a good agreement between their observations and simulated merger rates from \cite{Rodriguez15} for galaxies with $\mathrm{log}(M_{*}/M_{\odot})>10.3$, while they also obtained higher values than those predicted for lower mass galaxies. However, the results of that simulation depend significantly on the adopted mass ratio. Indeed, assuming $\mu>2$, the Illustris simulation is able to reproduce our observations, at least at $z\gtrsim4$. Moreover, \cite{Mantha18} also found merger rates comparable to those predicted by \cite{Rodriguez15} up to $z\sim3$ by dividing their observed pair fraction by the \cite{Snyder17} merger timescale.  

These results suggest that to reconcile observations and simulations, a redshift-evolving timescale should be employed when converting the pair fraction into a merger rate, with a $T_{\text{MM}}$ decreasing toward early epochs likely being the most suitable choice. However, many other factors can affect the $R_{\text{MM}}$ estimates, like the probability of merging $C_{\text{merg}}$ or the dependency of the merger timescale on several properties like the stellar mass of the galaxies, the on-going phase of the merger or the selection criteria of the close pairs. For these reasons, further investigation is needed in order to obtain more robust conclusions on the incidence of merger through cosmic time.  

\subsection{The contribution of major mergers to the galaxy mass-assembly}
The relative contribution of the different mechanisms driving the galaxy mass assembly through cosmic time has yet to be ascertained. The most favored scenario predicts cold accretion and in situ star formation as the principal sources of stellar mass increase in galaxies, with major mergers only playing a minor role in the galaxy build-up (e.g., \citealt{Dekel09,Conselice13,Kaviraj13,Sanchez14}). However, during the last years several works started to suggest a prominent contribution of mergers within this context, opening again the debate on the importance of these events in the framework of galaxy evolution (e.g., \citealt{Tasca14,Mantha18,Duncan19}).

Depending on the merger timescale adopted, with our new ALPINE data we can find large merger rates up to high redshifts which could imply a substantial contribution of these events to the galaxy mass accretion at early epochs. We find that the stellar mass accretion rate density due to major mergers has a similar redshift evolution as the cosmic SFRD \citep{Madau14} and that, assuming the \cite{Snyder17} merger timescale, they become comparable with each other at redshifts approaching the Reionization epoch. However, the lack of further constraints on the most suitable choice for the merger timescale prevent us from providing firm conclusions. Indeed, looking at Figure \ref{fig:merger_ssfr} (right panel), the importance of major mergers in the assembly of galaxies could be very different, especially at high redshifts, depending on all the above-mentioned caveats and also on the adopted fitting function. It is worth noting that the comparison between the SFRD and the mass accretion rate density in Figure \ref{fig:merger_ssfr} is not intended to provide the contribution of major mergers to the global star formation through cosmic time, rather it represents the relative importance of the two processes to the galaxy mass assembly at a given epoch.

Despite this, we can certainly affirm that major mergers are frequent processes in the early Universe and must be accounted for when considering the mechanisms of galaxy assembly across cosmic time.

\section{Summary and conclusions}
In this work, we investigate the role of major mergers in the stellar mass assembly of galaxies at $z\sim5$. To this aim, we take advantage of the recent [CII] observations of a significant statistical sample of normal star-forming galaxies in the early Universe, as carried out by the ALMA large program ALPINE. The large amount of data (both spectroscopic and photometric) available for these sources combined with their new morphological and kinematic characterization made possible by the 3D [CII] information, allow us to put one of the first constraints on the fraction and rate of mergers shortly after the end of the Epoch of Reionization. Indeed, only a handful of works have been undertaken so far about the study of the mergers contribution to the build-up of galaxies at $z>4$, among which some are affected by large uncertainties due, for instance, to the lack of spectroscopic redshifts which properly identify close pairs. In the following, we summarize our major results:

\begin{list}{$\bullet$}{}
\item We identify 23 mergers, corresponding to $\sim31\%$ of the 75 ALPINE [CII]-detected galaxies, with an average stellar mass of the most massive galaxies of $\mathrm{log}(M_{*}/M_{\odot})\sim10$. We find major merger fractions $f_{\text{MM}}=0.44^{+0.11}_{-0.16}$ and $f_{\text{MM}}=0.34^{+0.10}_{-0.13}$ at $z\sim4.5$ and $z\sim5.5$, respectively, after taking into account the presence of minor mergers and correcting for completeness. These results are in good agreement with morphological studies by \cite{Conselice09} at the same redshifts and, when combined with previous works down to the local Universe, suggest a cosmic merger fraction evolution with a rapid increase from $z=0$ to $z\sim2$, a peak at $z\sim2-3$, and a possible slow decline for $z>3$. This trend is well described by a combined power-law and exponential function and is quite comparable with the outputs of the EAGLE hydrodynamical simulations from which \cite{Qu17} claim that major mergers contribute less than $10\%$ to the stellar mass gain in galaxies less massive than log$(M_{*}/M_{\odot})=10.5$, attributing a leading role to the in situ star formation.  

\item We convert the merger fraction into the merger rate per galaxy by adopting different redshift scaling for the typical time of merging ($T_{\text{MM}}$). The results we obtain are strongly dependent on the choice of $T_{\text{MM}}$. For instance, \cite{Kitzbichler08} found a merger timescale evolution which diverges more and more from that of \cite{Snyder17} moving to high redshift. When using the first timescale prescription, we obtain $R_{\text{MM}}\sim0.09$ and $R_{\text{MM}}\sim0.07$ Gyr$^{-1}$ at $z\sim4.5$ and $z\sim5.5$, respectively. However, these numbers are more than an order of magnitude larger if $T_{\text{MM}}\propto(1+z)^{-2}$ is used. Accordingly, this uncertainty propagates to all our subsequent outcomes that involve the choice of a merger timescale. 

\item By integrating the merger rate over time we obtain the average number of mergers a galaxy undertakes during its cosmic history from the early to the local Universe which goes from less than 1 to $\sim8$ mergers at $z=0$, depending on the adopted merger timescale prescription. Our results are quite in agreement with the outputs of several cosmological simulations, especially when adopting the redshift-dependent \cite{Snyder17} merger timescale with which we can match the large merger rates predicted at high redshifts by the Illustris simulations \citep{Rodriguez15} and the halo-halo merger rate by \cite{Dekel13}. 

\item Another important quantity we estimate is the volume-averaged merger rate $\Gamma_{\text{MM}}$, that is the merger rate $R_{\text{MM}}$ multiplied by the number density of galaxies at a given redshift and for a specific stellar mass range. We find a nearly constant $\Gamma_{\text{MM}}$ at intermediate redshifts in agreement with previous works in the literature \citep{Mundy17,Duncan19}, and a possible decrease at both $z<1$ and $z>4$. 

\item We take advantage of the two observationally-estimated merger rates to provide a constraint on the contribution of major mergers to the galaxy mass assembly through the cosmic epochs. From the merger rate per galaxy, we obtain the average stellar mass accreted per major merger per unit of mass and compare it with the sSFR computed from different authors. The latter seems to dominate over the specific mass accretion rate at all times. However, when considering a rapidly evolving merger timescale like $T_{\text{MM}}\propto(1+z)^{-2}$, the two quantities become comparable within the uncertainties at $z>4$. We then estimate the mass accretion rate density from $\Gamma_{\text{MM}}$ and compare this quantity to the well-known SFRD cosmic evolution \citep{Madau14}. We note that the contribution of major mergers to the global star formation rate ranges between being approximately equal to the SFRD to less than 1\% of it, depending on the exact choice of the merging timescale, the parametric form used to fit the data, and the redshift.
\end{list}

To conclude, major mergers could have played a significant role in the galaxy mass assembly through cosmic time, especially when merger timescales from recent literature, in which $T_{\text{MM}}$ decreases rapidly with increasing redshift, are taken into account. However, future investigation is needed in order to finally establish the importance of such events in the complex picture of the galaxy evolution. In particular, a larger statistical sample of spectroscopically confirmed galaxies, observed with deeper resolution, will allow us to confirm the large fraction of mergers at early times and their incidence at different redshifts, unveiling the relative contribution of each process to the build-up of galaxies through the Universe.

\begin{acknowledgements}
We warmly thank the referee for her or his careful reading of the paper and constructive suggestions that contributed to improve the quality of this work. This paper is based on data obtained with the ALMA Observatory, under Large Program 2017.1.00428.L. ALMA is a partnership of ESO (representing its member states), NSF(USA) and NINS (Japan), together with NRC (Canada), MOST and ASIAA (Taiwan), and KASI (Republic of Korea), in cooperation with the Republic of Chile. The Joint ALMA Observatory is operated by ESO, AUI/NRAO and NAOJ. A.C. acknowledges the support from grant PRINMIUR 2017 - 20173ML3WW$\_$001. The Cosmic Dawn Center is funded by the Danish National Research Foundation under grant No. 140. G.C.J. acknowledges ERC Advanced Grant 695671 ``QUENCH'' and support by the Science and Technology Facilities Council (STFC). This work was supported by the Programme National Cosmology et Galaxies (PNCG) of CNRS/INSU with INP and IN2P3, co-funded by CEA and CNES. E.I.\ acknowledges partial support from FONDECYT through grant N$^\circ$\,1171710. This paper is dedicated to the memory of Olivier Le Fèvre, PI of the ALPINE survey.
\end{acknowledgements}

%
%

\bibliographystyle{aa} 
\bibliography{mybiblio.bib} 

\begin{appendix}
\section{Individual description of mergers}\label{app:mergers}
We present here the characterization of the mergers analyzed in this work, as illustrated in Section \ref{sec:mergers_charact} and Figure \ref{fig:merger_example}. For each source, we show the moment maps, the PVDs, and the integrated [CII] spectrum. Due to the peculiarity of these objects, we briefly comment on them in the following, also highlighting possible differences and similarities with the previous morpho-kinematic classifications.

\noindent
\textbf{CG$\_$38}: \cite{LeFevre20} classified this galaxy as a merger. Although the low S/N and small spatial extent, we include this source in our analysis as it shows multiple peaks in the PVDs that are associated to different components in the [CII] spectrum.

\noindent
\textbf{DC$\_$308643}: this source was classified as a merger by \cite{LeFevre20}. The moment-1 map shows a clear velocity gradient suggesting the presence of a rotating disk. However, a similar velocity map could also be reproduced by two merging components in an advanced phase of merging. This is suggested by the two close components visible in the PVD along the major axis and by the shape of the [CII] line which deviates from that of a single Gaussian.

\noindent
\textbf{DC$\_$372292}: we confirm the merger classification of this object by \cite{LeFevre20}. As for DC$\_$308643, a possible velocity gradient is present in the moment-1 map. However, two components separated in velocity by $\sim200$~km~s$^{-1}$ are clearly visible in the PVDs and in the spectrum.

\noindent
\textbf{DC$\_$378903}: we classify this source as a merger as in \cite{LeFevre20}. Despite the low S/N and spatial extent, we manage to decompose the [CII] spectrum in two guassian components that are also visible at $3\sigma$ in the PVDs and in the channel maps.

\noindent
\textbf{DC$\_$417567}: this source was classified as a merger by \cite{LeFevre20} because of the disturbed [CII] morphology and the presence of multiple peaks in the PVDs. These are reproduced by different Gaussian components in the overall [CII] spectrum. This target was also analyzed by \cite{Jones21} which classified it as uncertain because of its low S/N. However, the poor $\mathrm{^{3D}Barolo}$ fit supports the merger classification for this object.

\noindent
\textbf{DC$\_$422677}: as in \cite{LeFevre20}, we classify this source as a merger, mainly because of the presence of multiple components in the PVDs and in the spectrum.

\noindent
\textbf{DC$\_$434239}: this object was classified as a merger both by \cite{LeFevre20} and \cite{Jones21}. The [CII] morphology is extended and disturbed, as well as that from the UVista $K_s$ band. The PVDs show multiple peaks and we need at least three components to reproduce the global shape of the [CII] spectrum. 

\noindent
\textbf{DC$\_$493583}: we classify this source as a merger as in \cite{LeFevre20}. Two separate components are visible both in the PVDs and in the spectrum, with a fainter minor object emerging from the noise at $v\sim-200$~km~s$^{-1}$.

\noindent
\textbf{DC$\_$519281}: this source was classified as a merger by \cite{LeFevre20}. The complex velocity map and PVDs sustain this interpretation. Further evidence for ongoing merging activity is provided by the analysis of \cite{Jones21} (even if they classify this object as uncertain because of the low S/N and spatial resolution).

\noindent
\textbf{DC$\_$536534}: the [CII] emission from this galaxy is spatially resolved in different clumps, allowing \cite{LeFevre20} to classify it as a merger. The PVDs are clearly disturbed and multipeaked, and the integrated spectrum shows the presence of at least three different emitting components.

\noindent
\textbf{DC$\_$665509}: previously classified as a merging system \citep{LeFevre20}, this source shows multiple components both in the spectrum and in the PVDs with the two major emissions at $v\sim0$ and $v\sim150$~km~s$^{-1}$.

\noindent
\textbf{DC$\_$680104}: this object was classified as a merger by \cite{LeFevre20}. It shows a faint secondary component at $3\sigma$ in the moment-0 map which is also visible in the PVD along the major axis at a spatial offset of $\sim-1$~arcsec with respect to the main target. Although the S/N of this source is low, we consider it as a merger, mainly because of two peaks of emission in the PVDs at the same spatial position but separated by $\sim150$~km~s$^{-1}$ in velocity. These are also visible in the [CII] spectrum. 

\noindent
\textbf{DC$\_$814483}: the presence of multiple components in the optical images and the quite disturbed morphology of the [CII] emission led \cite{LeFevre20} to classify this galaxy as a merger. We confirm the previous classification also noting the complex velocity field and the multiple components in the spectrum, possibly associated to peaks of emission in the PVDs. 

\noindent
\textbf{DC$\_$818760}: this source was studied in detail by \cite{Jones20} and then further analyzed by \cite{Jones21}. It is likely a triple merger, with the two main components visible in the optical maps and close in velocity (as shown by the PVD along the major axis), and another fainter source completely obscured in the optical and separated both in space and velocity from the other two objects. We thus confirm the previous merger classification by \cite{LeFevre20} and \cite{Jones21}. In this analysis, we use a pseudo-slit 7.5~arcsec wide to recover the full [CII] emission, and we consider only the two close components with the larger [CII] fluxes and optical emission.  

\noindent
\textbf{DC$\_$834764}: this source was classified as an extended dispersion-dominated galaxy by \cite{LeFevre20}. However, the peaks of emission in the PVDs, the shape of the emission line, and the quite disturb velocity field suggest the presence of multiple components. We thus classify this object as a merger.

\noindent
\textbf{DC$\_$842313}: this galaxy is quite peculiar and was classified as a merger by \cite{LeFevre20}. The optical maps show the presence of two or three components. Indeed, the ALPINE target is at the center of the maps, clearly visible in the HST F814W image, while a brighter and larger component also detected in [CII] is present toward the north, particularly visible in the UVista $K_s$-band map. This is the well-studied submillimeter galaxy J1000+0234 (e.g., \citealt{Capak08,Schinnerer08,Jones17}), and it was serendipitously observed in ALPINE \citep{Loiacono21}. This galaxy has a large line width (i.e., $\sim1000$~km~s$^{-1}$) and is separated in velocity by $\sim750$~km~s$^{-1}$ from the main ALPINE target. As we are considering only those systems with a velocity separation $\Delta v\leq500$~km~s$^{-1}$ (assuming these components as gravitationally bound; \citealt{Patton2000}), we should exclude it from our analysis. However, considering only the emission arising from the spectral channels associated to the main target DC$\_$842313, we find further evidence of an ongoing merging. In particular, we note a disturbed [CII] morphology elongated toward the extended $K_s$-band contours in the southwest region of the optical image, and peaks of emission in the PVDs coinciding with multiple components in the spectrum. For these reasons, we think that this could be a triple merging system, and we keep it in the analysis of the merger fraction. In this case, the moment maps and PVDs show the emission from both the main target and J1000+0234, obtained considering all the spectral channels including the [CII] emission from the two sources. The bottom panel displays instead the individual normalized integrated spectra of the two components.

\noindent
\textbf{DC$\_$859732}: this galaxy was classified as a merger by \cite{LeFevre20}. Two distinct components are present in the [CII] spectrum, partially visible in the PVDs due to the low S/N.

\noindent
\textbf{DC$\_$873321}: this galaxy was classified as a merger both by \cite{LeFevre20} and \cite{Jones21}. The [CII] emission is elongated toward the two bright sources visible in the UVista $K_s$ band. These two components are at the same velocity, being indistinguishable in the PVDs and their spectra are shown in the bottom panel of Fig. \ref{fig:merger_appendix11} (right).

\noindent
\textbf{vc$\_$5100541407}: the extended [CII] morphology coincident with the two optical components visible in the maps, allowed \cite{LeFevre20} to characterize this source as a merger. \cite{Jones21} classified it as uncertain instead, mainly because of their adopted classification criteria. Given the morphological and kinematic information on this galaxy, we also consider it as a merger of two components emitting at the same velocity. A pseudo-slit 4.5~arcsec wide is used to compute the PVDs.
 
\noindent
\textbf{vc$\_$5101209780}: this galaxy was classified as a merger both by \cite{LeFevre20} and \cite{Jones21}, and it was the subject of a further and in-depth analysis by \cite{Ginolfi20}. The [CII] moment-0 map clearly shows the presence of two major components, as also confirmed by the optical images. Two fainter components are also found in the spectrum and PVDs, which are computed by adopting a pseudo-slit 4.5~arcsec wide. 

\noindent
\textbf{vc$\_$5180966608}: \cite{LeFevre20} classified this object as a merger while \cite{Jones21} considered it as uncertain, mainly because of the [CII] spectrum which resembles that of a single, extended source. We find more evidence for a merging system. Indeed, at least two optical components are visible both from HST and UVista within the $3\sigma$ [CII] emission contours. The velocity gradient is quite disturbed and several peaks of emission are present in the PVDs, more visible along the minor axis (which is aligned with the two optical sources).   

\begin{figure*}[t]
\begin{center}
	\subfigure{\includegraphics[width=\columnwidth,height=15.1cm]{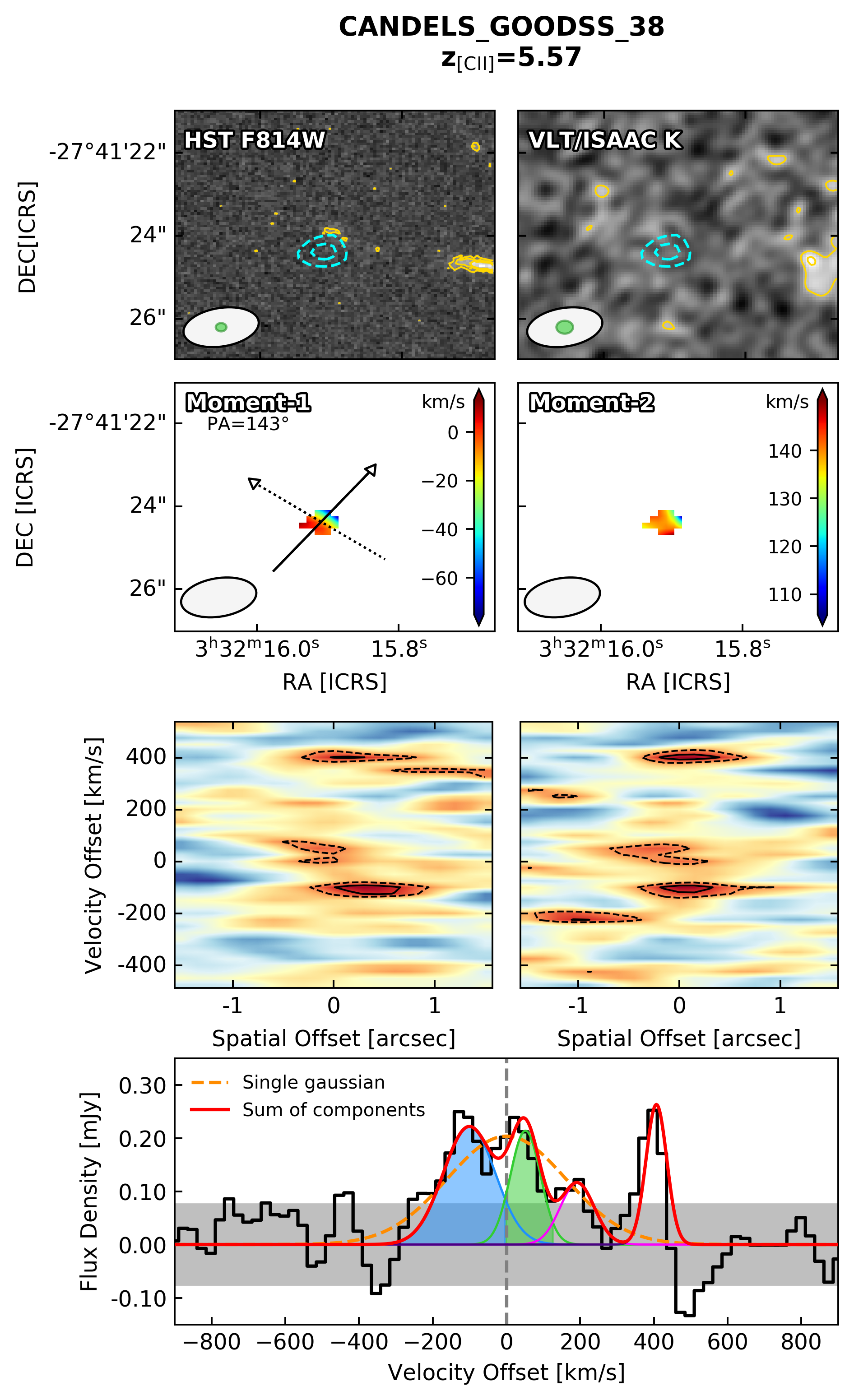}}
	\subfigure{\includegraphics[width=\columnwidth,height=15.1cm]{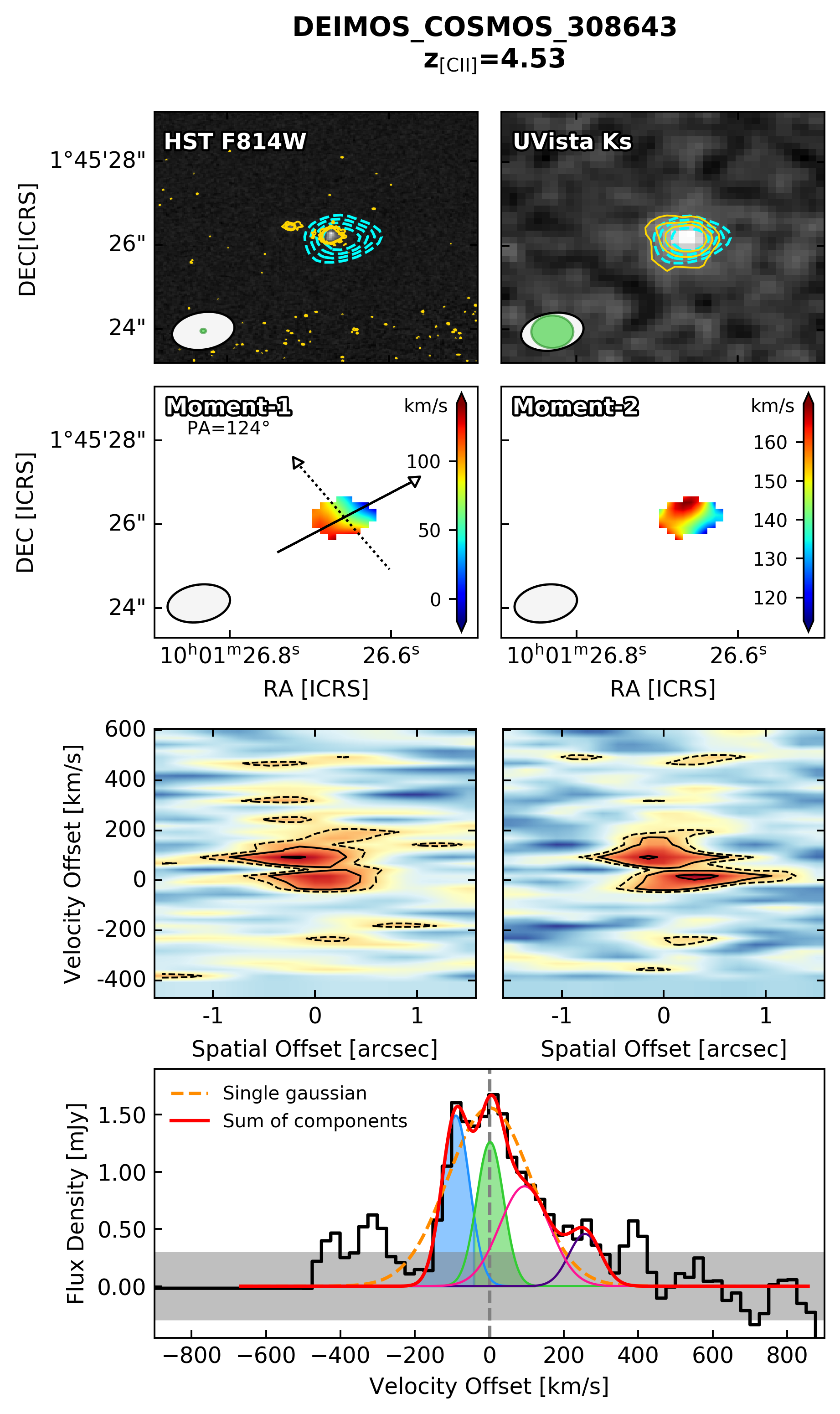}}
    \caption{Morpho-kinematic analysis of the ALPINE targets. \textit{First row}: HST/ACS F814W (left) and UVista DR4 $K_s$-band (right) images centered on the UV rest-frame position of the target. Each cutout is $6'' \times 6''$ wide. The cyan contours show the [CII] ALMA emission starting from $3\sigma$ above the noise level. Yellow contours in the optical maps represent 3, 5 and 7$\sigma$ emission. In the lower left corner, the ALMA beam (white) and HST or UltraVISTA resolutions (green) are displayed. \textit{Second row:} moment-1 (left) and moment-2 (right) maps color-coded for the velocity and velocity dispersion in $\mathrm{km~s^{-1}}$. The velocity map reports the direction of the major (solid) and minor (dashed) axis (centered on the coordinates returned by the best-fit 2D Gaussian model on the moment-0 map) along which the PVDs are computed. \textit{Third row:} PVDs along the major (left) and minor (right) axis color-coded for the flux intensity in each pixel. Dashed contours include the $2\sigma$ emission in the maps while 3, 5 and 7$\sigma$ emission is represented by solid lines. \textit{Fourth row:} [CII] spectrum (black histogram) extracted within the 3$\sigma$ contours of the intensity map. The gray shaded band marks the 1$\sigma$ level of the spectrum while the dashed vertical line shows the zero velocity offset computed with respect to the redshift of the [CII] line. Colored thin lines show individual possible components of the merging system, resulting in the global profile in red. The shaded areas under the curves represent the channels used to compute the [CII] intensity maps of the major and minor individual components. A single Gaussian fit is also visible with a dashed-orange line.}
    \label{fig:merger_appendix1}
\end{center}
\end{figure*}

\begin{figure*}[t]
\begin{center}
	\subfigure{\includegraphics[width=\columnwidth,height=15.1cm]{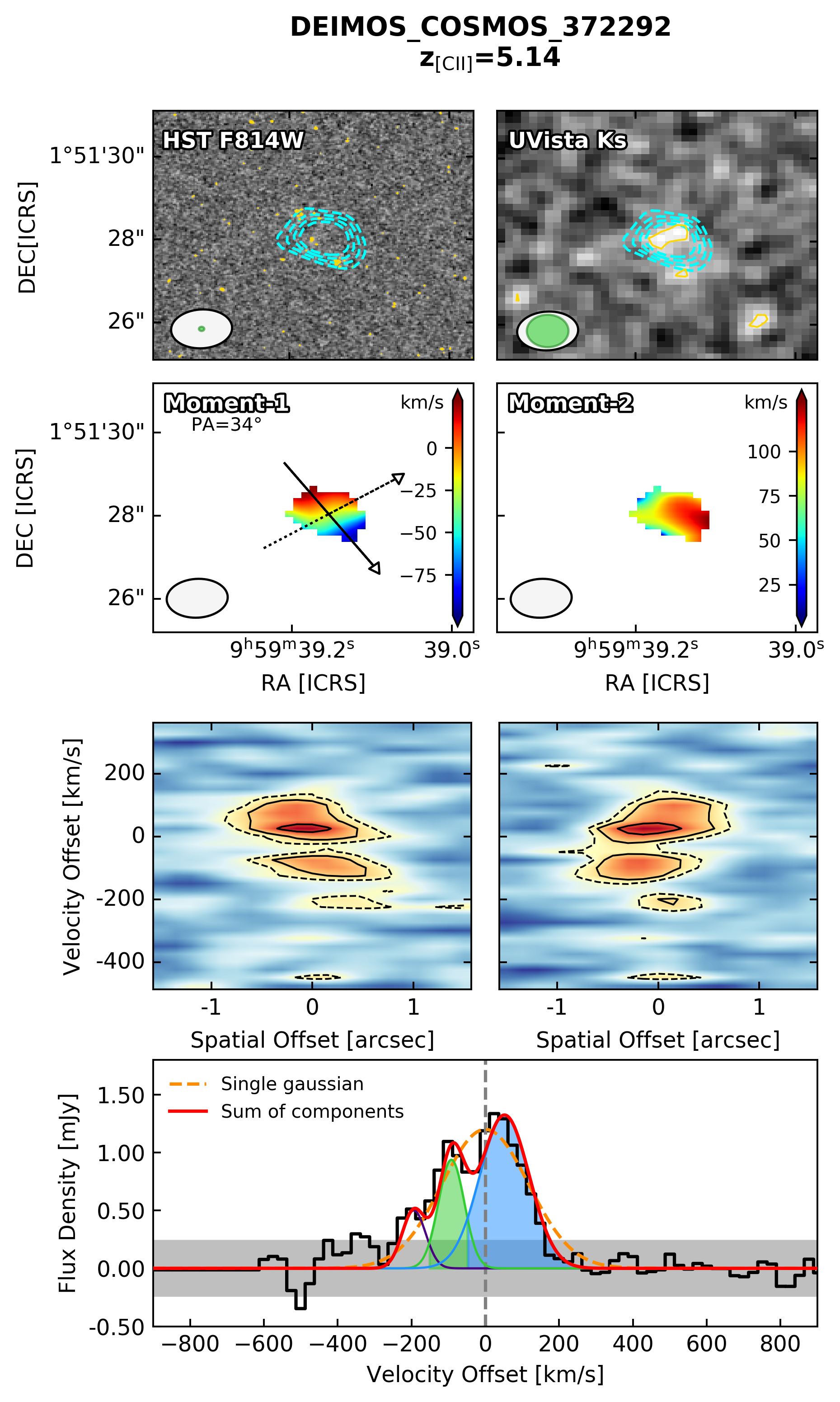}}
	\subfigure{\includegraphics[width=\columnwidth,height=15.1cm]{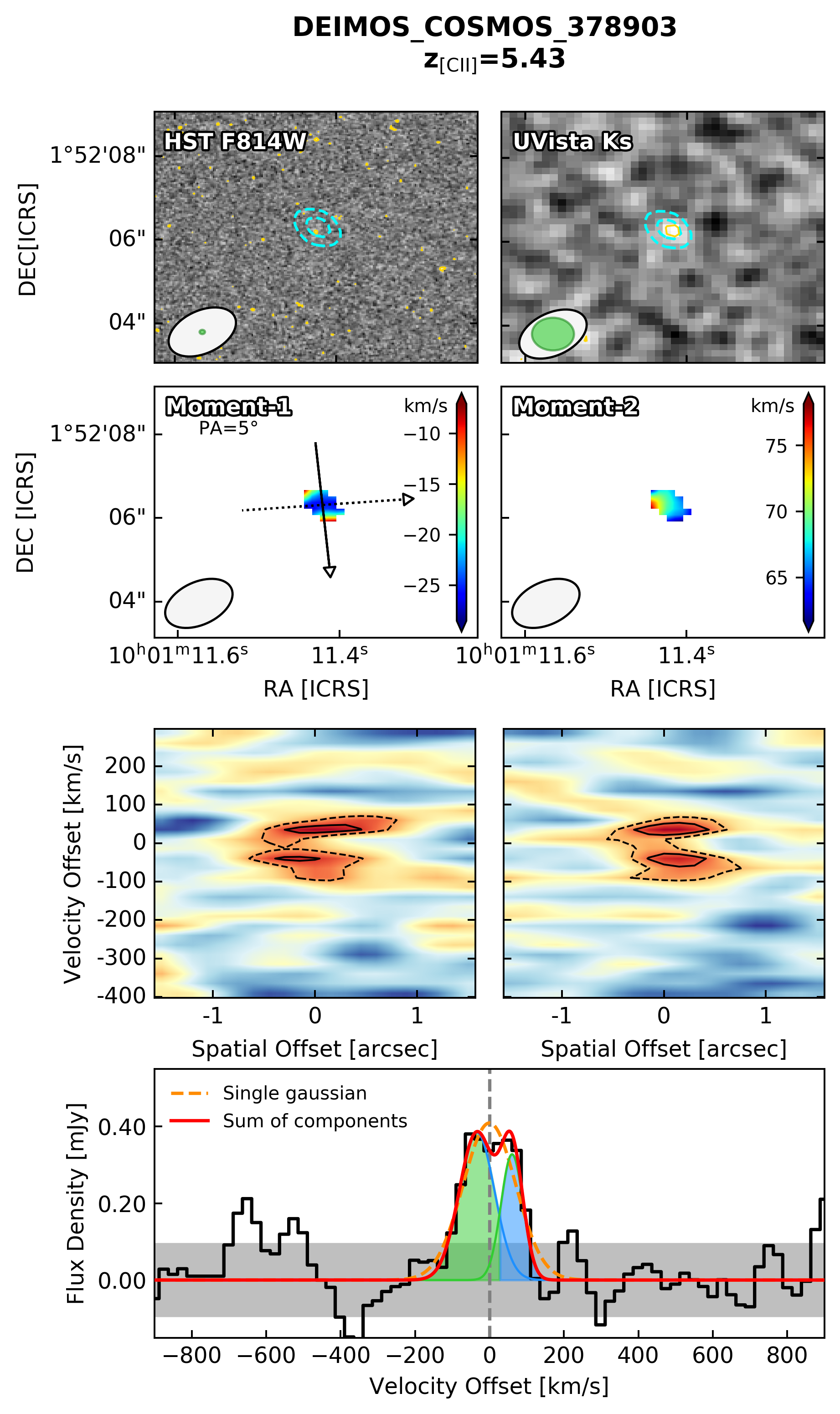}}
    \caption{Moment maps, PVDs and spectral decomposition for the ALPINE mergers, as described in Figure \ref{fig:merger_appendix1}.}
    \label{fig:merger_appendix2}
\end{center}
\end{figure*}

\begin{figure*}[t]
\begin{center}
	\subfigure{\includegraphics[width=\columnwidth,height=15.1cm]{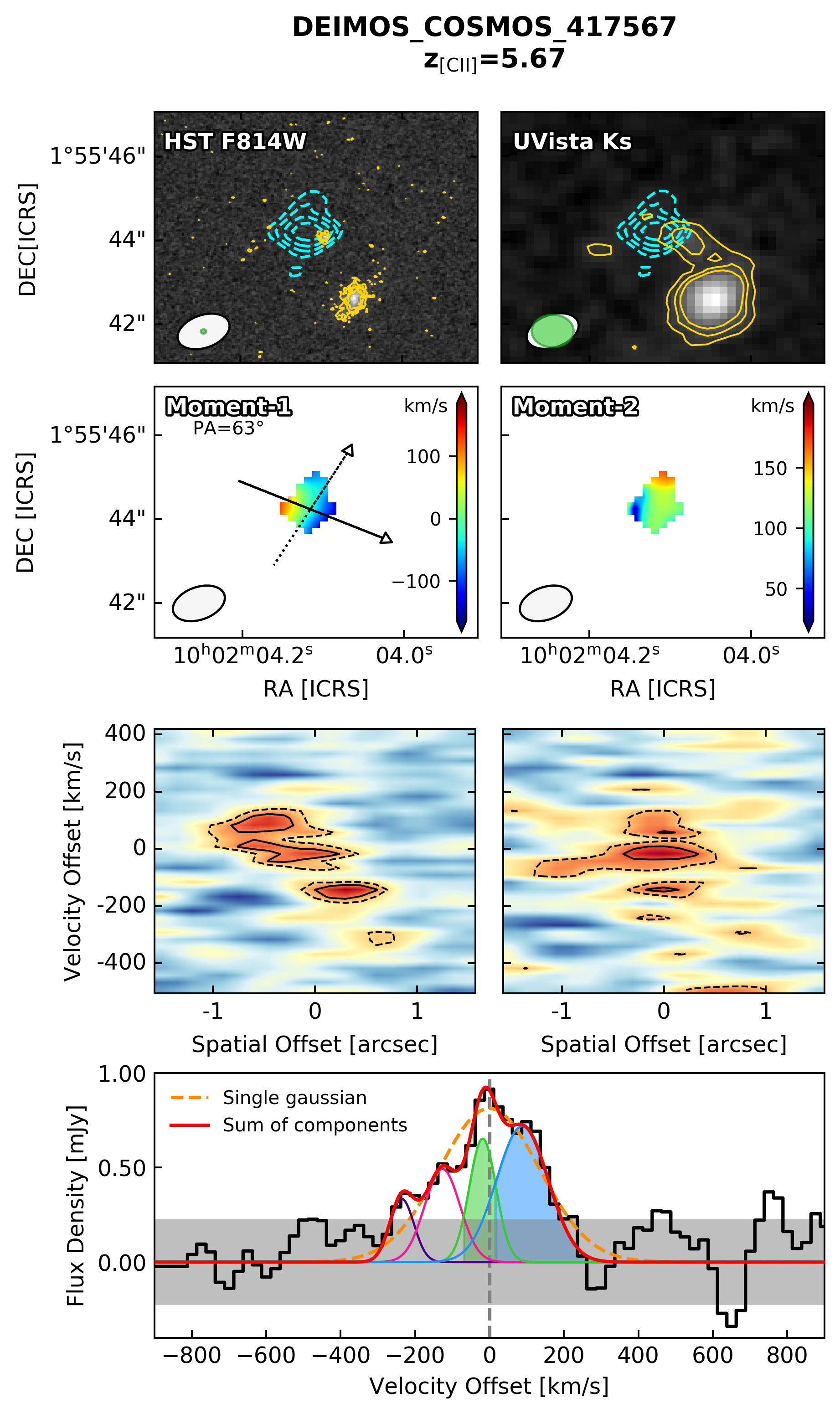}}
	\subfigure{\includegraphics[width=\columnwidth,height=15.1cm]{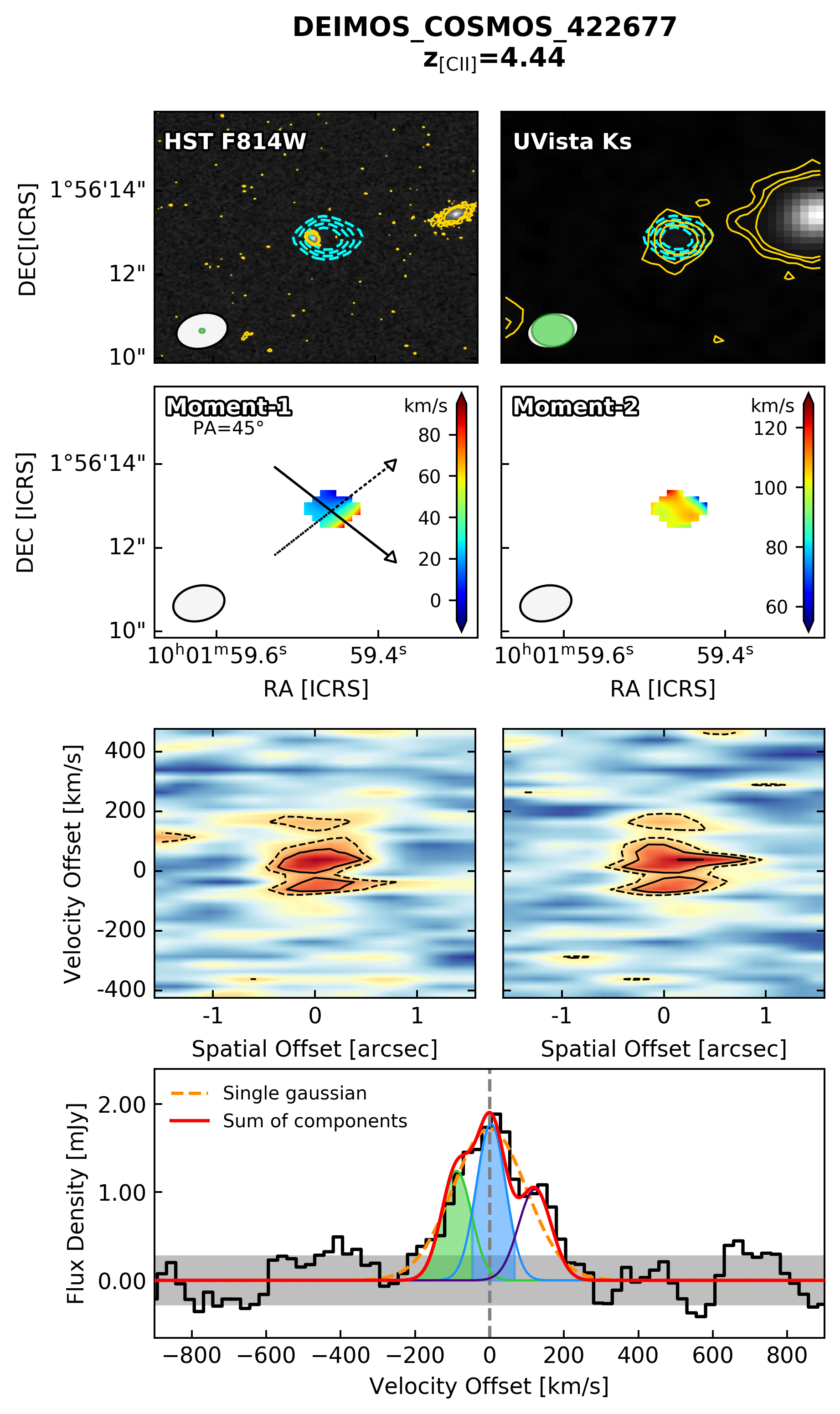}}
    \caption{Moment maps, PVDs and spectral decomposition for the ALPINE mergers, as described in Figure \ref{fig:merger_appendix1}.}
    \label{fig:merger_appendix3}
\end{center}
\end{figure*}

\begin{figure*}[t]
\begin{center}
	\subfigure{\includegraphics[width=\columnwidth,height=15.1cm]{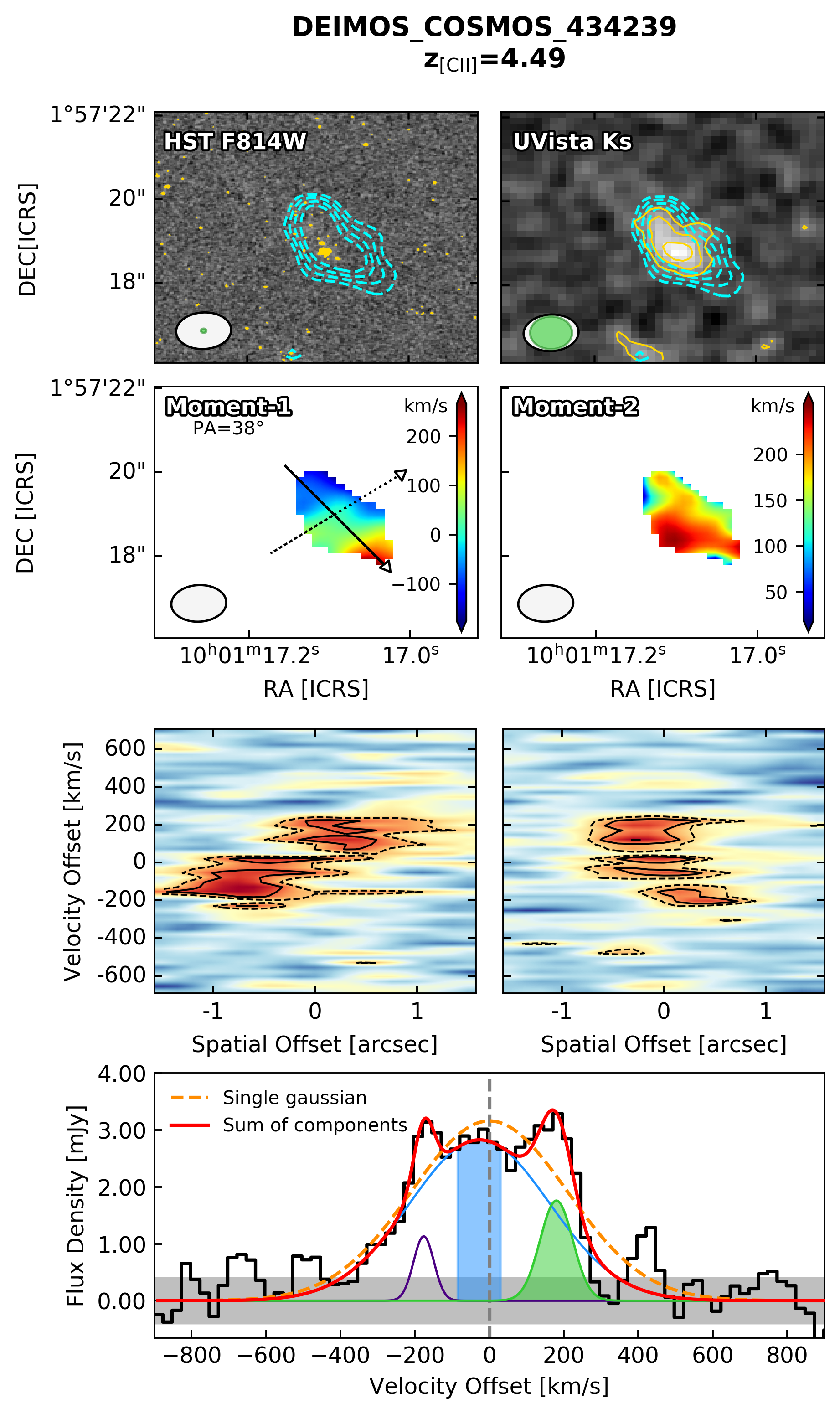}}
	\subfigure{\includegraphics[width=\columnwidth,height=15.1cm]{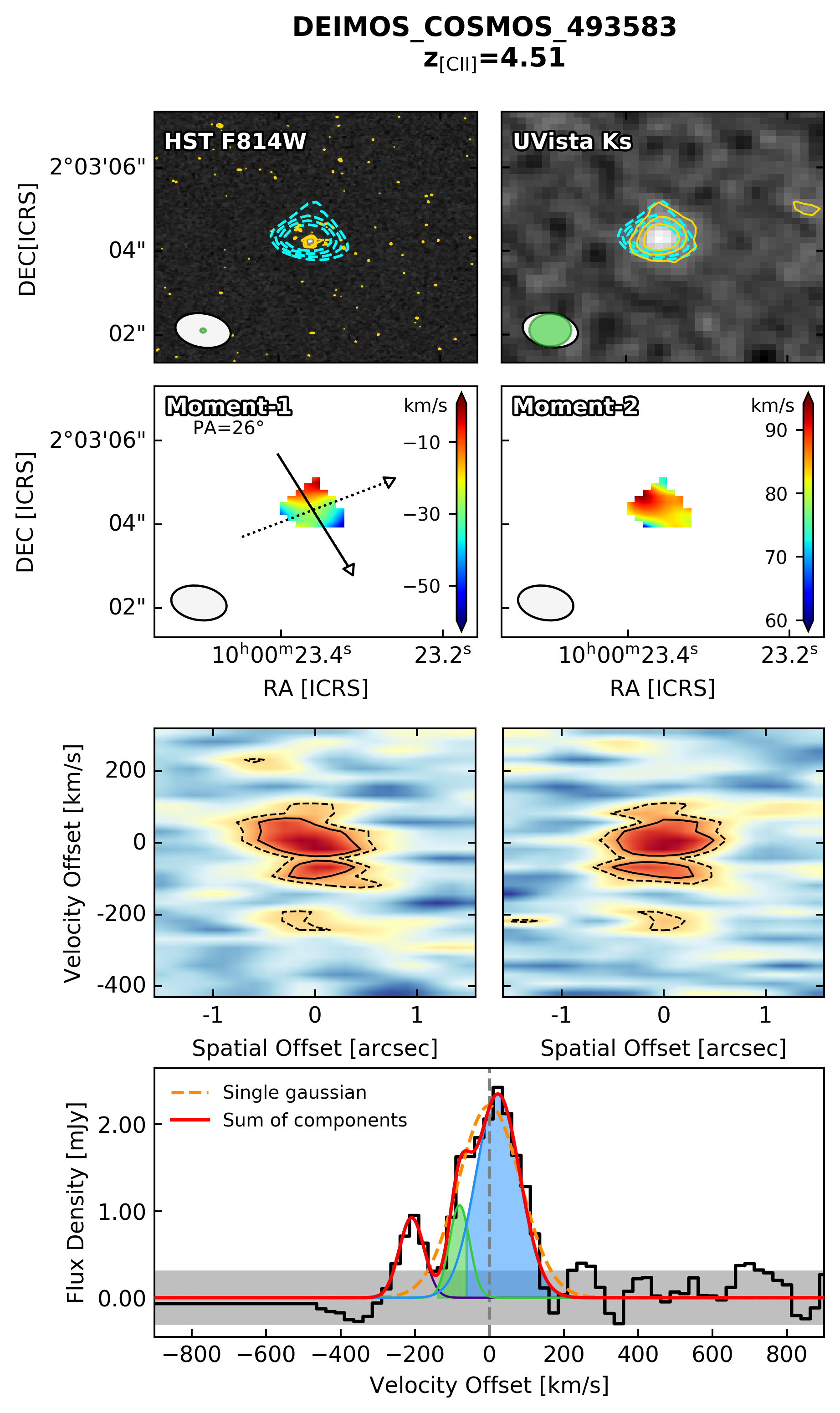}}
    \caption{Moment maps, PVDs and spectral decomposition for the ALPINE mergers, as described in Figure \ref{fig:merger_appendix1}.}
    \label{fig:merger_appendix4}
\end{center}
\end{figure*}

\begin{figure*}[t]
\begin{center}
	\subfigure{\includegraphics[width=\columnwidth,height=15.1cm]{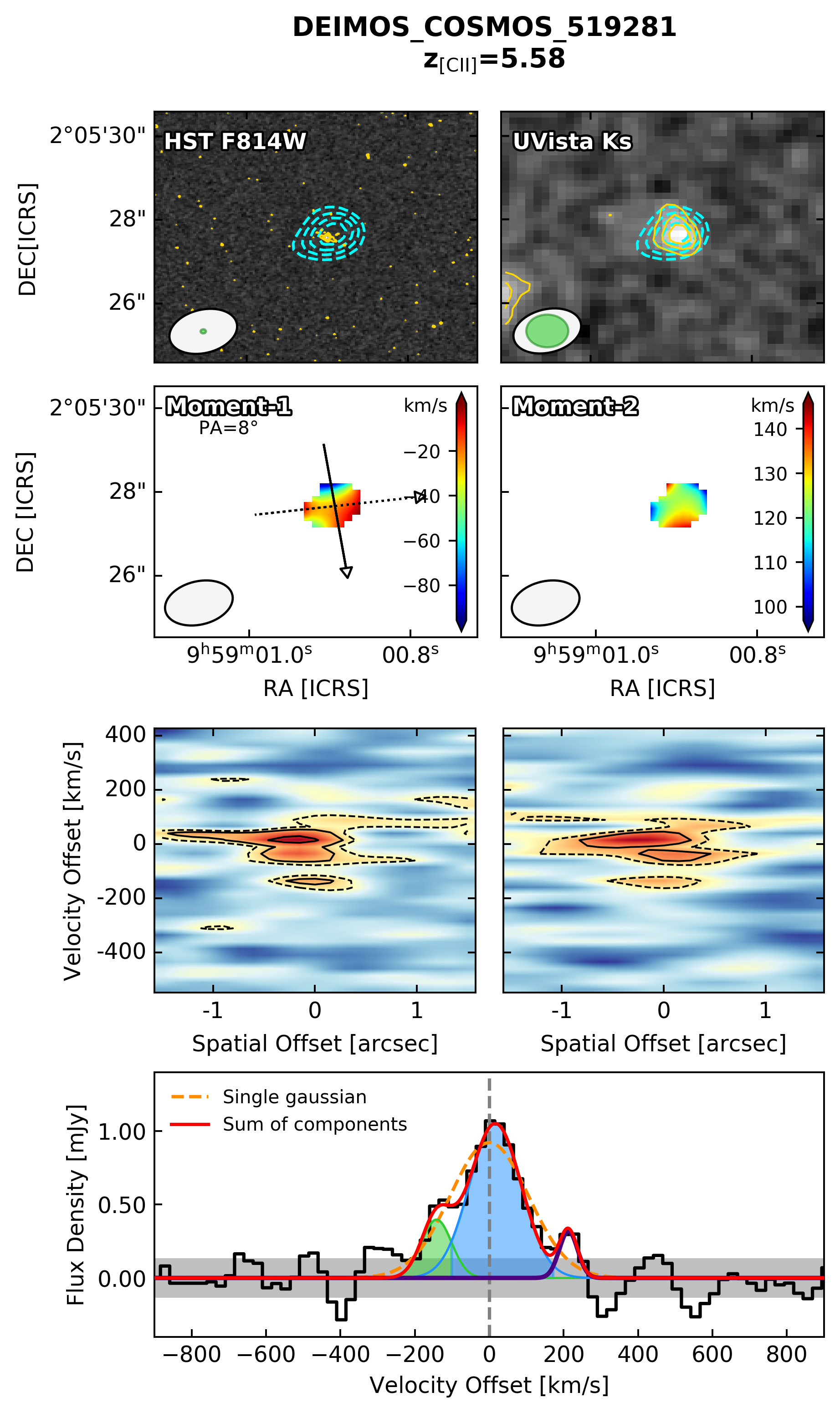}}
	\subfigure{\includegraphics[width=\columnwidth,height=15.1cm]{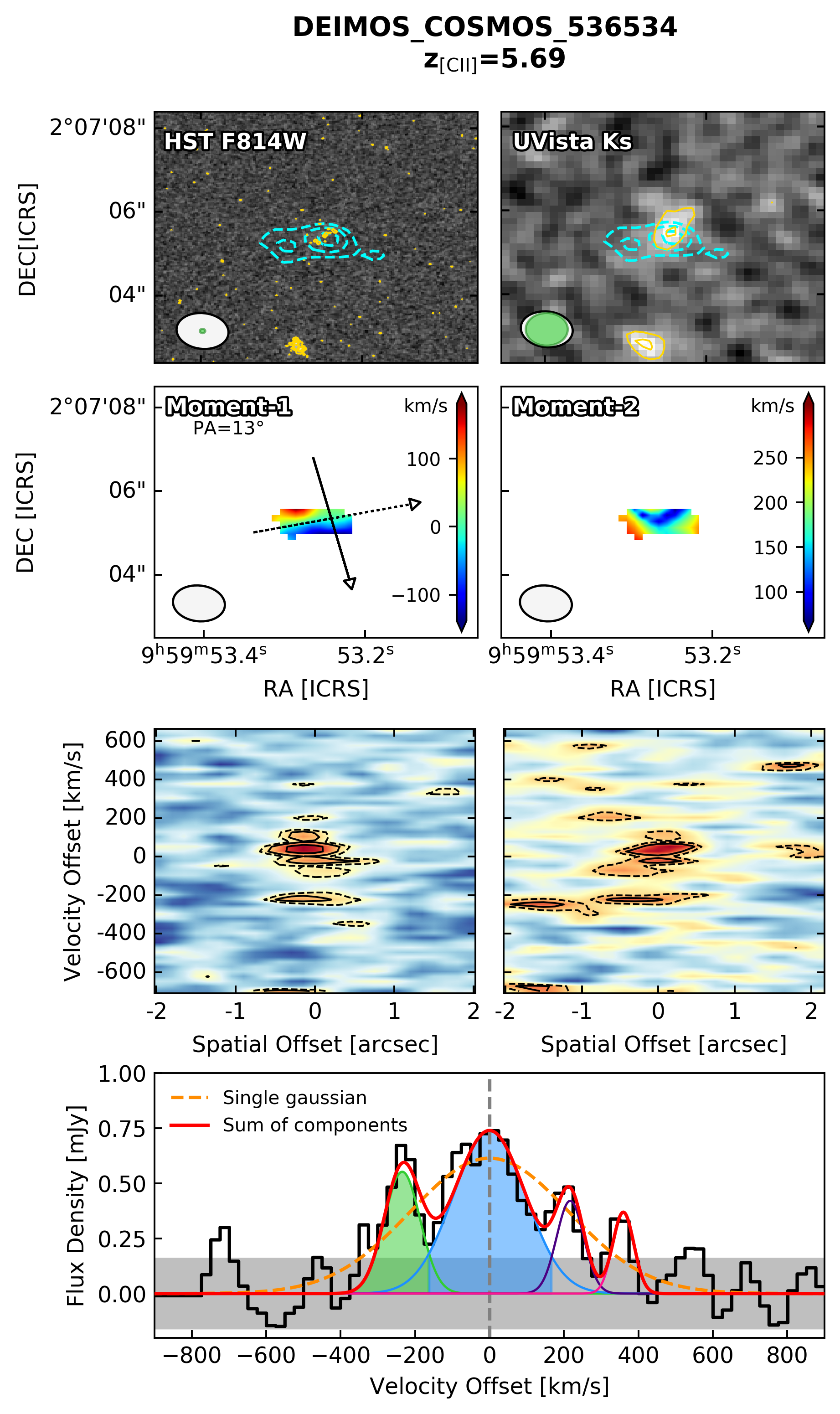}}
    \caption{Moment maps, PVDs and spectral decomposition for the ALPINE mergers, as described in Figure \ref{fig:merger_appendix1}.}
    \label{fig:merger_appendix5}
\end{center}
\end{figure*}

\begin{figure*}[t]
\begin{center}
	\subfigure{\includegraphics[width=\columnwidth,height=15.1cm]{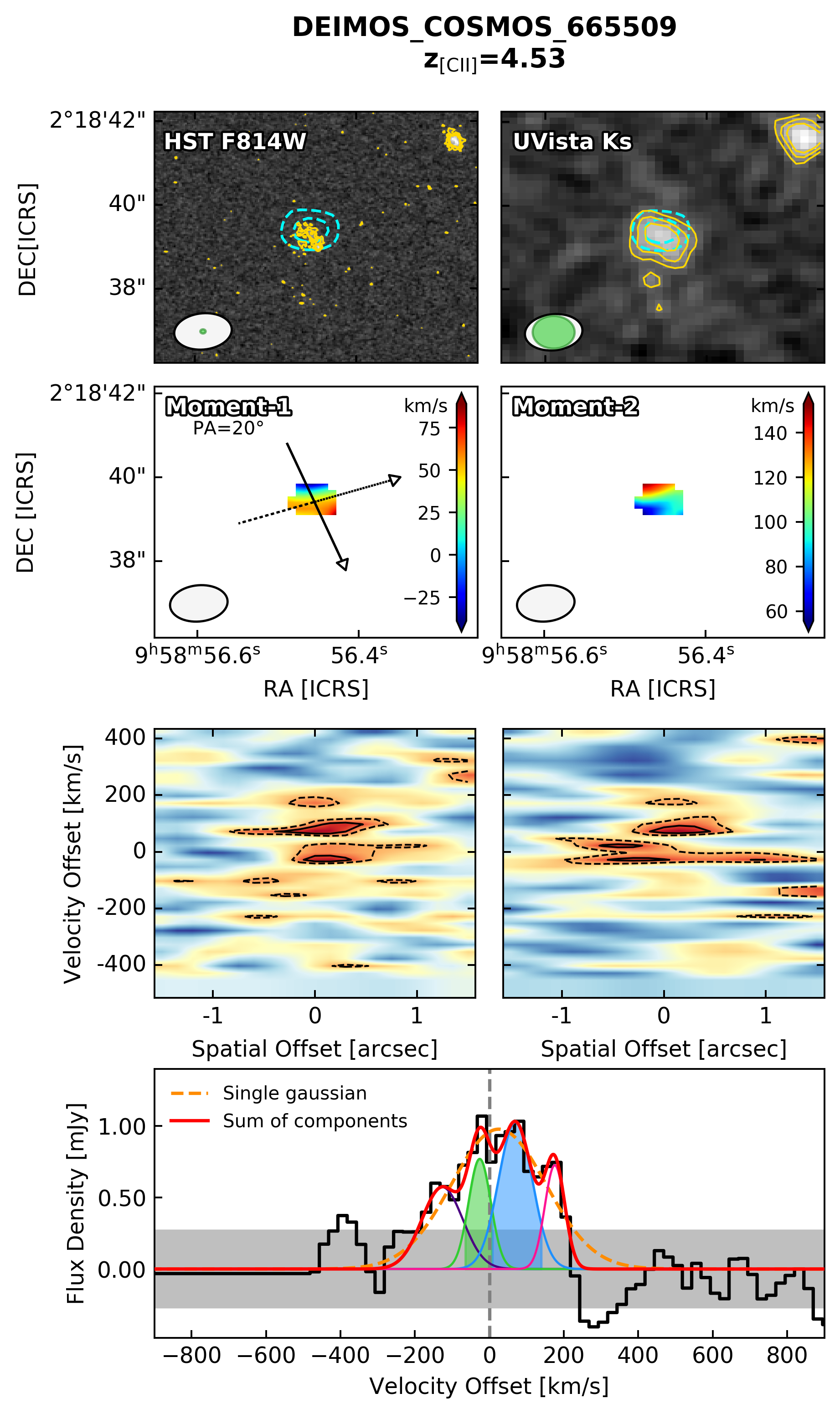}}
	\subfigure{\includegraphics[width=\columnwidth,height=15.1cm]{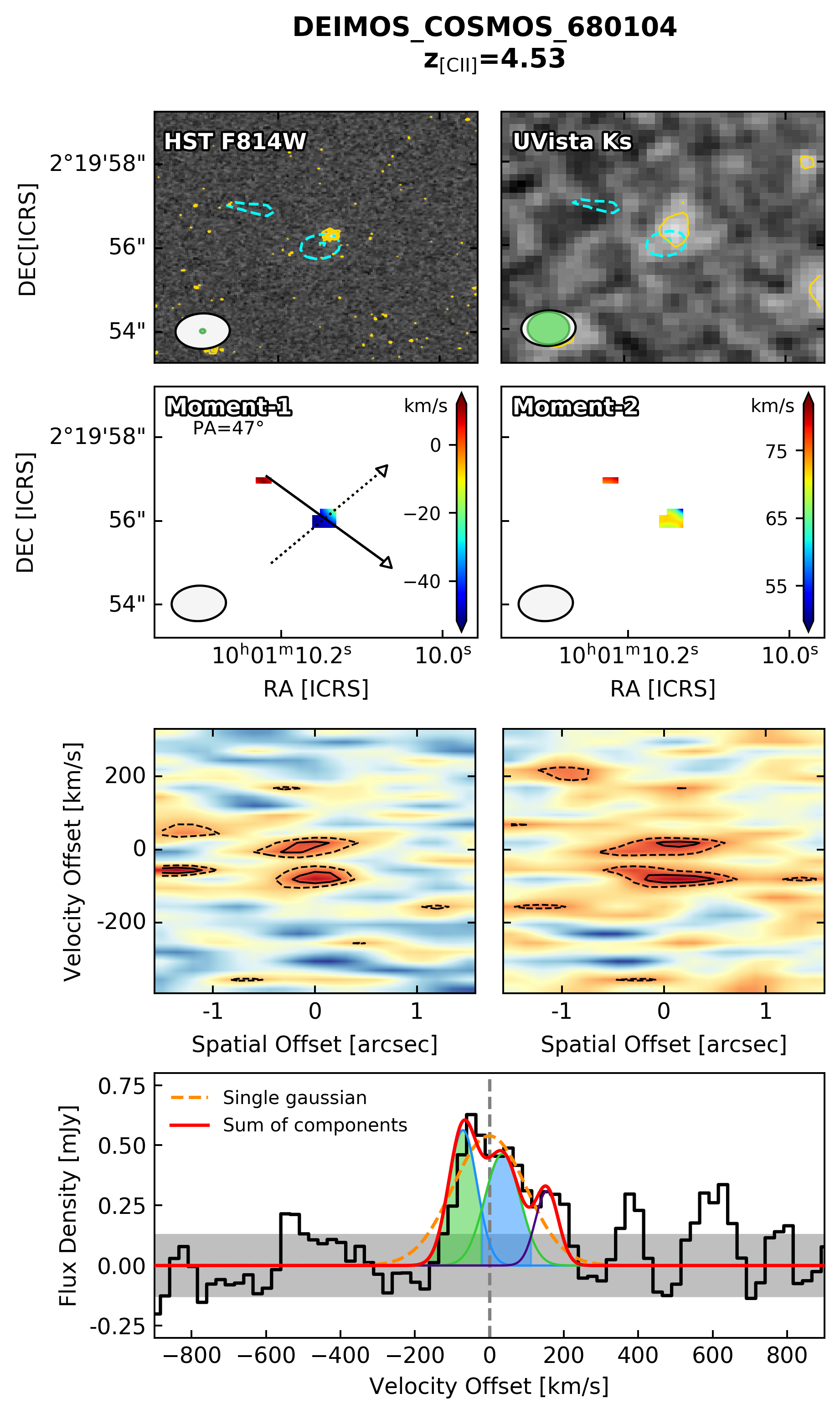}}
    \caption{Moment maps, PVDs and spectral decomposition for the ALPINE mergers, as described in Figure \ref{fig:merger_appendix1}.}
    \label{fig:merger_appendix6}
\end{center}
\end{figure*}

\begin{figure*}[t]
\begin{center}
	\subfigure{\includegraphics[width=\columnwidth,height=15.1cm]{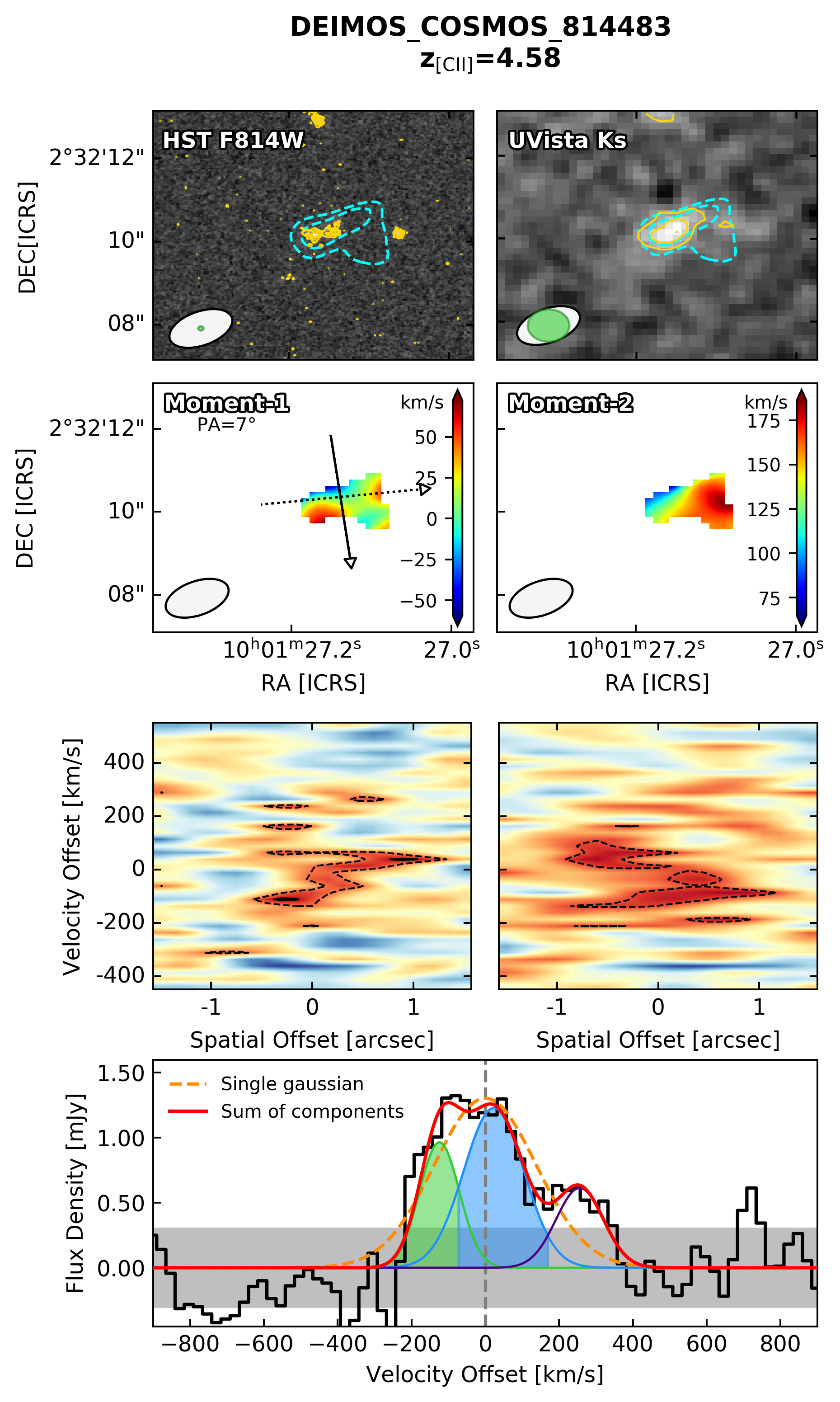}}
	\subfigure{\includegraphics[width=\columnwidth,height=15.1cm]{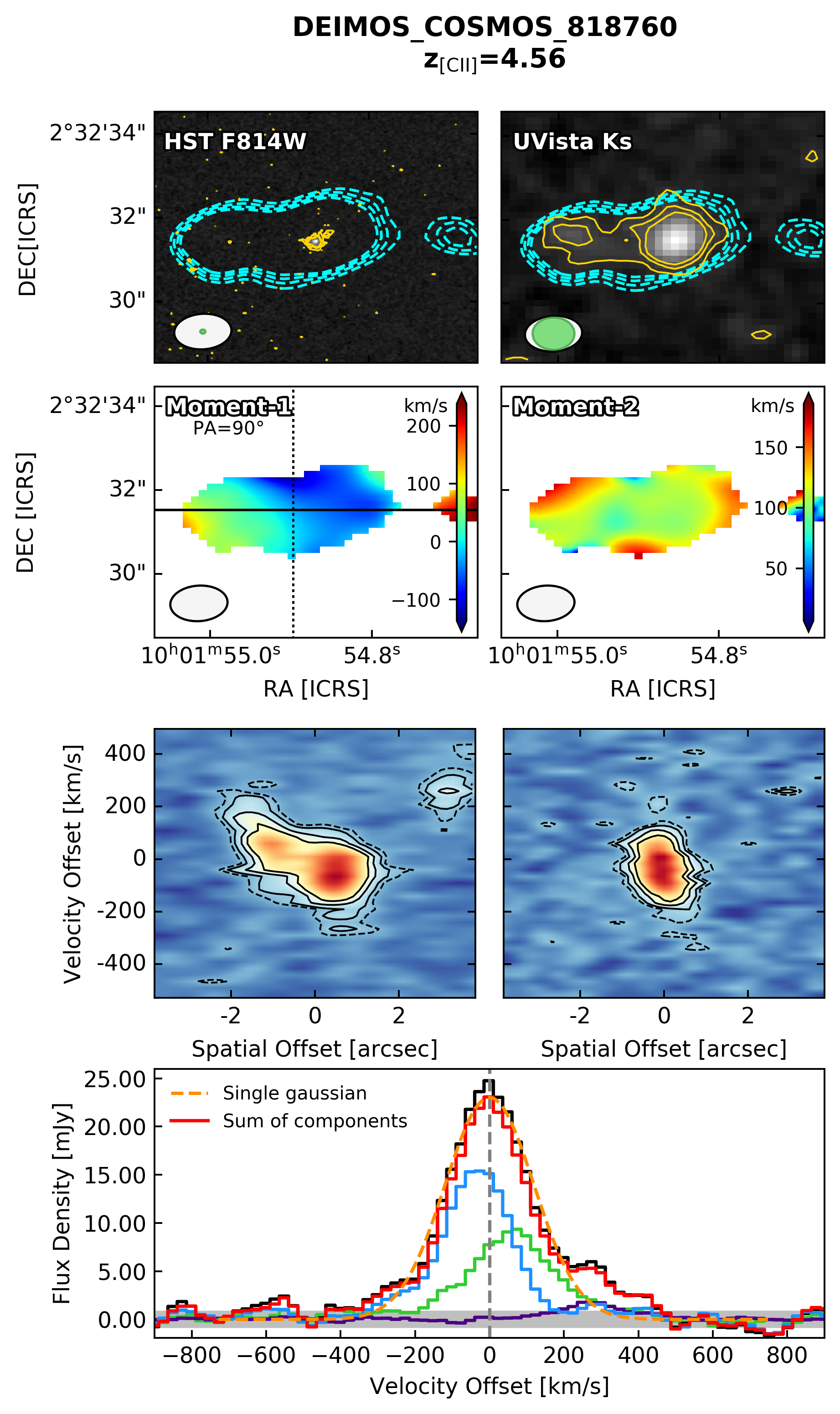}}
    \caption{Moment maps, PVDs and spectral decomposition for the ALPINE mergers, as described in Figure \ref{fig:merger_appendix1}.}
    \label{fig:merger_appendix7}
\end{center}
\end{figure*}

\begin{figure*}[t]
\begin{center}
	\subfigure{\includegraphics[width=\columnwidth,height=15.1cm]{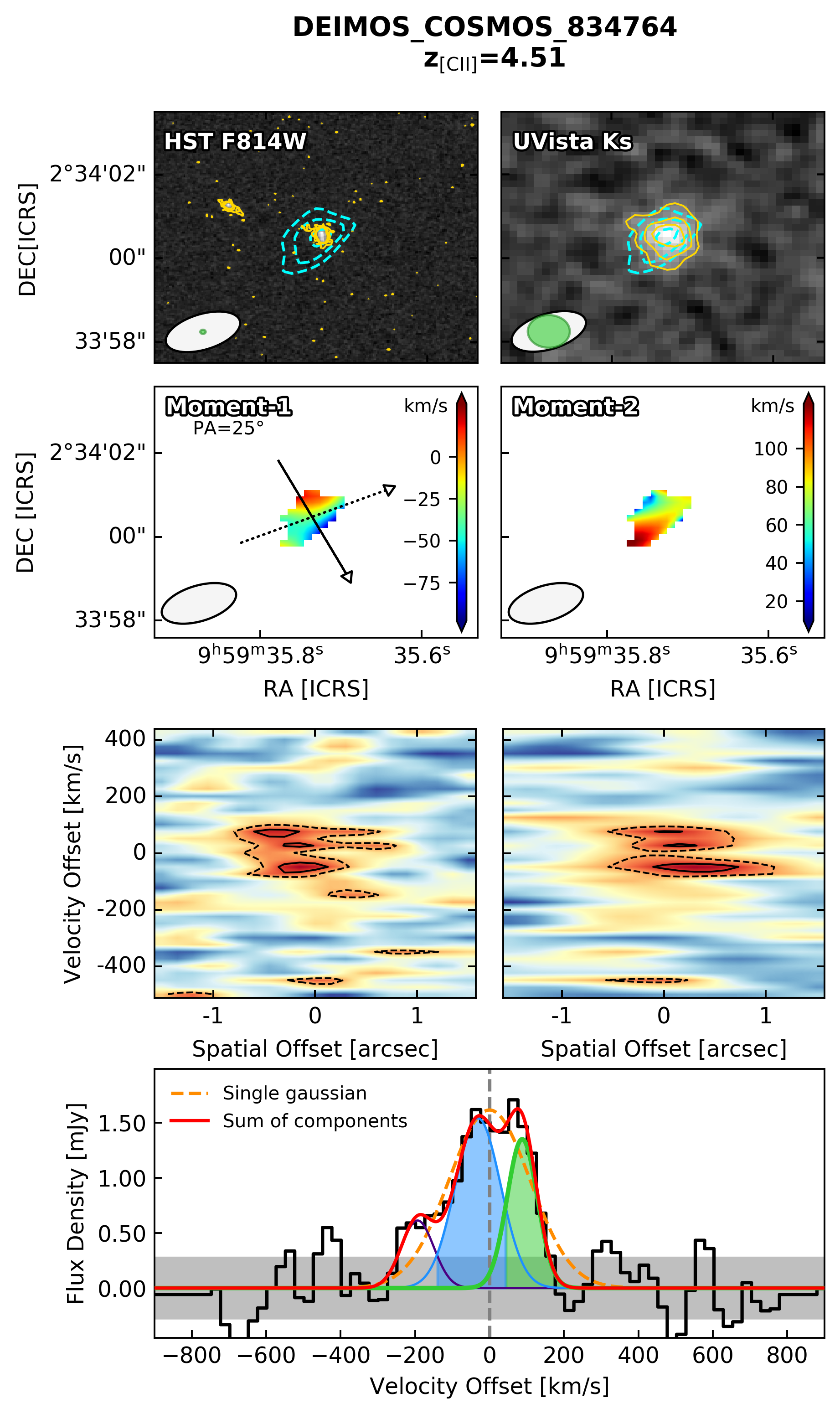}}
	\subfigure{\includegraphics[width=\columnwidth,height=15.1cm]{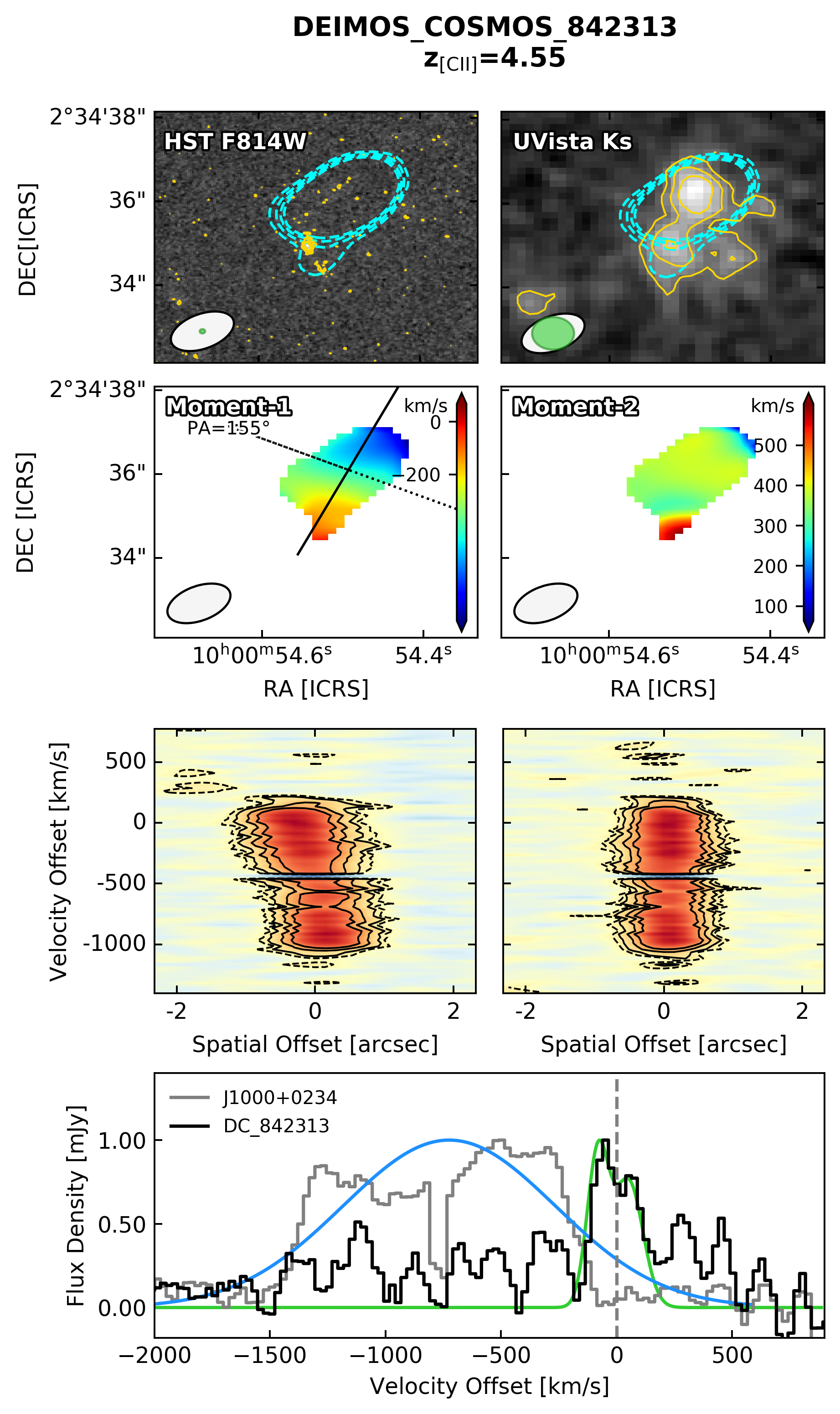}}
    \caption{Moment maps, PVDs and spectral decomposition for the ALPINE mergers, as described in Figure \ref{fig:merger_appendix1}.}
    \label{fig:merger_appendix8}
\end{center}
\end{figure*}

\begin{figure*}[t]
\begin{center}
	\subfigure{\includegraphics[width=\columnwidth,height=15.1cm]{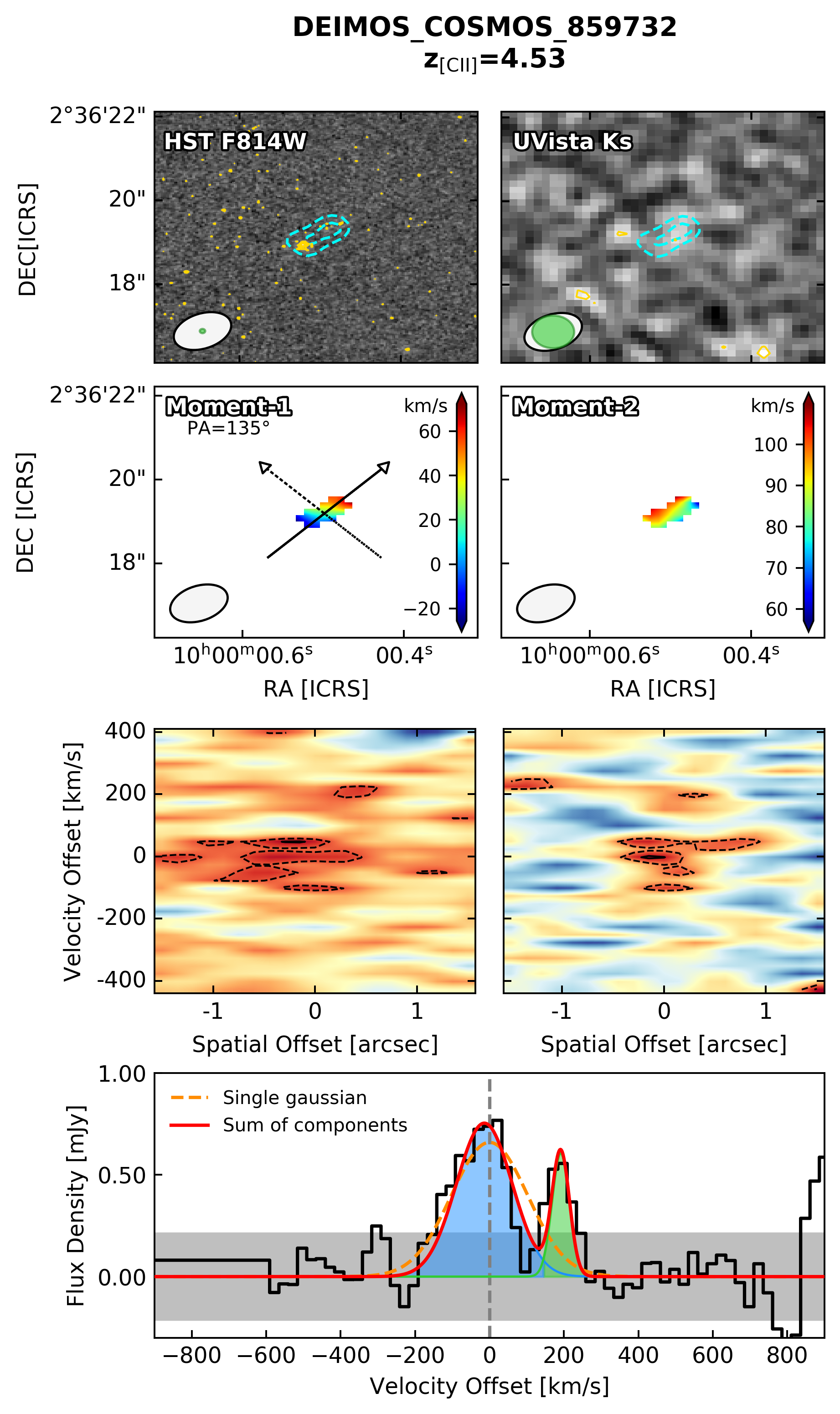}}
	\subfigure{\includegraphics[width=\columnwidth,height=15.1cm]{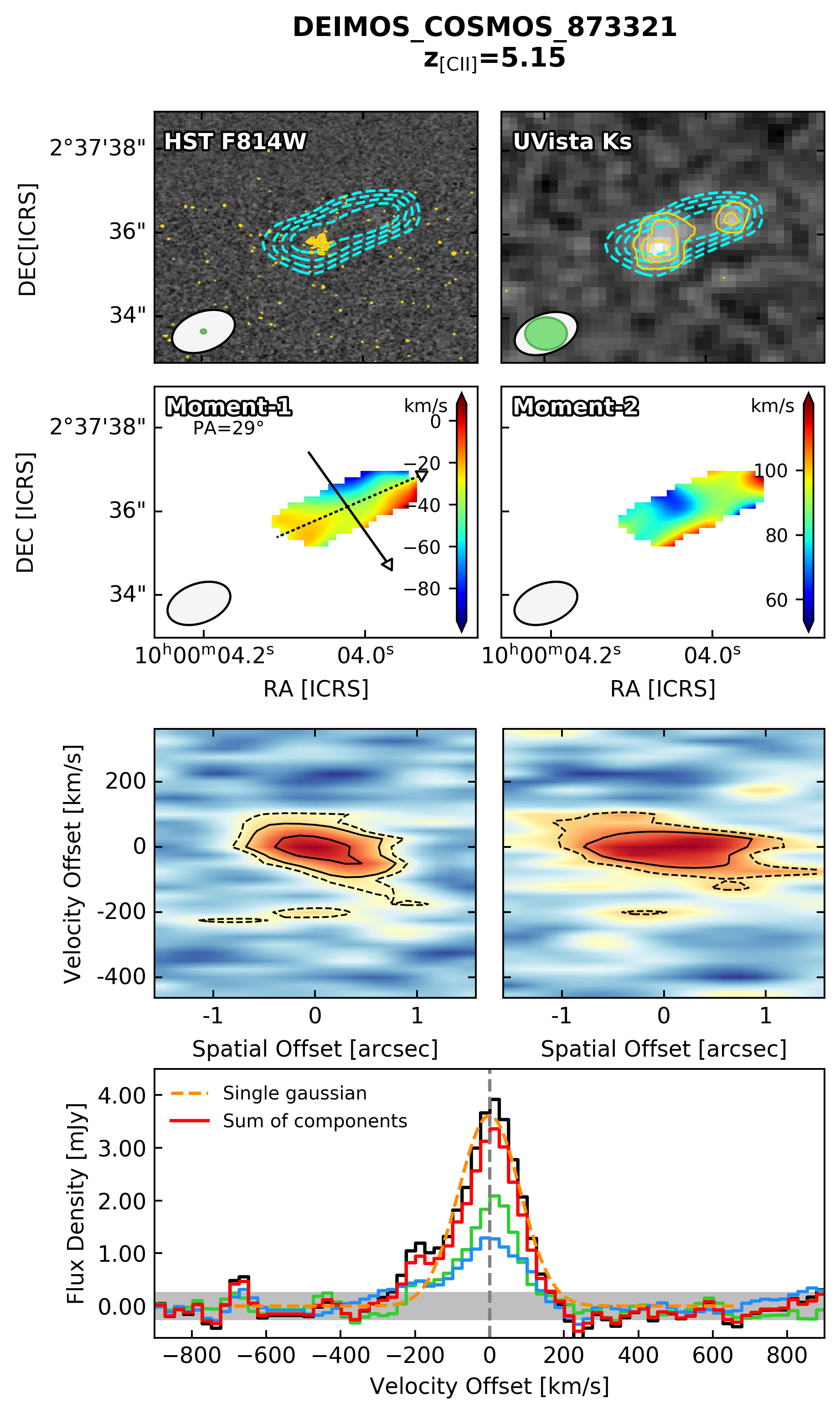}}
    \caption{Moment maps, PVDs and spectral decomposition for the ALPINE mergers, as described in Figure \ref{fig:merger_appendix1}.}
    \label{fig:merger_appendix9}
\end{center}
\end{figure*}

\begin{figure*}[t]
\begin{center}
	\subfigure{\includegraphics[width=\columnwidth,height=15.1cm]{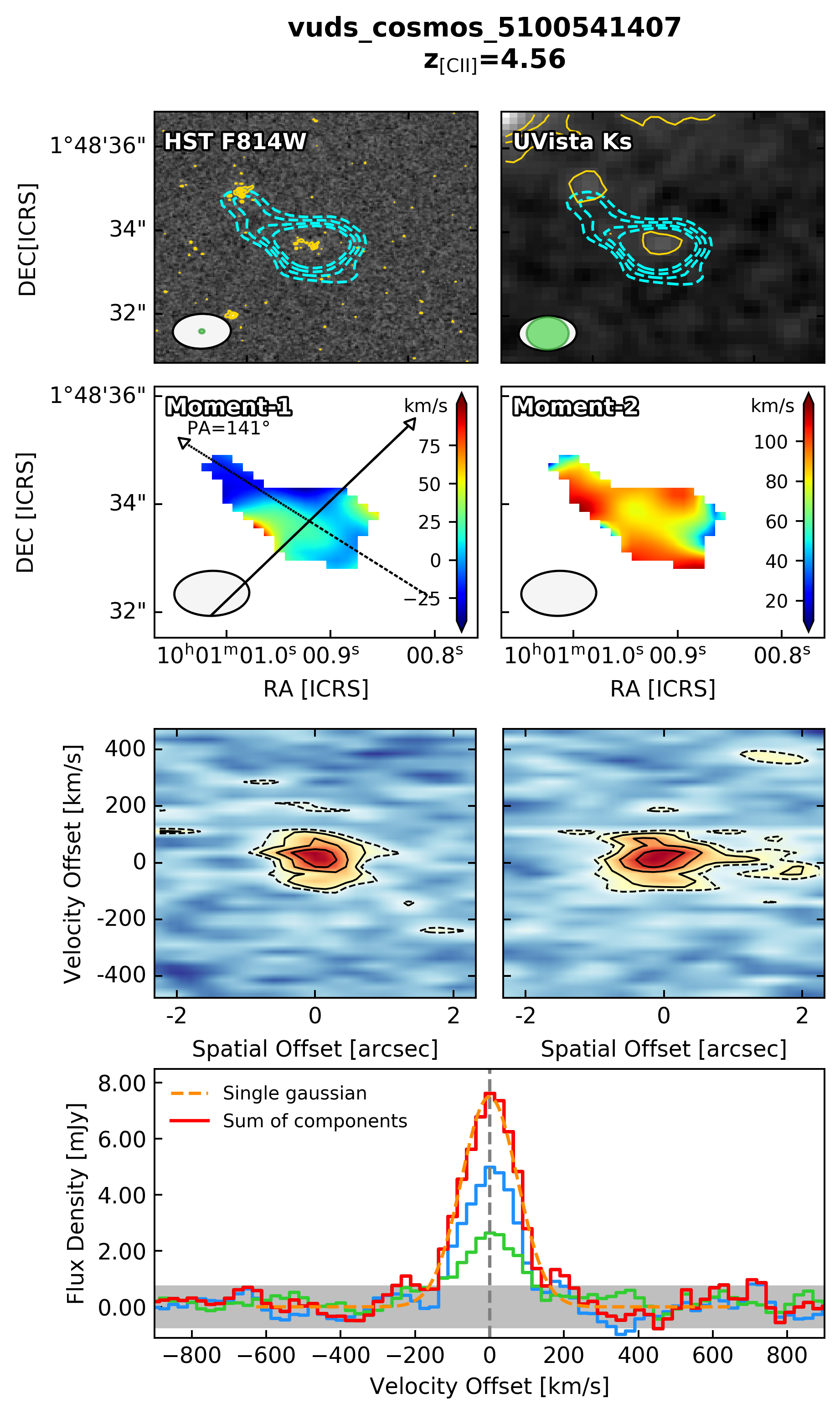}}
	\subfigure{\includegraphics[width=\columnwidth,height=15.1cm]{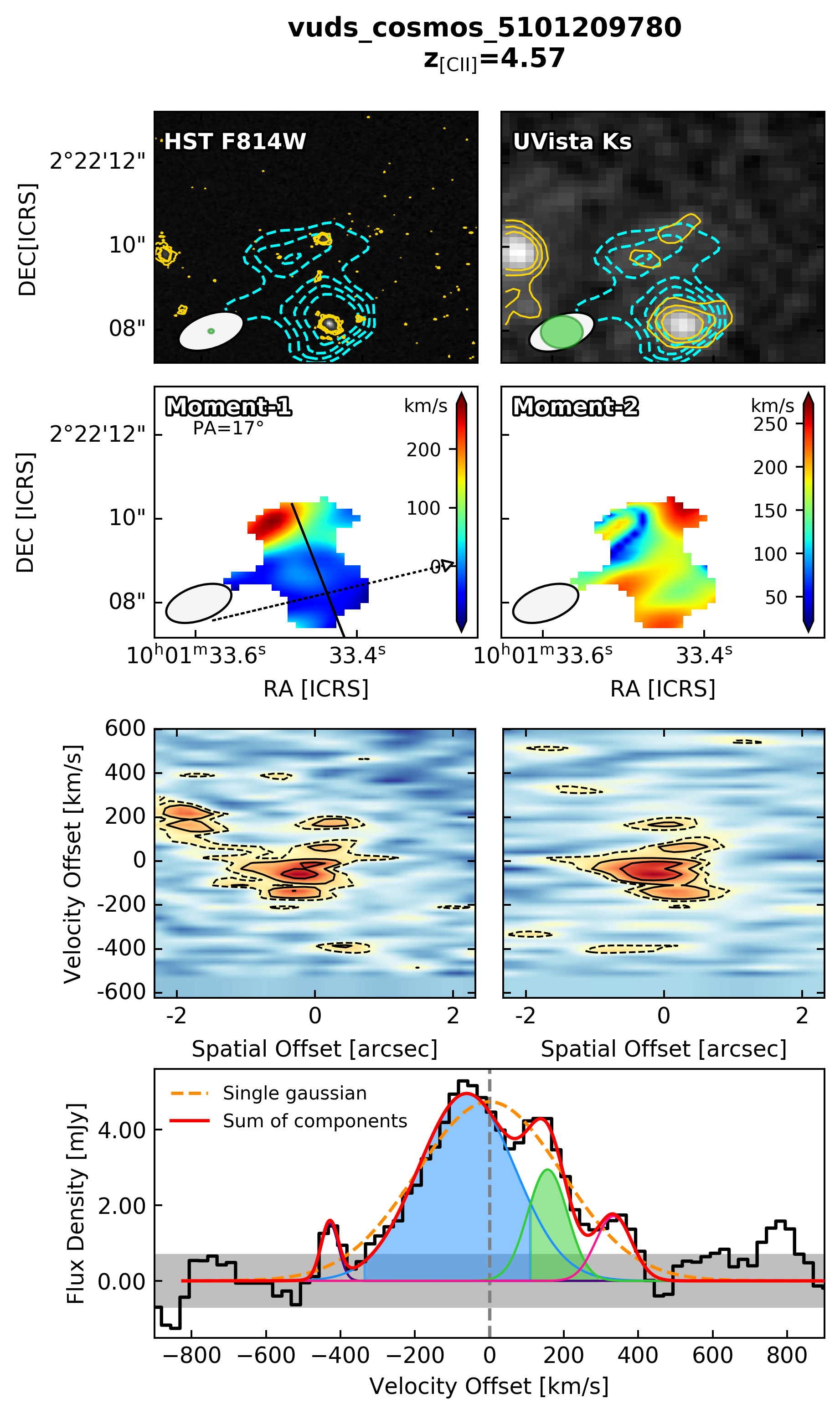}}
    \caption{Moment maps, PVDs and spectral decomposition for the ALPINE mergers, as described in Figure \ref{fig:merger_appendix1}.}
    \label{fig:merger_appendix10}
\end{center}
\end{figure*}

\begin{figure}[t]
\begin{center}
	\subfigure{\includegraphics[width=\columnwidth,height=15.1cm]{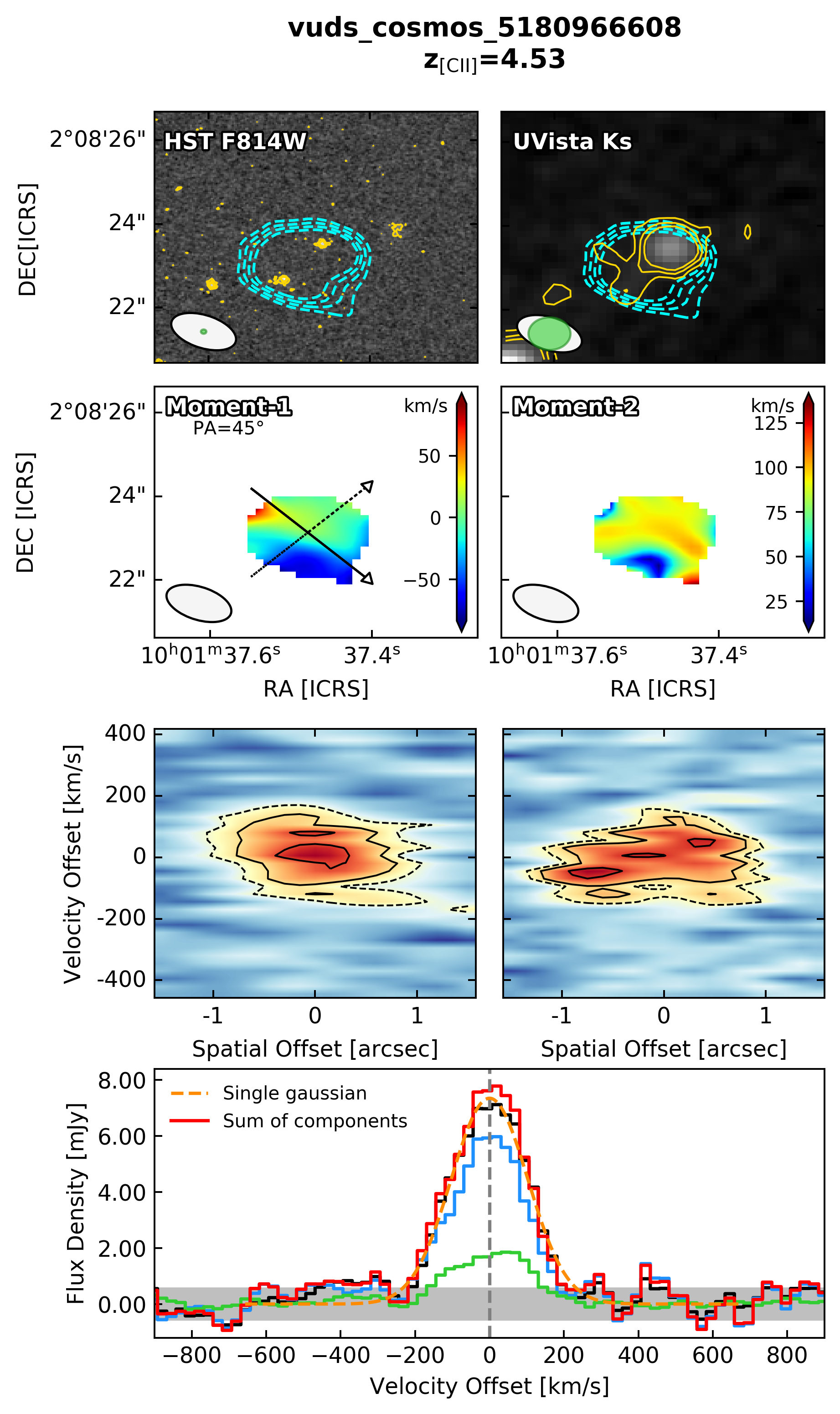}}
    \caption{Moment maps, PVDs and spectral decomposition for the ALPINE mergers, as described in Figure \ref{fig:merger_appendix1}.}
    \label{fig:merger_appendix11}
\end{center}
\end{figure}

\end{appendix}


\end{document}